\documentclass[aps,prd,a4paper,twocolumn,10pt]{revtex4-1}
\usepackage[colorlinks,linkcolor=blue,anchorcolor=blue,citecolor=blue,urlcolor=blue]{hyperref}
\usepackage{amsmath,amsfonts,amssymb}
\usepackage{color,graphicx,epsfig,subfigure}
\usepackage{diagbox,txfonts,acronym,indentfirst,booktabs}
\allowdisplaybreaks
\newcommand{\bea}{\begin{eqnarray}}
\newcommand{\eea}{\end{eqnarray}}
\newcommand{\bal}{{\bar{l}}}
\newcommand{\baz}{{\bar{z}}}
\newcommand{\cL}{{\cal L}}
\newcommand{\cM}{{\cal M}}
\newcommand{\cO}{{\cal O}}
\newcommand{\cP}{{\cal P}}
\newcommand{\cU}{{\cal U}}
\newcommand{\mSun}{{\rm M}_{\odot}}
\newcommand{\nn}{\nonumber}
\newcommand{\pd}{\partial}
\newcommand{\td}{\tilde}

\def\brk#1{\left\langle#1\right\rangle}
\newcommand{\TRC}{MOE Key Laboratory of TianQin Mission, TianQin Research Center for
Gravitational Physics \& School of Physics and Astronomy, Frontiers
Science Center for TianQin, Gravitational Wave Research Center of
CNSA, Sun Yat-sen University (Zhuhai Campus), Zhuhai 519082, China.}
\begin{document}
%%%%%%%%%%%%%%%%%%%%%%%%%%%%%%%%%%%%%%%%%%%%%%%%%%%%%%%%%%%%%%%%
\title{Detecting the gravitational wave memory effect with TianQin}

\author{Shuo Sun}
\author{Changfu Shi}
\email{Corresponding author. Email: shichf6@mail.sysu.edu.cn}
\author{Jian-dong Zhang}
\author{Jianwei Mei}

\affiliation{\TRC}
%%%%%%%%%%%%%%%%%%%%%%%%%%%%%%%%%%%%%%%%%%%%%%%%%%%%%%%%%%%%%%%%
\newacro{GR}{general relativity}
\newacro{GW}{gravitational wave}
\newacro{MBHB}{massive black hole binary}
\newacro{BH}{black hole}
\newacro{ASD}{amplitude spectral density}
\newacro{PSD}{power spectral density}
\newacro{BMS}{Bondi-Metzner-Sachs}
\newacro{SNR}{signal-to-noise ratio}
\newacro{PN}{post-Newtonian}
\newacro{CCE}{Cauchy-characteristic extraction}
\newacro{IMR}{Inspiral-Merger-Ringdown}
%%%%%%%%%%%%%%%%%%%%%%%%%%%%%%%%%%%%%%%%%%%%%%%%%%%%%%%%%%%%%%%%
\date{\today}
%%%%%%%%%%%%%%%%%%%%%%%%%%%%%%%%%%%%%%%%%%%%%%%%%%%%%%%%%%%%%%%%
\begin{abstract}
The gravitational wave memory effect is a prediction of general relativity. The presence of memory effect in gravitational wave signals not only provides the chance to test an important aspect of general relativity, but also represents a potentially non-negligible contribution to the waveform for certain gravitational wave events. In this paper, we study the prospect of detecting the gravitational wave memory effect directly with the planned space-based gravitational wave detector --- TianQin. We find that during its 5 years operation, for the gravitational wave signals that could be detected by TianQin, about $0.5\sim2.0$ signals may contain displacement memory effect with signal-to-noise ratios (SNRs) greater than 3. This suggests that the chance for TianQin to detect the displacement memory effect directly is low but not fully negligible. In contrast, the chance to detect the spin memory is negligible. We also study that in which parameter space, the memory effect is expected to be significant in waveform modeling.
\end{abstract}
%%%%%%%%%%%%%%%%%%%%%%%%%%%%%%%%%%%%%%%%%%%%%%%%%%%%%%%%%%%%%%%%
\maketitle

\section{Introduction}
%%%%%%%%%%%%%%%%%%%%%%%%%%%%%%%%%%%%%%%%%%%%%%%%%%%%%%%%%%%%%%%%

The observations of \acp{GW} from binary black hole mergers  \cite{LIGOScientific:2018mvr,LIGOScientific:2020ibl,LIGOScientific:2021djp} have opened a new window to observe the universe, which not only promise a deeper understanding of compact objects in the universe \cite{LIGOScientific:2018jsj,LIGOScientific:2020kqk,LIGOScientific:2021psn} but also provide a new way to test general relativity \cite{LIGOScientific:2019fpa,LIGOScientific:2020tif,LIGOScientific:2021sio}. Both the detection of \acp{GW} and the testing of general relativity require accurate waveforms of \acp{GW}. Among all the contributions to the \ac{GW} waveforms, the \ac{GW} memory effect is particularly interesting not only in that it can potentially be directly observed and thus serves as an important test of general relativity, but also in that it depends on the entire past history of the gravitational systems \cite{Blanchet:1992br,Blanchet:2013haa}.

Under real physical conditions, the spacetime background seen by an observer near the null infinity before and after a flux of \ac{GW} radiation is not purely vacuum, but contains \ac{GW} sources at a long distance. Due to the loss of energy and other conserved charges of the \ac{GW} sources, the radiation of \acp{GW} typically leads to a permanent change to the spacetime, and this has been known as the {\it linear} memory effect since the early 1970s \cite{zel1974,throne1987}. A second type of memory effect was later discovered by Christodoulou \cite{PhysRevLett.67.1486} and has been called the {\it nonlinear} memory effect, for it involves the contribution of the \ac{GW} strain at the quadratic order. The Christodoulou effect does not appear as a change of the charges of the \ac{GW} sources but is still related to the permanent change of the background spacetime. It has been suggested to rename the {\it linear} and {\it nonlinear} memory effect as the {\it ordinary} and {\it null} memory effect, respectively, to be more clear on what they truly represent \cite{Bieri:2013ada}.

Since the memory effect is a persistent change of the spacetime, it is natural to relate it to the symmetries and the corresponding charges that characterize different spacetimes (see \cite{Nichols:2018qac} for some early references). However, it was only in the past few years that the relations among the various types of memory effect and the various types of \ac{BMS} transformations \cite{Bondi:1962px,Sachs:1962wk,deBoer:2003vf,Barnich:2009se,Barnich:2010ojg,Kapec:2014opa,Kapec:2016jld,He:2017fsb} were elucidated and established \cite{Strominger:2014pwa,Pasterski:2015tva,Nichols:2018qac,Compere:2019gft}. Such understanding has enabled systematic calculations of the memory effect using the so called \ac{BMS} flux-balance laws \cite{Barnich:2011mi,Flanagan:2015pxa,Nichols:2017rqr,Compere:2019gft,Mitman:2020pbt,Mitman:2020bjf}.

The detection of memory effect has attracted much attention in recent years. Attempts have been made to calculate the memory effect for cosmic strings \cite{Jenkins:2021kcj} and core-collapse supernovae \cite{Mukhopadhyay:2021zbt}. There have also been calculation of memory effect in alternative theories of gravitation, such as scalar tensor theory \cite{Du:2016hww}, Brans-Dicke theory \cite{Seraj:2021qja,Tahura:2021hbk} and Chern-Simons modified gravity \cite{Hou:2021bxz}. Using the memory effect to test alternative theories of gravitation has also been studied, such as for the Brans-Dicke theory with screening \cite{Koyama:2020vfc}.

The strategies for detecting the \ac{GW} memory effect were first proposed by Braginsky and Thorne in the 1980s \cite{zel1985,throne1987}. Since the first detection of \acp{GW} by LIGO in 2015 \cite{LIGOScientific:2016aoc}, many works about detecting memory effect with current and future \ac{GW} detectors have appeared \cite{10.1111/j.1745-3933.2009.00758.x,vanHaasteren:2009fy,Pshirkov:2009ak,Cordes:2012zz,madison,NANOGrav:2015xuc,Lasky:2016knh,McNeill:2017uvq,Nichols:2017rqr,Divakarla:2019zjj,Boersma:2020gxx,Khera:2020mcz,Hubner:2019sly,Hubner:2021amk,Islam:2021old,Zhao:2021hmx}. It has been found that the memory effect is difficult to detect directly with the current ground-based \ac{GW} detectors, but the future ground-based detectors such as Cosmic Explorer \cite{LIGOScientific:2016wof} and Einstein Telescope \cite{Punturo:2010zz} may be able to detect the memory effect  \cite{Yang:2018ceq}. The memory effect produced by the massive BBH systems can fall in the most sensitive frequency band of space-based detectors, and thus have better chance at revealing the memory effect with space-based detectors. Indeed, the prospect of detecting memory effect with the planned space-based \ac{GW} detector LISA has been studied in \cite{Islo:2019qht}, and it has been shown that LISA can directly detect the displacement memory from the merger of \acp{MBHB}. It is also possible to detect memory effect with pulsar timing array \cite{10.1111/j.1745-3933.2009.00758.x,vanHaasteren:2009fy,Pshirkov:2009ak,Cordes:2012zz,madison,NANOGrav:2015xuc}.

TianQin is a space-based \ac{GW} detector targeting \acp{GW} in the frequency band $10^{-4}$ Hz$\sim 1$ Hz \cite{TianQin:2015yph,TianQin:2020hid}. Among the promising scientific discovery potential of TianQin in many different directions \cite{Hu:2017yoc,Fan:2020zhy,Liu:2020eko,Liu:2021yoy,Huang:2020rjf,Liang:2021bde,Shi:2019hqa,Bao:2019kgt,Zi:2021pdp,Zhu:2021bpp,Shi:2022qno,Xie:2022wkx}, the detection of \acp{MBHB} is an outstanding one \cite{haitian,Feng:2019wgq}. Motivated by the known result on LISA, it is likely that TianQin can also detect memory effect directly. But the orbit, the constellation size and motion of TianQin are quite different from those of LISA, and so are the sensitive frequency band and the response of the detector to \acp{GW}. So it is necessary to study the related problems for TianQin independently, especially when it comes to the question of when the memory effect will become a must-consider factor during science and data analysis.

In this paper, we study the prospect of directly detecting the gravitational wave memory effect with TianQin and also determine the region of parameter space in which the contribution of memory effect is potentially non-negligible.

The paper is organized as following. In Sec. \ref{BMSBL}, we recall the basic results on memory effect and on the calculation of the corresponding waveform corrections. In Sec. \ref{SNR}, we study the potential of detecting memory effect directly with TianQin. In Sec. \ref{MisM}, we study the parameter space in which the memory effect could become an adverse factor if not properly taken into consideration. Finally, we summarize in Sec. \ref{sum}.

\section{memory effect}\label{BMSBL}

In this section, we recall some of the basics of memory effect and the calculation of the corresponding waveform and we set $G=c=1$. The purpose is to have all the relevant nomenclature clearly defined and to keep the logic in a self-contained manor. There is nothing particularly new in this section and we mainly follow the treatment and notation of \cite{Flanagan:2015pxa,Mitman:2020pbt}.

Physically, there are three different types of memory effect: the displacement memory \cite{zel1974,throne1987}, which causes the relative displacement between two test masses to change after the passage of \acp{GW}, the spin memory \cite{Pasterski:2015tva}, which causes the change in the relative time delay of two free-falling test masses that are initially on anti-orbital trajectories, and the center-of-mass memory effect \cite{Nichols:2018qac}, which causes the change in the relative time delay of two free-falling test masses that are initially on anti-parallel trajectories. All three types of memory effect contain the previously mentioned {\it linear} ({\it ordinary}) and {\it nonlinear} ({\it null}) contributions.

The physical consequences of the memory effect described above are easy to visualize, but they do not constitute the best ways for detection. Due to the intrinsic weakness of \acp{GW}, the permanent change in the displacement and time delays in various detector dependent setup is also very small and cannot be directly measured with any currently known technology. The practically more relevant consequence of memory effect is the shifts and corrections to the \ac{GW} waveforms, and it is the knowledge about the waveform correction that offers the best chance for a detection.

Earlier methods for computing the waveform of memory effect include the \ac{PN} approximation and various kinds of postprocessing techniques based on existing numerical waveforms \cite{Favata:2008yd,Favata:2009ii,Favata:2010zu,Talbot:2018sgr,Nichols:2017rqr,Nichols:2018qac}. We will follow \cite{Mitman:2020pbt} to calculate the memory waveforms in this paper.

Following \cite{Flanagan:2015pxa,Mitman:2020pbt}, the Bondi-Sachs metric of a generic asymptotically flat spacetime containing \acp{GW} can be written as \cite{Bondi:1962px,Sachs:1962wk},
%%%
\bea ds^2&=&-e^{2\beta}\Big(Udu^2+2dudr)\nn\\
&&+r^2\gamma_{AB}(d\theta^A-\cU^Adu)(d\theta^B-\cU^Bdu)\,,\label{Bondi-Sachs}\eea
%%%
where $u=t-r$ is the retarded time, $\theta^A\in\{\theta^1,\theta^2\}$ are coordinates on the two-sphere, and $U$, $\beta$, $\cU^A$ and $\gamma_{AB}$ are functions of $u$, $r$ and $\theta^A$. A special feature of the Bondi-Sachs metric is
%%%
\bea g_{rr}=g_{rA}=0\,.\label{BMSBC1}\eea
%%%
The radial coordinate $r$ is determined by requiring that the determinant of $\gamma_{AB}$ is the same of that of the fixed round metric, denoted as $q_{AB}$, on the unit two-sphere,
%%%
\bea \det(\gamma_{AB})=\det(q_{AB})\,,\label{BMSBC2}\eea
%%%
where, if written in the usual spherical coordinates,
%%%
\bea q_{AB}d\theta^Ad\theta^B=d\theta^2+\sin^2\theta d\phi^2\,.\eea
%%%
Note the metric $q_{AB}$ will be used to raise and lower all capital Latin indices (e.g., $A,B$) throughout this paper.

The functions $U$, $\beta$, $\cU^A$ and $\gamma_{AB}$ in metric (\ref{Bondi-Sachs}) can be determined by source-less Einstein equations as
%%%
\bea U&=&1-\frac{2m}r-\frac{2M}{r^2}+\cO(r^{-3})\,,\nn\\
\beta&=&-\frac{C_2}{32r^2}-\frac{C_2^2}{48r^3}+\cO(r^{-4})\,,\nn\\
\gamma_{AB}&=&q_{AB}+\frac{C_{AB}}r+\cO(r^{-3})\,,\nn\\
\cU^A&=&-\frac{D_BC^{AB}}{2r^2}+\frac1{r^3}\Big[-\frac23N^A+\frac1{16}D^AC_2\nn\\
&&\qquad\qquad\qquad+\frac12C^{AC}D^BC_{BC}\Big]+\cO(r^{-4})\,,\label{BMSBC3}\eea
%%%
where all coefficient functions on the right-hand sides are functions of $(u,\theta^A)$ only, $D^A$ is the covariant derivative associated with $q_{AB}$, and
%%%
\bea C_2&=&C_{AB}C^{AB}\,,\quad q^{AB}C_{AB}=0\,,\nn\\
\dot{m}&=&-\frac18\dot{C}_{AB}\dot{C}^{AB}+\frac14D_AD_B\dot{C}^{AB}\,,\nn\\
\dot{N}_A&=&D_Am+\frac14D_BD_AD_CC^{BC}-\frac14D_BD^BD^CC_{CA}\nn\\
&&+\frac14D_B(\dot{C}^{BC}C_{CA})+\frac12D_B\dot{C}^{BC}C_{CA}\,,\nn\\
\dot{M}&=&-\frac12D_AD^A\Big(m+\frac38C_{BC}\dot{C}^{BC}\Big)+\frac13D_A\dot{N}^A\nn\\
&&-\frac14(D_A\dot{C}_{BC})\Big(D^AC^{BC}-2D^BC^{AC}\Big)\nn\\
&&+\frac18\dot{C}_{AB}\Big(D_CD^C+\frac32\Big)C^{AB}\,,\eea
%%%
where $C_{AB}$ is shear tensor, $m$ is the Bondi mass aspect, $M$ is the function of $(u, \theta^{A})$ and $N_{A}$ is Bondi angular-momentum aspect. Here an overdot means a derivative with respect to $u$. Follow \cite{Mitman:2020bjf,Mitman:2020pbt,Flanagan:2015pxa}, we rewrite $N_{A}$ in terms of $\hat{N}_{A}$, $\hat{N}_{A}$ is
%%%
\bea
\hat{N}_A &=& N_A-u D_A m \nn\\
&-&\frac{1}{16} D_A\left(C_{B C} C^{B C}\right)-\frac{1}{4} C_{A B} D_C C^{B C} .
\eea
%%%
The solution solves Einstein's equations to the $\cO(r^{-3})$ order, i.e.,
%%%
\bea R_{\mu\nu}dx^\mu dx^\nu\leq\cO(r^{-4})\,,\label{EFE-O3}\eea
%%%
where $R_{\mu\nu}$ is the Ricci tensor of the Bondi-Sachs metric (\ref{Bondi-Sachs}), $dx^\mu\in\{du,dr,d\theta^1,d\theta^2\}$, and we take
%%%
\bea du,dr\sim\cO(r^0)\,,\quad d\theta^1,d\theta^2\sim\cO(r^{-1})\,.\eea
%%%
All dynamical properties of the solution are encoded in the two unconstrained functions of the shear $C_{AB}\,$.

Solutions with all possible values of $C_{AB}$ form the solution space that preserves the structure of the Bondi-Sachs metric with the boundary conditions (\ref{BMSBC1}), (\ref{BMSBC2}) and (\ref{BMSBC3}). The solutions are related to each other through the \ac{BMS} transformations,
%%%
\bea \xi&=&f\pd_u+\Big[Y^A-\frac1rD^Af+\frac1{2r^2}C^{AB}D_Bf+\cO(r^{-3})\Big]\pd_A\nn\\
&&-\Big[\frac12rD_AY^A-\frac12D^AD_Af+\frac1{4r}\Big(D_AC^{AB}\Big)D_Bf\nn\\
&&\quad+\frac1{4r}D_A(D_BfC^{AB})+\cO(r^{-2})\Big]\pd_r\,,\label{BMS-xi}\eea
%%%
where
%%%
\bea f=\alpha(\theta^A)+\frac12uD_BY^B(\theta^A)\,.\eea
%%%
Here $\alpha(\theta^A)$ is unconstrained and $Y^A(\theta^B)$ must obey the conformal Killing equation on the unit two-sphere,
%%%
\bea D_AY_B+D_BY_A=q_{AB}D_CY^C\,.\eea
%%%
Under the action of (\ref{BMS-xi}), the shear transforms as
%%%
\bea\delta C_{AB}&=&f\dot{C}_{AB}-2D_AD_Bf+q_{AB}D_CD^Cf\nn\\
&&+\frac12(D_CY^C)C_{AB}+\cL_{\vec{Y}}C_{AB}\,,\eea
%%%
where $\cL_{\vec{Y}}$ is the Lie derivative with $\vec{Y}=\{Y^1,Y^2\}\,$. Other functions in the solution will transform accordingly, and we refer to \cite{Flanagan:2015pxa} for details on the transformation of the Bondi mass aspect $m$ and the angular-momentum aspect $N^A$.

The \ac{BMS} transformations (\ref{BMS-xi}) are examples of asymptotic symmetries which are associated with conserved charges that characterize different solutions in the solution space. In general, only the difference between the charges of a pair of solutions is defined (see \cite{Compere:2018aar} for a reader friendly exposition). In \ac{GW}-free and matter-free cases, however, it is possible to write the charges for each solution directly,
%%%
\bea Q&=&\frac1{16\pi}\int d^2\Omega\Big[4\alpha m-2u_0Y^AD_Am+2Y^AN_A\nn\\
&&\qquad\qquad-\frac18Y^AD_AC_2-\frac12Y^AC_{AC}D_BC^{BC}\Big]\,,\eea
%%%
where the integral is over a two-sphere near the null infinity and at the retarded time $u_0\,$.

In the usual spherical coordinates, the standard \ac{BMS} algebra is generated by
%%%
\bea \alpha&=&t^0-t^in_i+\sum_{\ell=2}^\infty\sum_{m=-\ell}^\ell\alpha_{\ell m}Y_{\ell m}\,,\nn\\
Y^A&=&w^{0i}e^A_i+w^{ij}e^A_{[i}n_{j]}\,,\label{BMS-xi0}\eea
%%%
where $t^\mu=\{t^0,t^i\}$ and $w^{0i}\,,\,w^{ij}$ are free parameters, $n_i=\{\sin\theta\cos\phi\,,\, \sin\theta\sin\phi\,,\, \cos\theta\}$, $e^A_i=D^An_i\,$ and $Y_{\ell m}$ is the usual spherical harmonics. The corresponding charges are
%%%
\bea Q=-P^\mu t_\mu+\frac12J^{\mu\nu}w_{\mu\nu}+\frac1{4\pi}\sum_{\ell=2}^\infty\sum_{m=-\ell}^\ell\alpha_{\ell m}\cP^\ast_{\ell m}\,.\eea
%%%
Here $P^\mu$ is the usual four-momentum, $J^{\mu\nu}$ is the usual angular momentum, and $\cP_{\ell m}$ are the {\it supermomenta} associated with the supertranslations, i.e., the $\ell\geq2$ terms in $\alpha$. So the standard \ac{BMS} algebra is simply the Poincar\'{e} algebra extended with the supermomenta.

The \ac{BMS} algebra can be extended if one allows for singular Killing vectors on the future null infinity \cite{Barnich:2010eb,Barnich:2011mi}. To do this, it is more convenient to use the complex stereographic coordinates ($z,\baz$) with $z=\cot(\theta/2)e^{i\phi}$ instead of the usual spherical coordinates. In this case, the conformal Killing equation requires $Y^z=Y^z(z)$ and $Y^\baz=Y^\baz(\baz)$, i.e., both to be meromorphic functions of their arguments. The corresponding Killing vector can be expanded with the set of vectors,
%%%
\bea l_m=-z^{m+1}\pd_z\,,\quad \bal_m=-\baz^{m+1}\pd_\baz\,,\eea
%%%
where $m\in\mathbb{Z}\,$. Among the vectors, $l_0\,,\,l_{\pm1}\,,\,\bal_0$ and $\bal_{\pm1}$ have already appeared in Eq. (\ref{BMS-xi0}), while the rest are called the {\it superrotations}. Although the Killing vectors are singular for the superrotations, the corresponding charges have been shown to be finite \cite{Flanagan:2015pxa}. The charges of the superrotations can be split to the spin parts, corresponding to operations analogues to the rotations in the Lorentz group, and the center-of-mass parts, corresponding to operations analogues to the Lorentz boost.

It has been realized that the displacement memory is related to changes in the supermomenta and the corresponding flux, the spin memory is related to changes in the spin parts of the super-angular momenta and the corresponding flux, and the center-of-mass memory effect is related to changes in the center-of-mass parts of the super-angular momenta and the corresponding flux \cite{Strominger:2014pwa,Pasterski:2015tva,Nichols:2018qac,Compere:2019gft}. As we have mentioned earlier, however, that for a true detection of the memory effect we have to focus on the waveforms. From this perspective, the displacement memory is characterized by a finite change in the \ac{GW} strain, the spin memory is characterized by a change in the time integral of the magnetic part of the \ac{GW} strain, while the center-of-mass memory effect is characterized by a change in the time integral of certain expression that has the dimension of the \ac{GW} strain \cite{Nichols:2018qac}.

It was only recently that the displacement and spin memorys are successfully captured in the SXS catalog \cite{sxs,Boyle:2019kee} by using the \ac{CCE}, in which one has to evolve a world tube produced by a Cauchy evolution to asymptotic infinity and then to extract the \ac{GW} strain \cite{Moxon:2020gha}. In \cite{Mitman:2020pbt}, the authors use the \ac{BMS} flux-balance law to calculate the memory effect from numerical waveforms known at the null infinity, and they find that the results agree with those from \ac{CCE} well.

In the treatment of \cite{Mitman:2020pbt}, the key quantity to calculate is the leading order spin-weight -2 \ac{GW} strain,
%%%
\bea h=\frac12\bar{q}^A\bar{q}^BC_{AB}=\sum_{\ell\geq2}\sum_{|m|\leq\ell}h_{\ell m} ~_{-2}Y_{\ell m}(\theta,\phi)\,,\label{gwstrain}\eea
%%%
where $q^A$ is the complex dyads \cite{Moxon:2020gha}, $q^A=-\{1,i\csc\theta\}\,$, $~_{-2}Y_{\ell m}(\theta,\phi)$ is spin-weighted -2 spherical harmonics and the angles $(\theta, \phi)$ are the inclination and the reference phase of the source respectively. In reverse, one has
%%%
\bea C_{AB}=\frac12\Big(q_Aq_Bh+\bar{q}_A\bar{q}_B\bar{h}\Big)\,.\eea
%%%
Splitting the shear into the electric and the magnetic components,
%%%
\bea C_{AB}=\Big(D_AD_B-\frac12q_{AB}D_CD^C\Big)\Phi+\epsilon_{C(A}D_{B)}D^C\Psi\,,\eea
%%%
where $\epsilon_{AB}$ is the Levi-Civita tensor on the two-sphere, $\Phi$ and $\Psi$ are the electric and magnetic parts of shear tensor $C_{AB}$.
one has for the electric and magnetic memory effect, respectively,
%%%
\bea
\Delta J^{(E)}&=&\frac12\bar{q}^A\bar{q}^B\Delta C^{(E)}_{AB}=\frac12\bar\eth^2\Delta\Phi\nn\\
&=&\frac{1}{2}\bar{\eth}^{2}\mathfrak{D}^{-1}\left[\Delta m+\frac{1}{4}\int^{u}_{-\infty}|\dot{h}|^{2}du\right]\,,\label{dis}\\
\Delta J^{(B)}&=&\frac12\bar{q}^A\bar{q}^B\Delta C^{(B)}_{AB}=-\frac12i\bar\eth^2\Delta\Psi\nn\\
&=&\frac{1}{2}i\bar{\eth}^{2}\mathfrak{D}^{-1}D^{-2}\text{Im}\Bigg\{\bar{\eth}(\partial_{u}\hat{N})\nn\\
&&+\left.\frac{1}{8}\left[\eth(3h\bar{\eth}\dot{\bar{h}}-3\dot{h}\bar{\eth}\bar{h}+\dot{\bar{h}}\bar{\eth}h-\bar{h}\bar{\eth}\dot{h})\right]\right\}.
\label{spin}
\eea
%%%
The $\Delta m = m(u)-m(-\infty)$ and $\hat{N}= q^{A}\hat{N}_{A}$. In Eq. (\ref{dis}) and Eq. (\ref{spin}), $\rm Im$ means the imaginary part of the function, $D^{2}=\bar{\eth}\eth$ is the usual Laplacian on the two-sphere, $\mathfrak{D}=\frac{1}{8}D^{2}(D^{2}+2)$, and $\eth$ and $\bar{\eth}$ are the spin-weight operators in the Newman-Penrose convention \cite{roger1},
%%%
\bea &&\eth_{s}Y_{\ell m}=+\sqrt{(\ell-s)(\ell+s+1)}_{s+1}Y_{\ell m}\,,\nn\\
&&\bar{\eth}_{s}Y_{\ell m}=-\sqrt{(\ell+s)(\ell-s+1)}_{s-1}Y_{\ell m}\,.\eea
%%%
For the spin weight 0 spherical harmonics $Y_{\ell m}\,$,
%%%
\bea D^{2}Y_{\ell m}&=&-\ell(\ell+1)Y_{\ell m}\,,\nn\\
\mathfrak{D}Y_{\ell m}&=&\frac{1}{8}(\ell+2)(\ell+1)\ell(\ell-1)Y_{\ell m}\,.\label{eq4b}\eea
%%%
The $\ell \leq1$ modes of $Y_{\ell m}$ are in the kernel of $\mathfrak{D}$, so $\mathfrak{D}^{-1}$ is defined by projecting out the $\ell \leq1$ modes \cite{Mitman:2020pbt},
%%%
\bea\mathfrak{D}^{-1}Y_{\ell m}=\left\{\begin{matrix}0&:&\ell\leq 1\,,\cr
\left[\frac{1}{8}(\ell +2)(\ell+1)\ell(\ell-1)\right]^{-1}Y_{\ell m}&:&\ell\geq2\,.\end{matrix}\right.\label{mkD}\eea
%%%

The first term in the right hand side of Eq. (\ref{dis}) is the ordinary part of the displacement memory, and the second term is the null part. The Bondi mass aspect $m$ can be written in terms of Weyl scalar $\Psi_2$ and $h$ as
%%%
\bea m=-{\rm Re}\left[\Psi_2+\frac{1}{4}\dot{h}\bar{h}\right]\,.\eea
%%%
Similarly, the first term in the right hand side of Eq. (\ref{spin}) is the ordinary part of spin memory mode and the second part is the null part. $\hat{N}$ can be written in terms of Weyl scalar $\Psi_1$ and $h$ as
%%%
\bea \hat{N}=2\Psi_1-u\eth m - \frac{1}{8}\eth(h\bar{h})-\frac{1}{4}\bar{h}\eth h\,.\eea
%%%

As discussed in \cite{Mitman:2020pbt}, $\Delta J^{(E)}$ is dominated by the displacement memory, $\int\Delta J^{(B)}(u)du$ is the spin memory, while the explicit formula for computing the center-of-mass memory by using the \ac{BMS} flux-balance law is not known yet. The detection of the center-of-mass memory effect using \ac{PN} waveform has been studied in \cite{Nichols:2018qac}. It has been found that, for a future ground-based detector such as the Einstein-Telescope, which can achieve a \ac{SNR} at the order $\cO(10^3)$ for a signal like GW150914, can only detect the ordinary part of the center-of-mass memory effect with a \ac{SNR} several orders of magnitude below unit. The prospect for space-based detector will be similar, so the chance for detection is negligible for TianQin. Therefore, in this work, we only consider the displacement memory and the spin memory mode.

\section{Waveform model}

We use the IMRPhenomXHM waveform \cite{Garcia-Quiros:2020qpx,Colleoni:2020tgc,Pratten:2020fqn}  to calculate the displacement memory strain with Eq. (\ref{dis}) and the spin memory strain with Eq. (\ref{spin}),
%%%
\bea h_{\rm dis}=\Delta J^{(E)}\,,\quad h_{\rm spin}=\Delta J^{(B)}\,.\eea
%%%
IMRPhenomXHM can fastly produce relatively accurate waveforms containing the most dominant modes, such as (2, 2), (2,1), (3,3), (3,2) and (4,4), which are expected to contribute to the memory effect. It should also be noticed that IMRPhenomXHM is an aligned-spin model and has only been calibrated with NR for mass ratios $q=m_1/m_2\in[1,18]$, but it has been noted in \cite{Garcia-Quiros:2020qpx} that the waveforms still have reasonable accuracy outside of the calibration region. So we will use waveforms with mass ratios in the region $q\in[1,23]$ in this work.

In our calculations, we use IMRPhenomXHM to generate the frequency domain waveform first, and then use the inverse Fourier transform method from LALsimulation\cite{LALSuite} to generate the time domain waveform, which is necessary for calculating the memory effect. We only keep dominant modes for the displacement memory and the spin memory mode, i.e., $h_{(2,0)}$ and $h_{(3,0)}$, respectively. This is reasonable, since the higher $\ell$ modes are strongly suppressed, as one can see from Eq. (\ref{mkD}).

To assess the detection potential of memory effects, we need to convert waveforms to the frequency domain by a Fourier transformation. We use Planck-taper window \cite{McKechan:2010kp} to window the time-domain waveform to remove edge effects. In the calculation, we choose $\epsilon=0.04$ (see Eq. (7) in McKechan et al. \cite{McKechan:2010kp} for the detail of window function).

Since the memory effect is always subdominant compared to the full \ac{GW} strain, one should search for memory effect in \ac{GW} signals with the highest possible \ac{SNR}. So we focus on \acp{MBHB} with masses in the range $10^4~\mSun\sim10^7~\mSun$ in this study. 
In the following part of this work, all the mass mentioned are redshifted mass.

It is important to determine how much data will be needed in the calculate the memory effect. For this we note that the vast majority of the \acp{SNR} produced by \acp{MBHB} on TianQin can be obtained within the last hour of data before merger \cite{haitian,Feng:2019wgq}. But as we will see below, we will not need a full hour of data for total masses near $10^4~\mSun\,$. We also note the memory effect is mainly accumulated during the merger phase when the most energy and angular momentum are radiated away, while the spin memory may get relatively more contribution from the inspiral phase \cite{Mitman:2020pbt,Nichols:2017rqr}. So we choose to calculate the memory effect by keeping a fixed number of waveform cycles (for any given mass ratio) before merger, where the number of cycles is defined as \cite{Maggiore:2007ulw}
%%%
\begin{equation}
\mathcal{N}=\frac{1}{32 \pi^{8 / 3}}\left(\frac{G M_{c}}{c^{3}}\right)^{-5 / 3}\left(f_{\min }^{-5 / 3}-f_{\max }^{-5 / 3}\right)\,.
\end{equation}
%%%
Here $f_{\min}$ and $f_{\max}$ are the frequencies at the beginning and the end points of the waveform,  $G$ is gravitational constant, $c$ is light speed, $M_{c}$ is the chirp mass,
%%%
\bea
M_{c}=\frac{(m_{1}m_{2})^{3/5}}{(m_{1}+m_{2})^{1/5}}\,.
\eea
%%%

\begin{figure}
\centering
\includegraphics[scale=0.56]{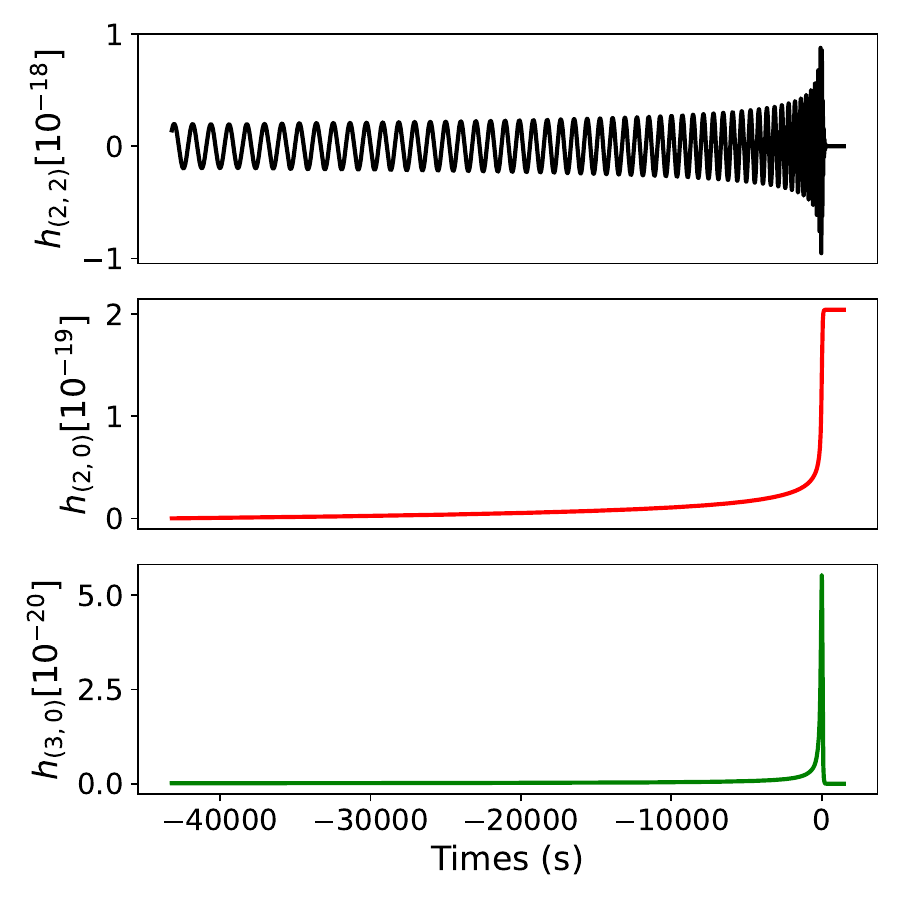}
\caption{Example of waveforms generated with IMRPhenomXHM. Plotted with a non-spinning MBHB with mass ratio $q=1$, total mass $M=10^{6}\mSun$ at luminosity distance $D_{\rm L}=2\rm Gpc$. The top panel is the real part of dominant GW mode $h_{(2,2)}$, the middle panel is the real part of dominant mode of displacement memory $h_{(2,0)}$, and the bottom panel is the imaginary part of dominant spin memory mode $h_{(3,0)}$. The cycles of the waveform for this plot is about 57.}
\label{waveformall}
\end{figure}

For the efficiency of calculation, we have generated the waveforms for a chosen total mass ($2\times10^6~\mSun$) with a given duration (one day) at different symmetric mass ratios, 
%%%
\bea
\eta = \frac{m_{1}m_{2}}{(m_{1}+m_{2})^{2}}\,,
\eea
%%%
ranging from $\eta=0.04$ to $\eta=0.25$ at a $0.01$ incremental rate, and the waveform is then used for all source masses through appropriate scaling (For sources with total mass $10^4~\mSun$ and $10^7~\mSun\,$, this corresponds to a duration of 0.12 hour and 5 day, respectively). The number of cycles involved in such waveforms increases monotonically when the symmetric mass ratio is lowered, varying from about 57 for $\eta=0.25$ to about 113 for $\eta=0.04$. The lost \ac{SNR} for cycles not included is expected to be negligible. We have explicitly checked that the difference between the memory \acp{SNR} from waveforms keeping the last 150 cycles and the last $50$ cycles before merger is no greater than $0.026\%$, for equal mass sources with different total masses. For unequal-mass sources, the number of cycles in the waveform is always greater than that of the equal-mass sources and so the lost \ac{SNR} is expected to be even smaller. In Fig. \ref{waveformall}, we give an example of the waveforms generated with IMRPhenomXHM. One can see that the accumulation of memory effect is most significant near the time of merger.

\section{Detection potential}\label{SNR}

In this section, we study the prospect of using TianQin to detect the memory effect.

The data stream output from a detector can be schematically written as
%%%
\bea d(t)=s(t)+n(t)\,,\eea
%%%
where $s(t)$ is the \ac{GW} signal registered by the detector and $n(t)$ is the noise in the data stream.

The registered signal depends on the detector through the antenna pattern functions. For space-based \ac{GW} detectors like TianQin and LISA, they are given by
%%%
\bea F_{+}(\theta,\phi,\psi)&=&\frac{\sqrt{3}}{2}\bigg[\frac{1}{2}(1+\cos^{2}\theta)\cos2\phi\cos2\psi\nn\\
&&-\cos\theta\sin2\phi\sin2\psi\bigg]\,,\nn\\
F_{\times}(\theta,\phi,\psi)&=&\frac{\sqrt{3}}{2}\bigg[\frac{1}{2}(1+\cos^{2}\theta)\cos2\phi\sin2\psi\nn\\
&&+\cos\theta\sin2\phi\cos2\psi\bigg]\,,\eea
%%%
where the $\theta$ and $\phi$ are the position of the source in the detector frame, and $\psi$ is the polarization angle. For any incoming \ac{GW} signal with the usual plus and cross modes, $h(t,\iota,\varphi_{c})=h_{+}(t,\iota,\varphi_{c})- i h_{\times}(t,\iota,\varphi_{c})$, where $\iota$ and $\varphi_{c}$ are the inclination and the reference phase, respectively, the registered signal is given by
%%%
\bea s(t)=F_{+}(\theta,\phi,\psi)h_{+}(t,\iota,\varphi_{c})+F_{\times}(\theta,\phi,\psi)h_{\times}(t,\iota,\varphi_{c})\,.\eea
%%%
For a preliminary estimation, we focus on the sky-averaged response of \acp{GW},
%%%
\bea\brk{s^\ast(t)s(t)}=\int_0^{+\infty}S_h(f)\bar{R}(f)df\,,\eea
%%%
where $S_h(f)$ is the \ac{PSD} of the incoming \ac{GW} signal and $\bar{R}(f)$ is the sky-averaged response of the detector. In this work, we use
%%%
\bea S_h(f)&=&\frac2T\Big|\td{h}_{+,\times}(f)\Big|^2\equiv\frac2T\Big|\td{h}(f)\Big|^2\,,\nn\\
\bar{R}(f)&=&\frac3{10}\left[1+\left(\frac{2fL_0}{0.41}\right)^{2}\right]^{-1}\,,\eea
%%%
where $T$ is the time span of the data stream, $\td{h}_{+,\times}(f)$ are the Fourier components of $h_{+,\times}(t)$, and $L_0$ is the arm-length of the detector.

The noise $n(t)$ is usually given in terms of its \ac{PSD},
%%%
\bea S_N(f)=\frac2T\Big|\td{n}(f)\Big|^2\,,\eea
%%%
where $\td{n}(f)$ is the Fourier component of $n(t)\,$. For a space-based detector like TianQin and LISA, the noise can be grossly grouped into two categories: those from the propagation of the laser used to measure the change of spacetime due to the passage of \acp{GW} and those from the irregular motion of the test masses serving as reflecting endpoints for the lasers. The former is quantified with the total displacement measurement noise in a single laser link, denoted by $S_x$, and the latter is quantified with the residue acceleration noise of a test mass in the sensitive direction, denoted by $S_a$. The structure of $S_N(f)$ and the constants $L_0\,$, $S_x\,$ and $S_a$ are different for different detectors. For TianQin, $L_{0}=\sqrt{3}\times 10^{8}{\rm~m}\,$, and the following noise model is used in the process of mission development \cite{TianQin:2015yph,TianQin:2020hid},
%%%
\bea S_N(f)&=&\frac1{L_0^2}\Big[S_{x}+\frac{4S_{a}}{(2\pi f)^4}\Big(1+\frac{10^{-4}{\rm Hz}}{f}\Big)\Big]\,,\nn\\
S_{x}^{1/2}&=&1\times10^{-12}{\rm~m}/{\rm Hz}^{1/2}\,,\nn\\
S_a^{1/2}&=&1\times10^{-15}{\rm~m/s}^2{\rm /Hz}^{1/2}\,.\eea
%%%
The huge number of Galactic compact binaries, most of which are ultra-compact double white dwarf systems, can generate a foreground confusion noise that may affect the detection of \acp{GW} from other sources. However, it has been shown that the expected foreground confusion noise for the TianQin is relatively weak \cite{Huang:2020rjf}. So we will not consider it in this work. For LISA, the parameters of the Galactic confusion noise \cite{Cornish:2017vip,lisasen} are for 4 years of data.

\begin{figure}
\centering
\includegraphics[scale=0.45]{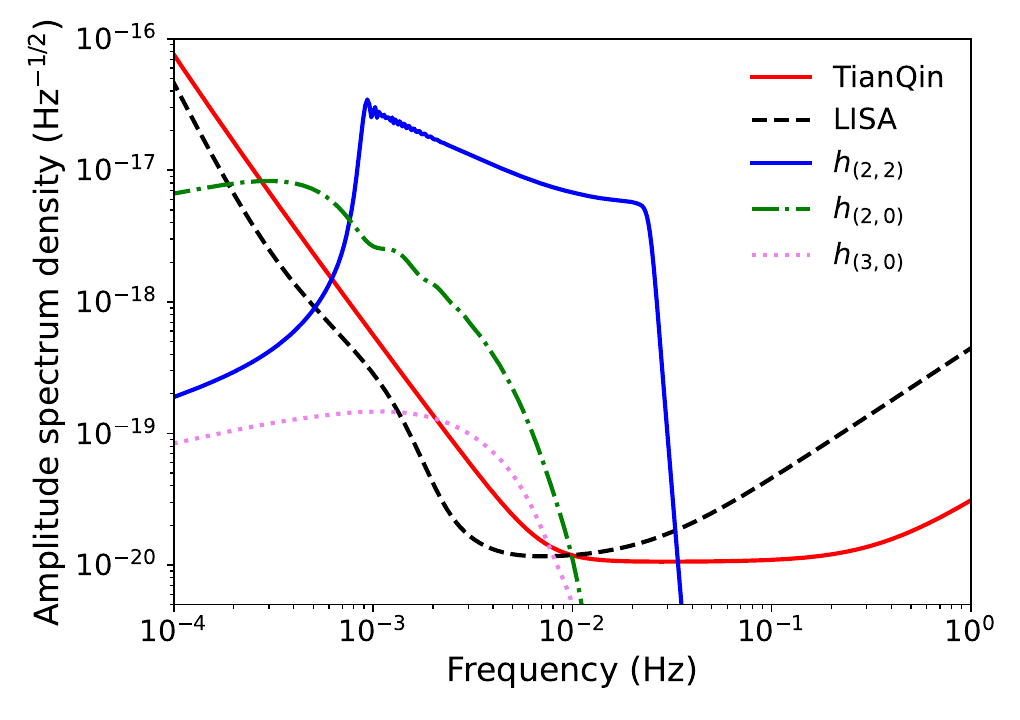}
\caption{The sensitivity curves of TianQin and LISA, together with the dominant modes, $h_{(2,2)}$, $h_{(2,0)}$ and $h_{(3,0)}$, for the usual \ac{GW} strain, the displacement memory and the spin memory mode, respectively, plotted with a non-spinning \ac{MBHB} with mass ratio $q=1$, total mass $M=  10^{6}~\mSun\,$, inclination $\iota=\pi/4$ and luminosity distance $D_{\text{L}}=2~\text{Gpc}\,$. }
\label{asdall}
\end{figure}

The sky averaged sensitivity, or equivalently the \ac{ASD}, of TianQin is define as
%%%
\bea h_{\rm eff}(f)=\sqrt{S_n(f)}\,,\quad S_n(f)\equiv\frac{S_N(f)}{\bar{R}(f)}\,.\eea
%%%
The numerical results are plotted in Fig. \ref{asdall}. As a comparison, we plot the sensitivity curve of LISA \cite{lisasen}, for which the contribution of the foreground confusion noise is non-negligible. We also plot the strains for the dominant mode $h_{(2,2)}$, the dominant mode of displacement memory  $h_{(2,0)}$, and the dominant spin memory mode $h_{(3,0)}$.

\subsection{SNR}

The \ac{SNR} is a key quantity one can use to assess the detection potentia of a signal, which is defined as
%%%
\bea \rho=\sqrt{4\int^{f_{max}}_{f_{min}}\frac{|\td{h}(f)|^2\bar{R}(f)}{S_N(f)}df}\,,\eea
%%%
where $f_{min}$, $f_{max}$ are the minimal and maximal frequency limits of the waveform.
For ground-based detectors, one may use an \ac{SNR} threshold as low as $\rho=3$ to claim a detection of the memory effect \cite{Kennefick:1994nw,Lasky:2016knh,Johnson:2018xly}. However, for space-based detectors, there could be several signals coexisting in the data, potentially making it more difficult to identify the memory effect. But since the actual \ac{SNR} threshold is still not well studied so far, we will use the set of thresholds $\rho=3,5,8$ to get an indicative result for the detection potential of TianQin. As an explicit example, for the source parameters used in Fig. \ref{asdall}, the \acp{SNR} are: $\rho_{(2,2)}=1112.2\,$, $\rho_{(2,0)}=14.7$ and $\rho_{(3,0)}=3.8\,$.

\begin{figure}
\centering
\includegraphics[scale=0.45]{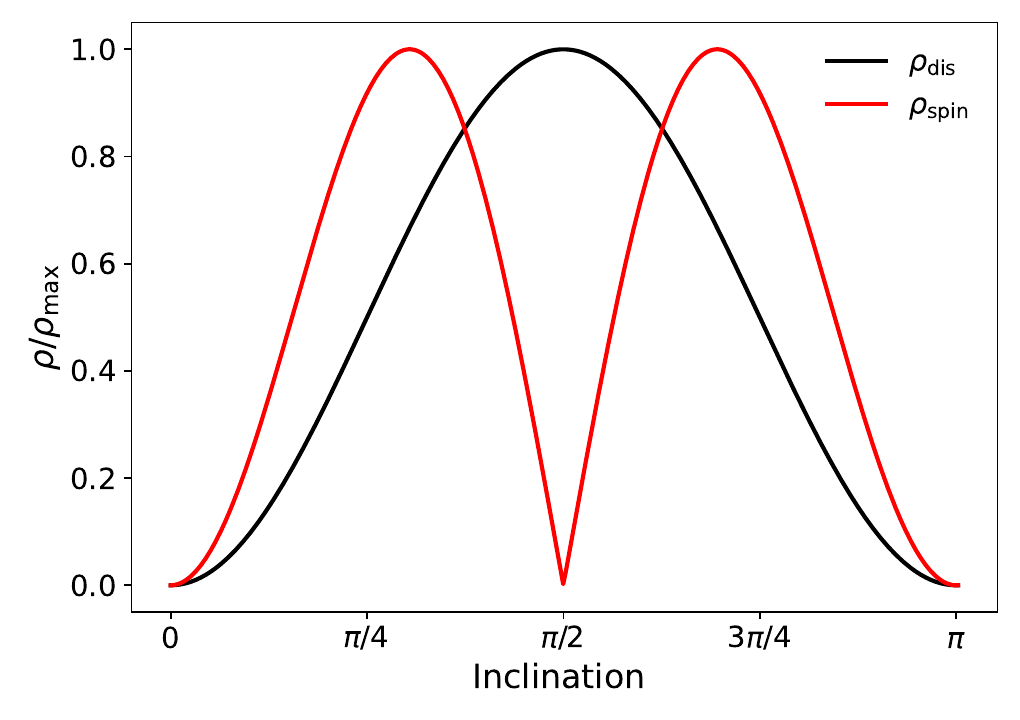}
\caption{Dependence of \ac{SNR} on the inclination angle $\iota$. Note different values of $\rho_{\rm max}$ are used for the displacement memory and the spin memory mode. Plotted using a \ac{MBHB} with mass ratio $q=1$, total mass $M=  10^{6}~\mSun\,$ and luminosity distance $D_{\text{L}}=2~\text{Gpc}\,$.}
\label{iota}
\end{figure}

There are several parameters that can strongly affect the \ac{SNR}. The most significant one is the inclination angle $\iota$, and the dependence of inclination has been well studied (see e.g. \cite{Favata:2009ii,Talbot:2018sgr,Islam:2021old}). We plot the SNR of memory effects as a function of $\iota$ in Fig. \ref{iota}. For the displacement memory, the SNR will get its maximum at $\iota=\pi/2$, and approaches to $0$ as $\iota$ approaches to $0$ and $\pi$. On the other hand, for the spin memory mode, the \ac{SNR} will be $0$ while $\iota=\pi/2$, and get its maximum near $\iota=3\pi/10$ and $\iota=7\pi/10$. For all the examples to be studied in the following part, we will choose $\iota=\pi/2$ for the displacement memory and $\iota=\pi/4$ for the spin memory mode.

\begin{figure}
\centering
\includegraphics[scale=0.45]{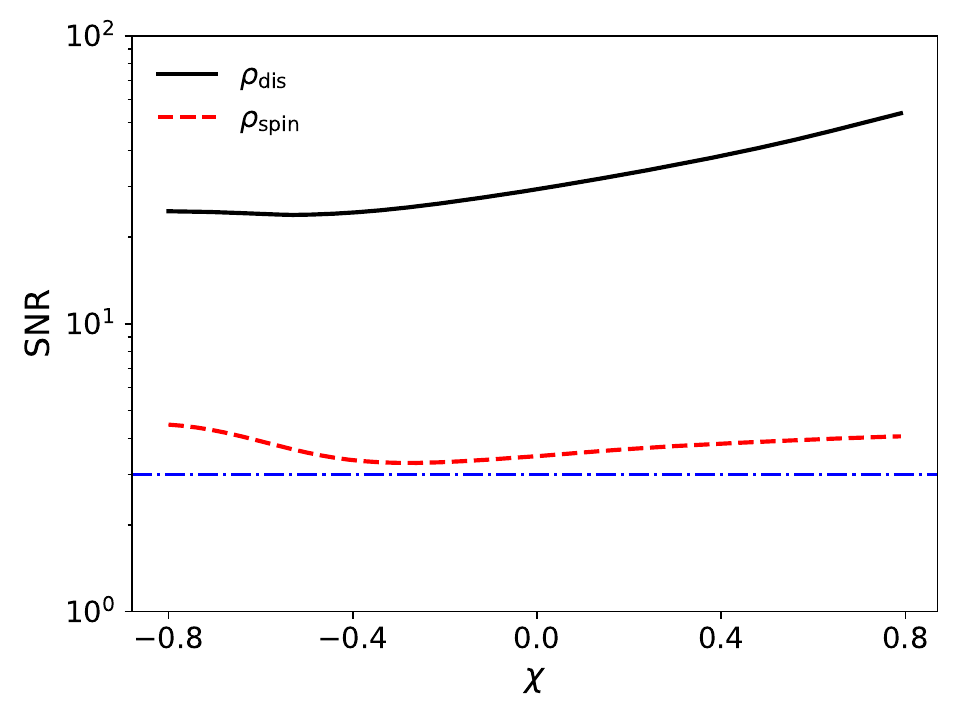}
\caption{Dependence of SNR on the effective spin $\chi$. Plotted using a MBHB with mass ratio $q = 1$, total mass $M = 10^{6} \mSun$ and luminosity distance $D_{L}=2\text{Gpc}$ . The blue horizontal line represents SNR equal to 3. We choose inclination $\iota=\pi/2$ for displacement memory, and inclination $\iota=\pi/4$ for spin memory mode.}
\label{effespin}
\end{figure}

Another important parameter is the spin of the components of the binaries, which can increase the efficiency of energy loss through \ac{GW} radiation \cite{Reisswig:2009vc,Lousto:2009mf}. In our situation, we consider the aligned spin systems. We use the effective spin defined by \cite{Ajith:2009bn,Santamaria:2010yb},
%%%
\bea
\chi = \frac{m_{1}\chi_{s_{1}}+m_{2}\chi_{s_{2}}}{m_{1}+m_{1}},
\eea
%%%
where $\chi_{s_{1},s_{2}}$ is aligned dimensionless spin components of two black holes and are defined as
%%%
\bea
\chi_{s_{1},s_{2}}=\frac{\vec{S}_{1,2}\cdot \vec{L}}{m_{1,2}^{2}|\vec{L}|},
\eea
%%%
where $\vec{S}_{1,2}$ are the spins (intrinsic angular momentum) of the two individual black holes, $\vec{L}$ is the orbital angular momentum and $m_{1,2}$ are the masses of two black holes.
The dependence of \ac{SNR} on the effective spin $\chi$ is plotted in Fig. \ref{effespin}. One can see that the \ac{SNR} of the displacement memory increases nearly monotonically with $\chi$, while the \ac{SNR} of the spin memory mode does not vary too much with $\chi\,$. 

\begin{figure}[t]
\includegraphics[scale=0.45]{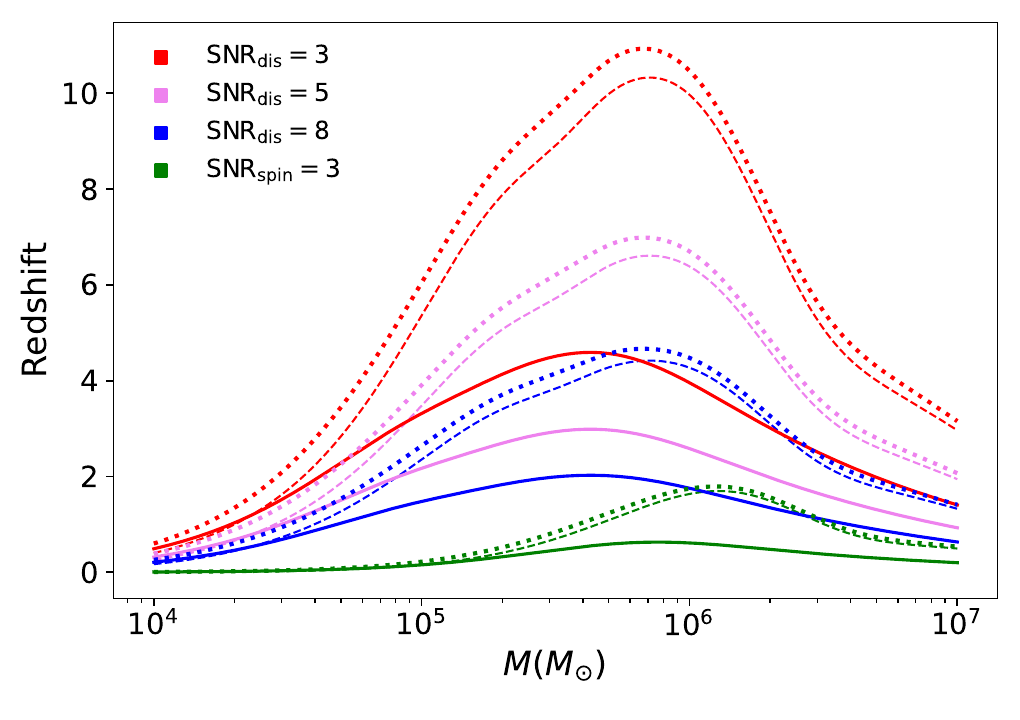}
\caption{The dependence of detection horizon for memory effect on the total mass, at \acp{SNR} equal to 3, 5, 8 for the displacement memory and the \ac{SNR} equal to 3 for the spin memory mode. Other source parameters are $q=1\,$ and $\chi=0.8\,$. Three detector configurations are considered: TianQin (solid), LISA (dashed) and TianQin + LISA (dotted).}
\label{detecthoriz}
\end{figure}

It is expected that the \ac{SNR} will strongly depend on the source distance, for it is inversely proportional to the magnitude of the \ac{GW} strain. In Fig. \ref{detecthoriz}, we plot the horizon distance for detection in terms of the redshift $z$ as a function of the total mass $M\,$. For the displacement memory, the maximum redshifts for \acp{SNR} equal to 3, 5, 8 are $z=4.6$, $z=2.98$ and $z=2.03$, respectively. For the spin memory mode, the maximum redshifts for \ac{SNR} equal to 3 is $z=0.48$. As a comparison, we also give the results on LISA and the joint detection with TianQin and LISA (denoted as TianQin + LISA). The horizon distance for LISA is almost twice of that for TianQin within a large part of the mass ranges, with the maximum redshifts reaching $z=10.3$, $z=6.6$, $z=4.4$ for the SNRs of the displacement memory equal to 3, 5 and 8, and the maximum redshift for detecting the spin memory mode with SNR equal to 3 is $z=1.7$. There is also obvious improvement of TianQin + LISA over LISA alone, with the contribution of TianQin being more significant near the lower mass end.

\begin{figure*}
\subfigure{
\includegraphics[scale=0.33]{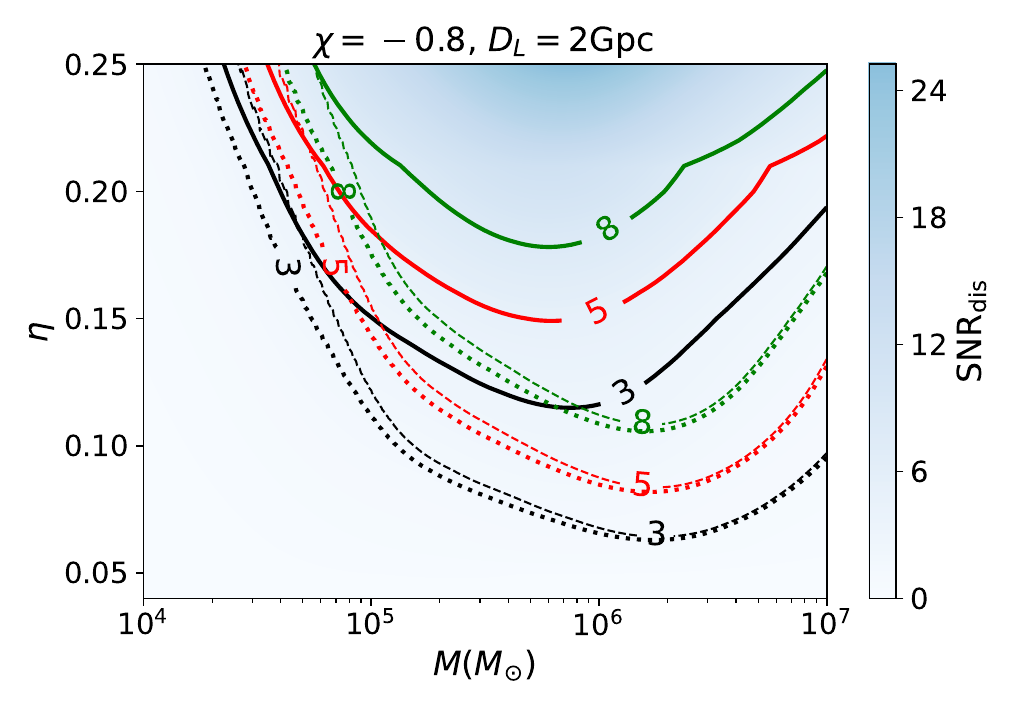}}
\subfigure{
\includegraphics[scale=0.33]{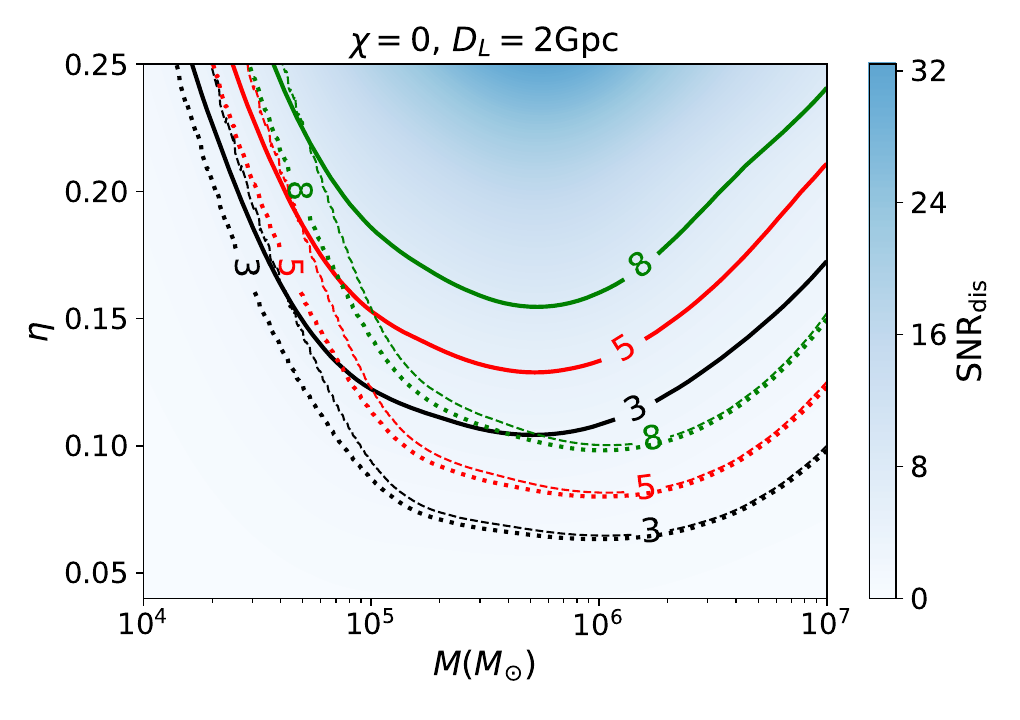}}
\subfigure{
\includegraphics[scale=0.33]{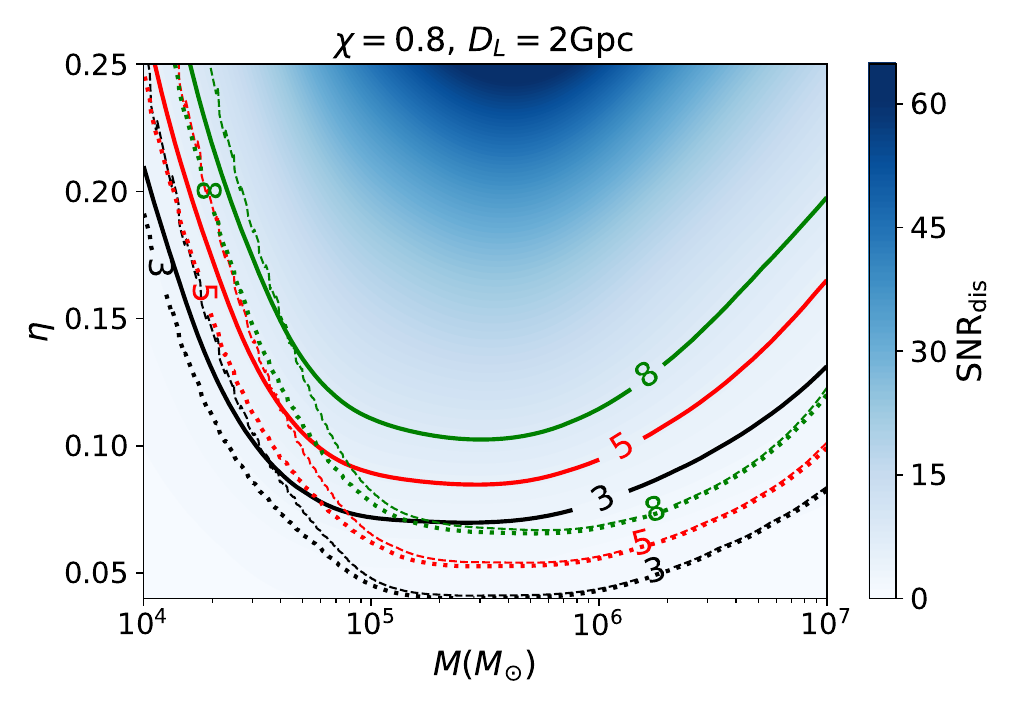}}
\subfigure{
\includegraphics[scale=0.33]{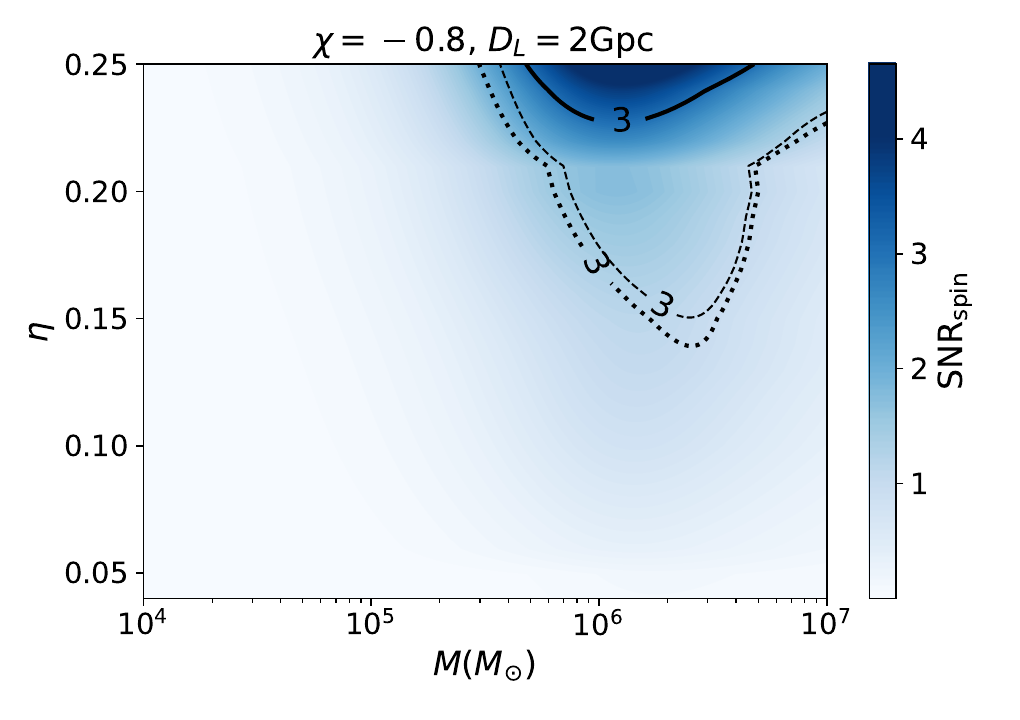}}
\subfigure{
\includegraphics[scale=0.33]{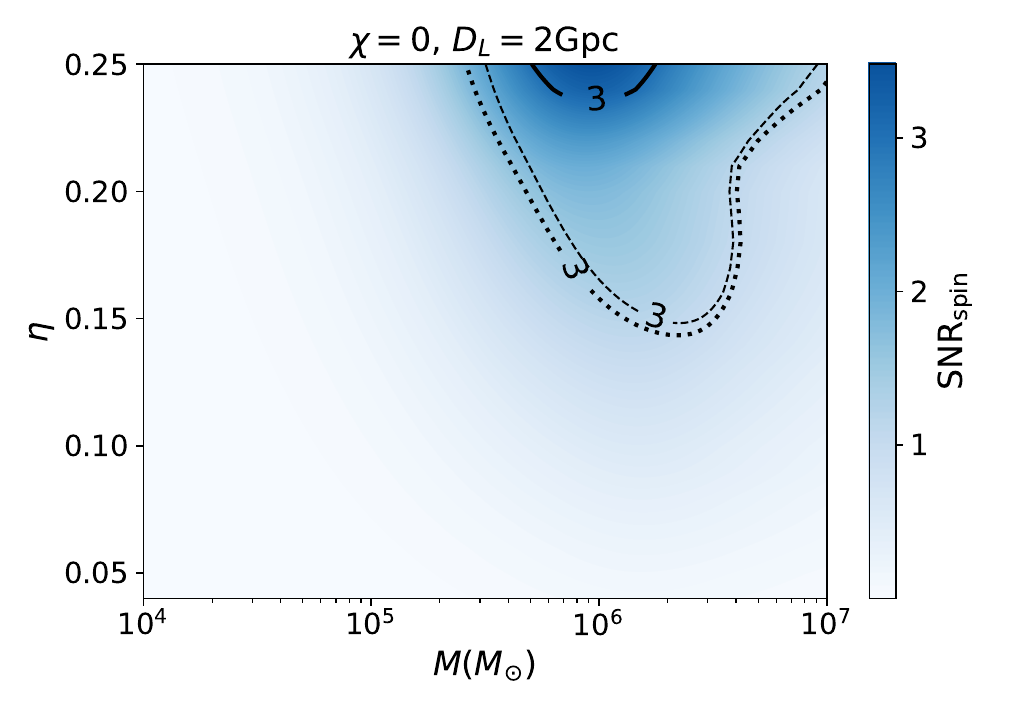}}
\subfigure{
\includegraphics[scale=0.33]{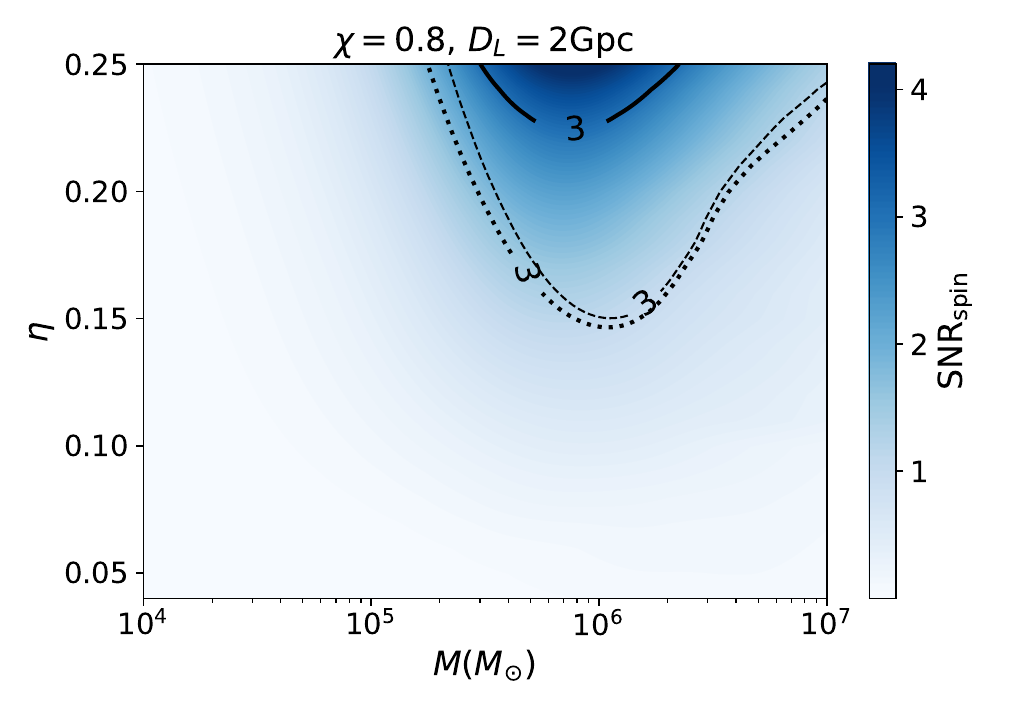}}
\caption{The dependence of \acp{SNR} on the symmetric mass ratio $\eta$ and the total mass $M\,$, for three values of the effective spin, $\chi=-0.8,0,0.8$, and fixed luminosity distance $D_{\text{L}}=2{\rm~Gpc}\,$. The upper panel is for the displacement memory and the lower panel is for the spin memory mode. The numbers on the contours are the corresponding \acp{SNR}. Three detector configurations are considered: TianQin (solid), LISA (dashed) and TianQin + LISA (dotted).}
\label{etamass}
\end{figure*}

The dependence of \ac{SNR} on the total mass $M$ and the symmetric mass ratio $\eta$ is plotted in Fig. \ref{etamass} for different values of $\chi$. The maximal \acp{SNR} for both displacement memory and spin memory mode appear around $\eta=0.25\,$, which is consistent with the expectation that systems of equal mass can radiate the most energy and angular momentum during \ac{GW} radiation and thus can incur the largest memory effect. For the displacement memory, it is possible to have \acp{SNR} passing $\rho=3\,$ for $\eta$ roughly in the range $0.06 \sim 0.25\,$, while for the spin memory mode, the allowed range for $\eta$ is narrowed to the small region of roughly $0.22\sim0.25\,$. Depending of the values of the $\chi$, the maximal \acp{SNR} for the displacement memory appear roughly in the range $4\times 10^5~\mSun \sim 8\times 10^5~\mSun\,$, while the maximal \acp{SNR} for the spin memory mode appear roughly in the range $8\times 10^5~\mSun \sim 1.5\times 10^6~\mSun\,$. As a comparison, we have also included the contours for LISA and TianQin + LISA in the same plots. One can see that in detecting the displacement memory, LISA is superior in most parameter space, while TianQin can be competitive in a small region near the lower mass end and such advantage grows bigger with $\chi\,$. There is obvious improvement of TianQin + LISA over LISA at the lower mass end of the plots. In the detection of the spin memory mode, LISA is always superior than TianQin while the improvement of TianQin + LISA is obvious for all mass ranges.

\subsection{Detection number}\label{sec4}

It is known that TianQin can detect several to dozens of \acp{MBHB} per year \cite{haitian}. So a natural question to ask is how many of the detected events will contain memory effect that is directly detectable.

There is still a lot of uncertainty in the astrophysical models that determine the population of \acp{MBHB} in the universe. In this work, we use three different models for merger history of massive black holes which have been used in \cite{haitian}. These three models are generated using the semi-analytic model proposed in \cite{Barausse:2012fy} and successively improved in \cite{Sesana:2014bea,Antonini:2015sza}. One model is referred as "popIII" which corresponds to a light seed model \cite{Madau:2001sc}, and the other two models are referred as "Q3\_d" and "Q3\_nod", which correspond to two heavy seed models \cite{Bromm:2002hb,Begelman:2006db,Lodato:2006hw} with and without the time delay between the merger of a \ac{MBHB} and that of their host galaxies.

In \cite{haitian}, 1000 mock catalogs have been generated for each of the astrophysical models. We search in each of these mock catalogs for events containing memory effect with \acp{SNR} exceeding $\rho=3\,$, counting the detection number, and then average over the 1000 mock catalogue for each of the astrophysical models.

\begin{table*}[t]
\setlength{\tabcolsep}{3mm}{
\begin{tabular}{lcccccccc}
\hline \hline
\  & IMR  &$\mathcal{M}>Threshold$ &$\rho_{dis}>3$ & $\rho_{dis}>5$ & $\rho_{dis}>8$ & $\rho_{spin}>3$ & $\rho_{spin}>5$ & $\rho_{spin}>8$\\
\hline
popIII TianQin & $56.8$ & 0.9&$0.5$ & 0.3 & 0.1 & $\sim0$ & $\sim0$ & $\sim0$ \\
\hline
Q3\_d TianQin & 18.1 & 0.9&0.6 & 0.3 & 0.2 & $\sim0$ & $\sim0$ & $\sim0$ \\
\hline
Q3\_nod TianQin & 271.4 & 3.6&2.0 & 1.2 & 0.7 & $0.2$ & 0.1 & $\sim0$ \\
\hline
popIII LISA & $148.35$ &3.3& $1.6$ & 0.7 & 0.4 & $0.1$ & $\sim0$ & $\sim0$ \\
\hline
Q3\_d LISA & 37.4 &4.9& 2.6 & 1.4 & 0.8 & 0.2 & 0.1 & $\sim0$ \\
\hline
Q3\_nod LISA & 295.5 &12.2& 5.8 & 2.6 & 1.4 & 0.4 &0.2  &0.1 \\
\hline
\end{tabular}}
\caption{The expected IMR detection number and both displacement memory and spin memory mode detection number on TianQin and LISA detectors for "popIII", "Q3\_d" and "Q3\_nod" astrophysical models.}
\label{pop}
\end{table*}

The result is given in TABLE. \ref{pop}.
One can see that, depending on the astrophysical models used, TianQin can in average detect $0.5\sim2.0$ events whose displacement memory has \acp{SNR} no less than 3 and about $0.1\sim0.7$ events whose displacement memory has \acp{SNR} no less than 8. On the one hand, these result suggests that the chance for TianQin to directly detect the displacement memory is not very high. But on the other hand, these results suggest that the chance for TianQin to directly detect the displacement memory is not fully negligible. In contrast, less than $0.2$ events are expected to be detected which will contain the spin memory mode whose \ac{SNR} can reach 3, meaning that the chance for TianQin to directly detect the spin memory mode from a single \ac{MBHB} event is likely negligible. The results on LISA have also been included in TABLE. \ref{pop}. One can see that the expected detection number of LISA is about $2\sim4$ times that of TianQin, depending on which astrophysical model and which \ac{SNR} value one is looking at.

\begin{figure*}
\subfigure{
\includegraphics[scale=0.335]{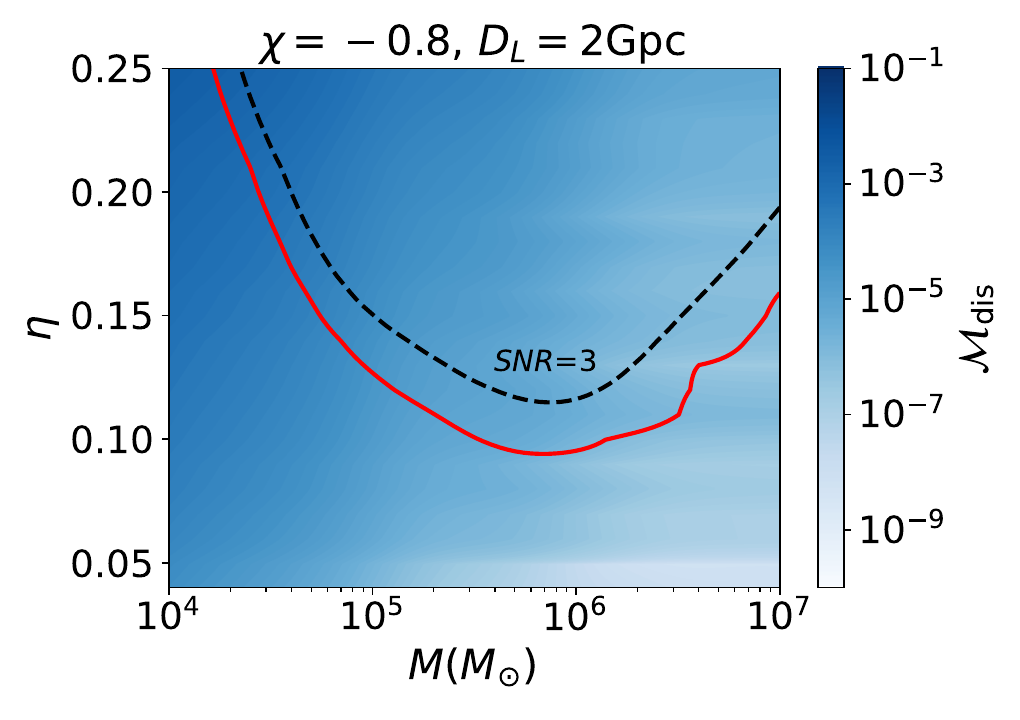}}
\subfigure{
\includegraphics[scale=0.335]{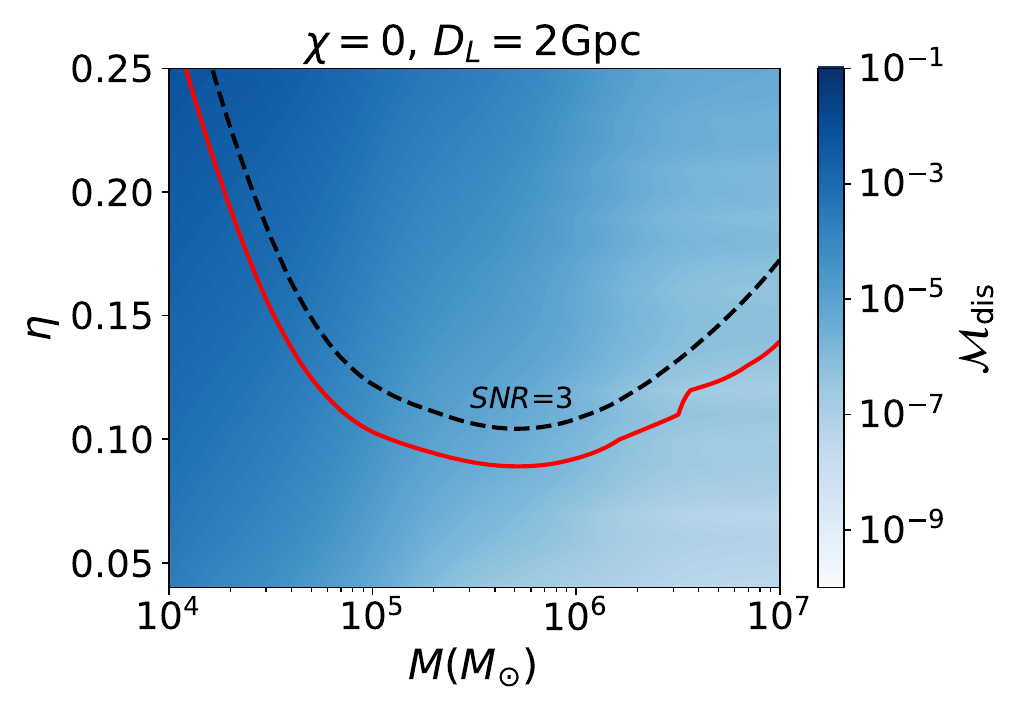}}
\subfigure{
\includegraphics[scale=0.335]{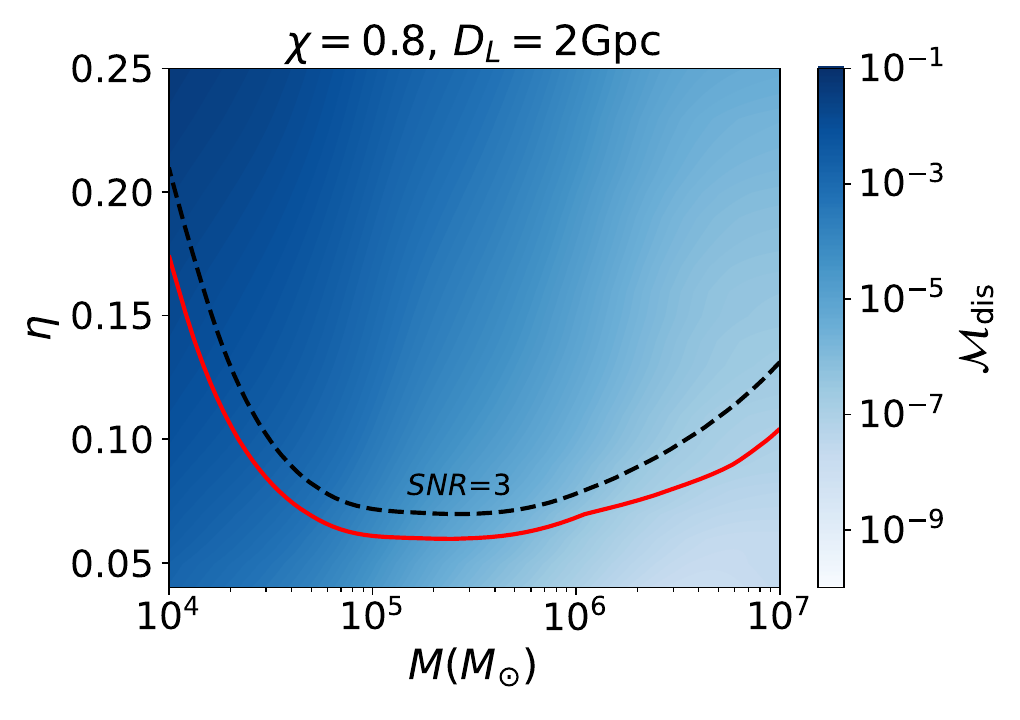}}
\subfigure{
\includegraphics[scale=0.335]{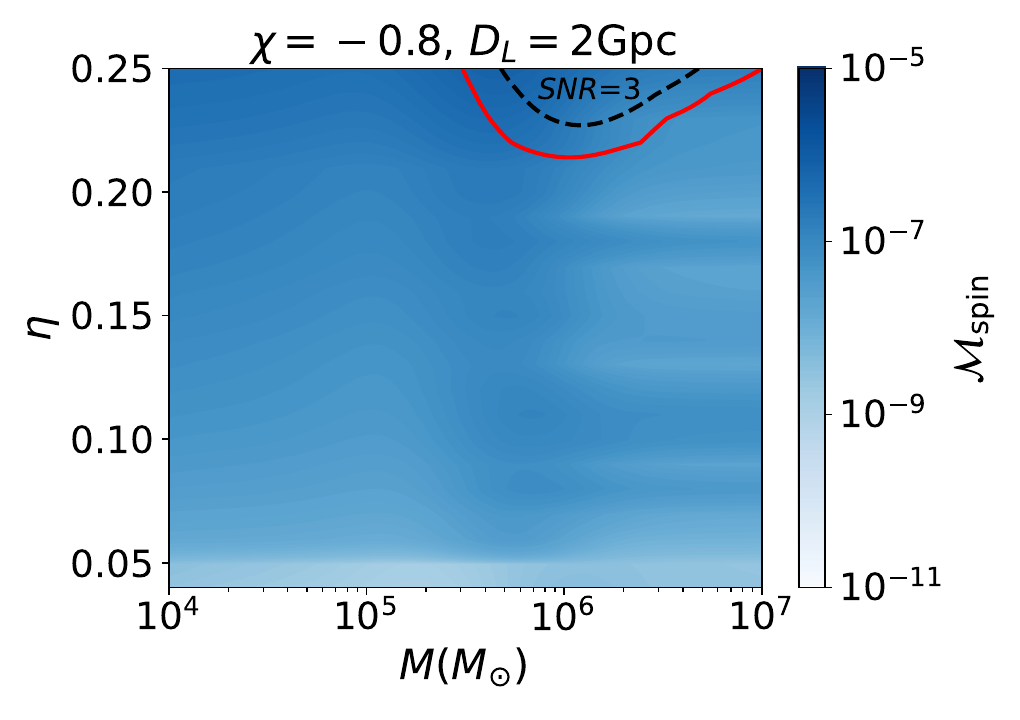}}
\subfigure{
\includegraphics[scale=0.335]{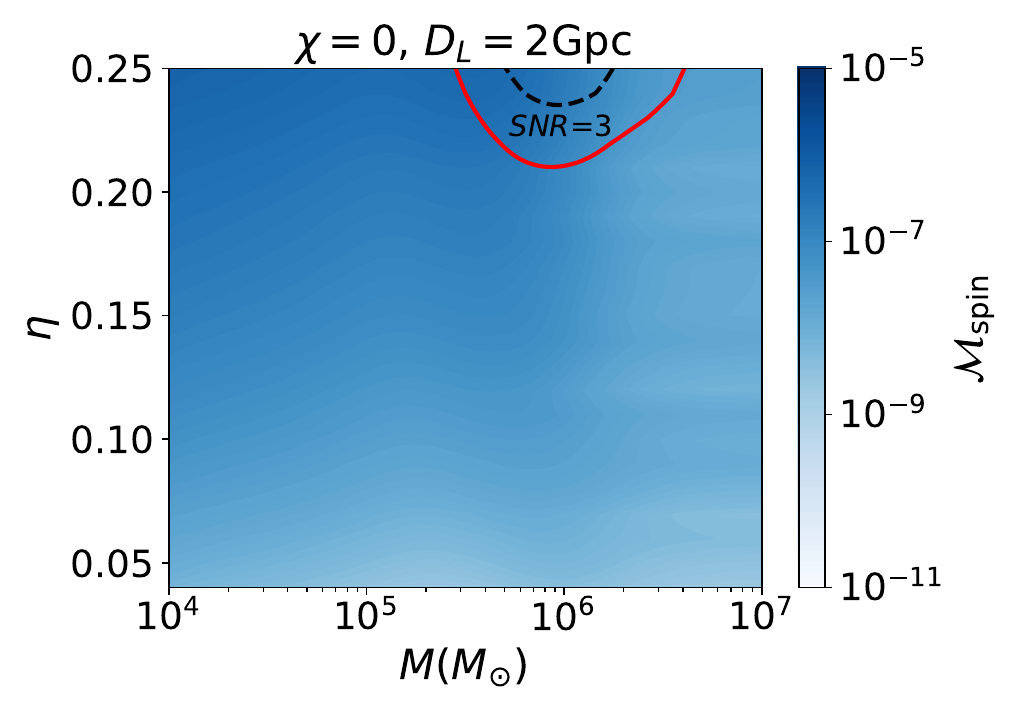}}
\subfigure{
\includegraphics[scale=0.335]{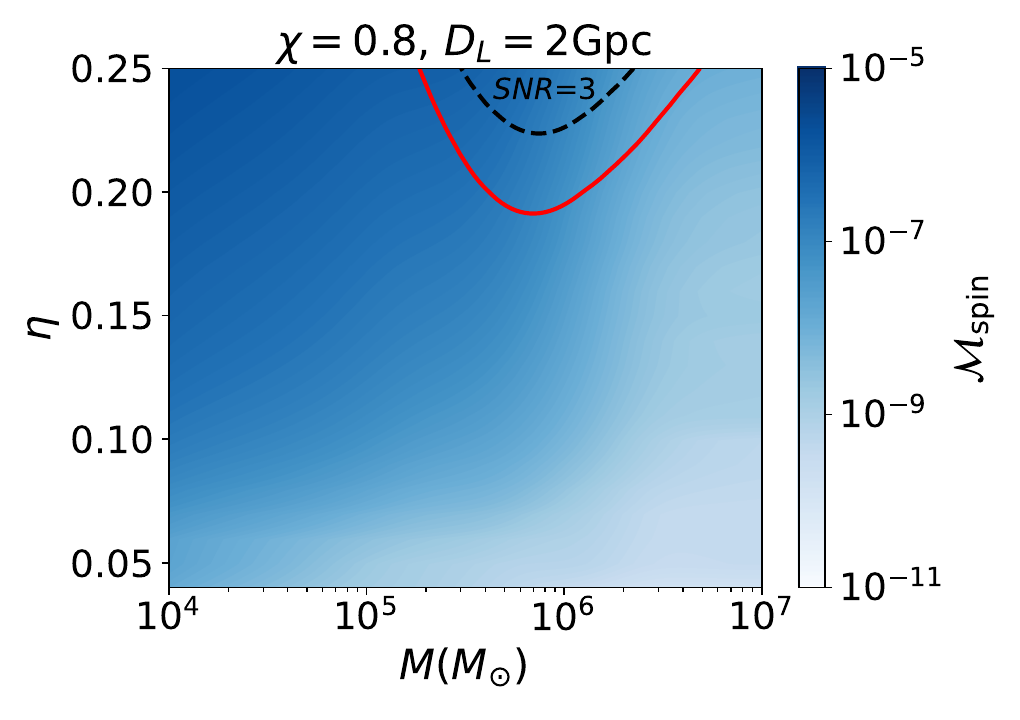}}
\caption{The dependence of mismatch on the symmetric mass ratio $\eta$ and the total mass $M\,$, for three values of the effective spin, $\chi=-0.8,0,0.8$, and fixed luminosity distance $D_{\text{L}}=2{\rm~Gpc}\,$. The upper panel is for the displacement memory while the lower panel is for the spin memory mode. The black dashed lines mark where the memory SNR equals 3. The red solid lines mark where the mismatch reaches the threshold determined by Eq. (\ref{thre}), in which $\rho$ is calculated using the complete waveform but containing no memory effect.}
\label{etamassmismatch}
\end{figure*}

\section{Relevant parameter space}\label{MisM}

Due to the important application of \ac{MBHB} events in fundamental physics \cite{Shi:2019hqa,Bao:2019kgt}, astrophysics \cite{haitian} and cosmology \cite{Zhu:2021aat}, each detected \ac{MBHB} event would be extremely precious and it is extremely important to have accurate waveform models to precisely measure the source parameters. As we have mentioned before, although the memory effect is physically interesting on its own right, it can also lead to systematic errors if not properly taken into consideration in waveform modeling. In this section, we study when the contribution of memory effect will become non-negligible. One can intuitively reason that this should be equivalent to when the memory effect will become detectable. So our key task here is to determine the relevant parameter space.

The effect of neglecting the memory effect in the waveforms can be quantified using the mismatch between the waveforms with and without the memory effect. For two different waveform models $\td{h}_1(f)$ and $\td{h}_2(f)$, the mismatch is defined as
%%%
\bea \cM=1-\frac{\brk{\td{h}_1,\td{h}_2}}{\sqrt{\brk{\td{h}_1,\td{h}_1}\brk{\td{h}_2,\td{h}_2}}}\,,\label{mismatch}\eea
%%%
where
%%%
\bea\brk{\td{h}_1,\td{h}_2}=4\text{Re}\int^{f_{max}}_{f_{min}}\frac{\td{h}_1(f)\td{h}_2^{*}(f)}{S_{n}(f)}df\,.\eea
%%%

The systematic errors introduced by a reliable waveform need to be lower than statistical errors. In high SNR regime, the statistical errors can be estimated from the inverse of Fisher matrix $\Gamma_{ij}=(\partial_{i}h|\partial_{j}h)$ (where i, j are the waveform's parameters), and these errors are decreased as $\text{SNR}^{-1}$ \cite{Vallisneri:2007ev}.
The threshold value of mismatch is given by \cite{Chatziioannou:2017tdw,Mangiagli:2018kpu,Baird:2012cu}
%%%
\bea\cM_{th}=\frac{D}{2\rho^2}\,,\label{thre}\eea
%%%
where $D$ is the number of parameters whose estimation is affected by the waveform model accuracy. In our case, we have $D=4$. One needs $\cM<\cM_{th}$ to ensure that the parameter estimation is unbiased. It should be noted that the results given by Eq. (\ref{thre}) become unreliable as the SNR decreases \cite{Vallisneri:2007ev,Baird:2012cu}.

The dependence of the mismatch on the symmetric mass ratio $\eta$ and the total mass $M$, for several values of the effective spin $\chi$ and the fixed luminosity distance $D_{\rm L}\,$, is plotted in Fig. \ref{etamassmismatch}. One can see the contours for the mismatch threshold and those for the \ac{SNR} equal to 3 are close to each other. 

\begin{figure}
\subfigure{
\includegraphics[scale=0.45]{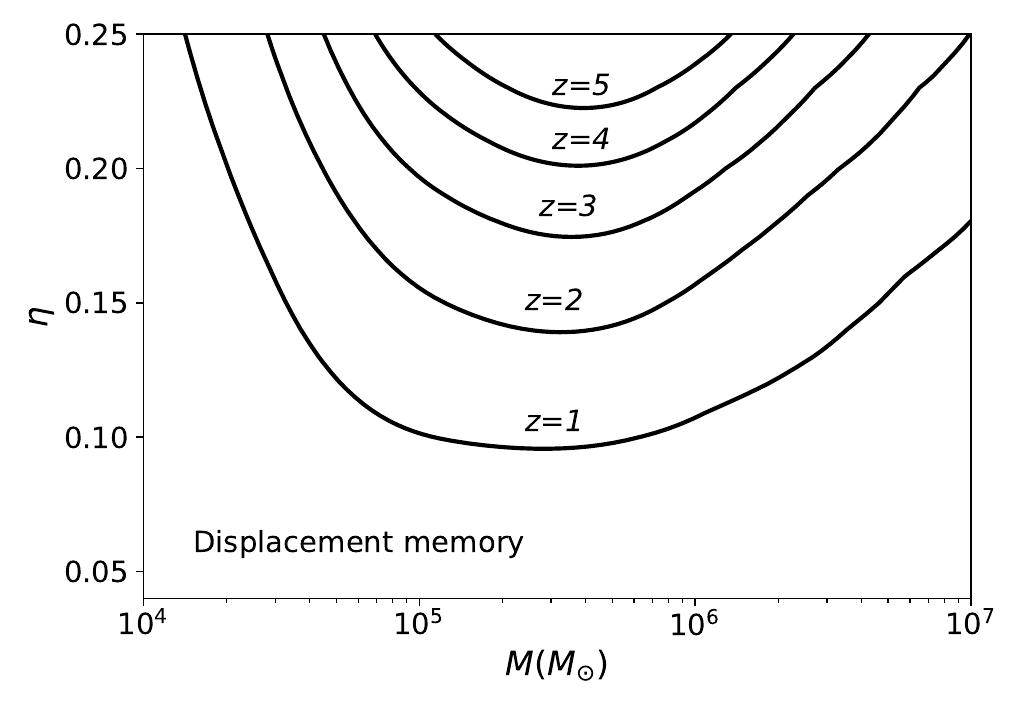}}
\subfigure{
\includegraphics[scale=0.45]{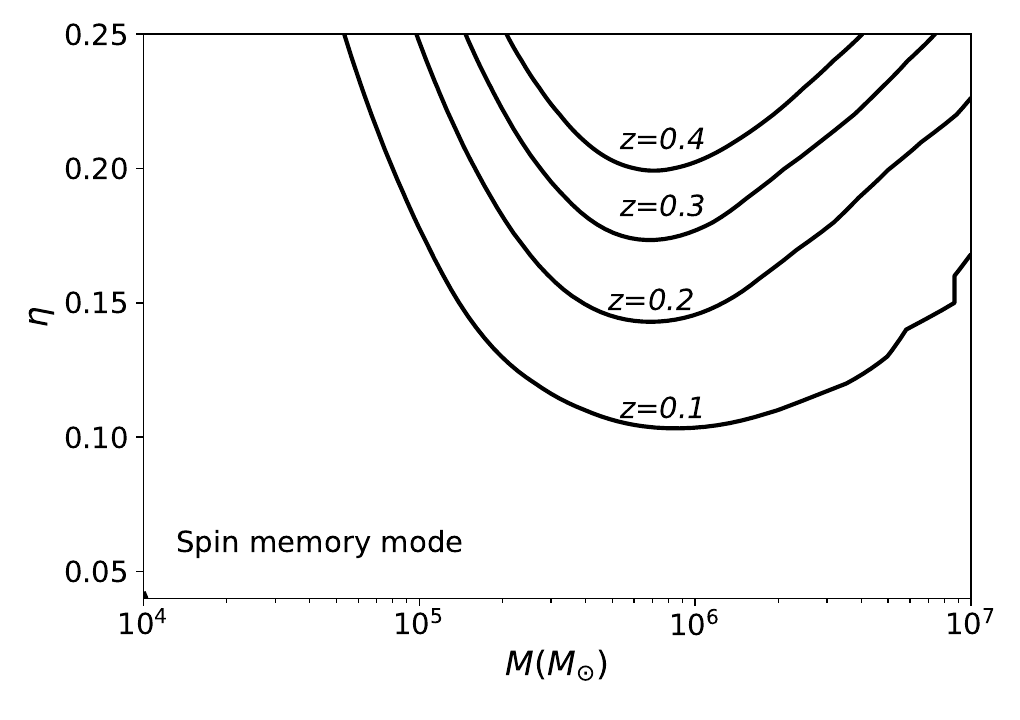}}
\caption{The critical contour of mismatch at different redshift values. We have used $\chi=0.8$ in the plots.}
\label{threred}
\end{figure}

We also plot the critical contours where the mismatch reaches its threshold for different redshift values on $M$-$\eta$ plane in Fig. \ref{threred}. One can see that the critical contours shrink when the redshift is increased. For a source with total mass near $4\times10^5~\mSun\,$, one should consider the contribution of displacement memory to the waveform even when the source is beyond redshift 5, while for the spin memory mode, it can only have influence for very small redshift, another indication that spin memory is unlikely to be directly detected with a single source.

\begin{figure}[t]
\centering
\includegraphics[scale=0.45]{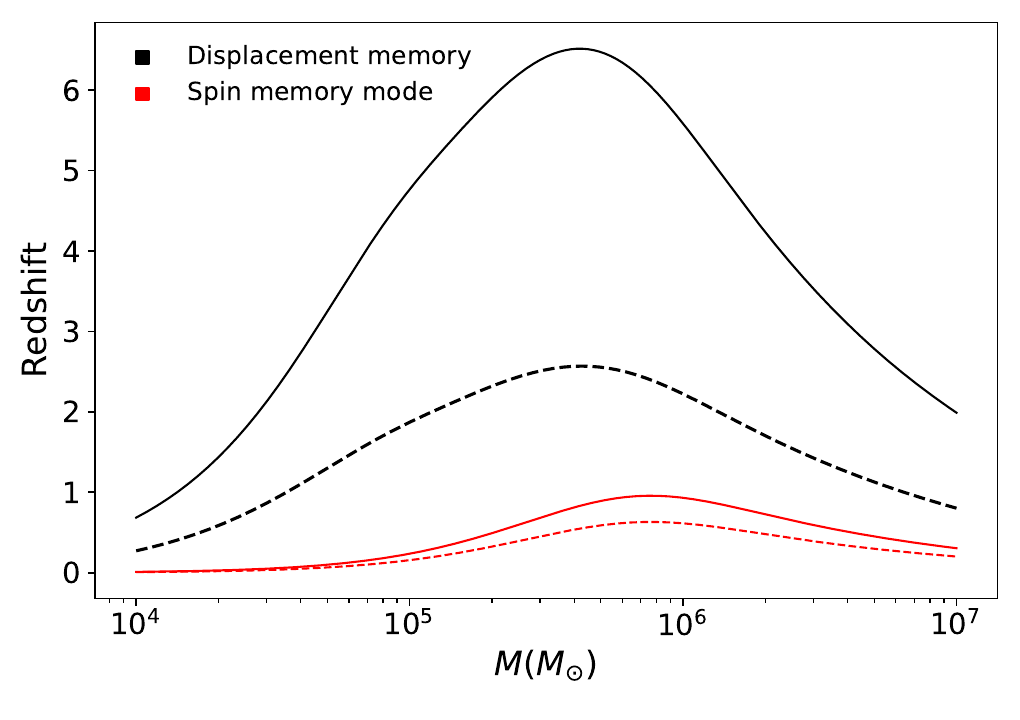}
\caption{The horizon distance in redshift for the detection with a criterion of mismatch and SNR in terms of reshifted total mass. The calculation is made by the \ac{MBHB} with $M=10^{6}~\mSun$, $q=1$ and $\chi=0.8$.  In calculation, we choose inclination $\iota=\pi/2$ for displacement memory, and inclination $\iota=\pi/4$ for spin memory mode.}
\label{redshifitthre}
\end{figure}

In Fig. \ref{redshifitthre}, we compare the horizon distance deduced by using the mismatch threshold and by requiring $\rho=3\,$. One can see that the horizon distance to pass the threshold of mismatch is always larger than requiring $\rho=3\,$. Using the criterion of mismatch, the maximum redshift is $z=6.37$ for the displacement memory and $z=0.58$ for the spin memory mode.

\section{Conclusion}\label{sum}
Among all the curious effects, the radiation of \acp{GW} also cause permanent changes in the background spacetime. Such changes are caused not only by the changes in the BMS charges in spacetime, but also by the energy and angular momentum fluxes that go to the null infinity. So the detection of memory effect will serve as an important portal to study the nature of gravity and spacetimes.

These permanent changes, which are called the displacement, spin and center-of-mass memory, are related to the various types of \ac{BMS} transformations and their extensions. Such connection has enabled the calculation of the displacement memory strain and the spin memory mode strain using the \ac{BMS} flux-balance laws.

In this paper, we study the prospect of using TianQin, a planned space-based \ac{GW} detector, to directly detect the memory effect.

The \acp{SNR} depend strongly on many of the source parameters. For the displacement memory, the maximal \acp{SNR} can be achieved with the inclination angle $\iota=\pi/2$, large effective spin, nearly equal mass and with total masses in the range $4\times 10^5~\mSun \sim 8\times 10^5~\mSun\,$. For the spin memory mode, the maximal \acp{SNR} can be achieved with the inclination angle near $\iota= 3\pi/10$ and $\iota= 7\pi/10$, nearly equal mass and with total masses in the range $8\times 10^5~\mSun \sim 1.5\times 10^6~\mSun\,$. With favorable source parameters, the memory \acp{SNR} can reach 3 for sources located as far as $z\approx4.6$ for detecting the displacement memory and $z\approx0.48$ for detecting the spin memory mode. By using currently available astrophysical models for \acp{MBHB}, For memory effect with a SNR equal to 3, we find that TianQin can detect about $0.5\sim2.0$ \ac{MBHB} events with the displacement memory effect, and less than $0.7$ events with spin memory mode during its 5 years observation. Thus TianQin will have a low but non-negligible chance to detect the displacement memory while the chance to detect the spin memory from a single \ac{MBHB} event is likely negligible.

We also study the question of when the contribution of memory effect will become non-negligible in the waveform modeling of \acp{MBHB}. We find that contour of SNR=3 are included in the criterion for the mismatch beyond the threshold. By using (nearly) optimal values for the inclination angle and the effective spin, we determine the parameter space in which the memory effect can be safely neglected.

We note LISA can do better than TianQin in detecting the memory effect for most sources while TianQin can become competitive for sources with lower total masses. With the current astrophysical models on \acp{MBHB}, the expected detection number of LISA is about $2\sim4$ times that of TianQin, depending on which astrophysical model and which \ac{SNR} value one is looking at. We also note TianQin + LISA can have obvious improvement over each of the individual detectors.

\section{Acknowledgement}
We are grateful to Keefe Mitman, Michael Boyle, Moritz H\"ubner, Oliver Boersma, Alberto Mangiagli and Neev Khera for their kind helps. We thank Yiming Hu, Qingfei Zhang and Han Wang for many useful discussions. We thank Alberto Sesana and Enrico Barausse for sharing their simulated catalogue of massive black holes. We also thank the anonymous referee for his valuable comments, which helped to improve the manuscript. We acknowledge the usage of the calculation utilities of LALsuite \cite{LALSuite}, SXS \cite{sxsgit}, NUMPY \cite{vanderWalt:2011bqk}, and SCIPY \cite{Virtanen:2019joe}, and plotting  utilities of MATPLOTLIB \cite{Hunter:2007ouj}. This work has been supported by Guangdong Major
Project of Basic and Applied Basic Research (Grant No.2019B030302001).

\bibliographystyle{apsrev4-1}
\bibliography{bibsun}

\begin{thebibliography}{116}%
\makeatletter
\providecommand \@ifxundefined [1]{%
 \@ifx{#1\undefined}
}%
\providecommand \@ifnum [1]{%
 \ifnum #1\expandafter \@firstoftwo
 \else \expandafter \@secondoftwo
 \fi
}%
\providecommand \@ifx [1]{%
 \ifx #1\expandafter \@firstoftwo
 \else \expandafter \@secondoftwo
 \fi
}%
\providecommand \natexlab [1]{#1}%
\providecommand \enquote  [1]{``#1''}%
\providecommand \bibnamefont  [1]{#1}%
\providecommand \bibfnamefont [1]{#1}%
\providecommand \citenamefont [1]{#1}%
\providecommand \href@noop [0]{\@secondoftwo}%
\providecommand \href [0]{\begingroup \@sanitize@url \@href}%
\providecommand \@href[1]{\@@startlink{#1}\@@href}%
\providecommand \@@href[1]{\endgroup#1\@@endlink}%
\providecommand \@sanitize@url [0]{\catcode `\\12\catcode `\$12\catcode
  `\&12\catcode `\#12\catcode `\^12\catcode `\_12\catcode `\%12\relax}%
\providecommand \@@startlink[1]{}%
\providecommand \@@endlink[0]{}%
\providecommand \url  [0]{\begingroup\@sanitize@url \@url }%
\providecommand \@url [1]{\endgroup\@href {#1}{\urlprefix }}%
\providecommand \urlprefix  [0]{URL }%
\providecommand \Eprint [0]{\href }%
\providecommand \doibase [0]{http://dx.doi.org/}%
\providecommand \selectlanguage [0]{\@gobble}%
\providecommand \bibinfo  [0]{\@secondoftwo}%
\providecommand \bibfield  [0]{\@secondoftwo}%
\providecommand \translation [1]{[#1]}%
\providecommand \BibitemOpen [0]{}%
\providecommand \bibitemStop [0]{}%
\providecommand \bibitemNoStop [0]{.\EOS\space}%
\providecommand \EOS [0]{\spacefactor3000\relax}%
\providecommand \BibitemShut  [1]{\csname bibitem#1\endcsname}%
\let\auto@bib@innerbib\@empty
%</preamble>
\bibitem [{\citenamefont {Abbott}\ \emph
  {et~al.}(2019{\natexlab{a}})\citenamefont {Abbott} \emph
  {et~al.}}]{LIGOScientific:2018mvr}%
  \BibitemOpen
  \bibfield  {author} {\bibinfo {author} {\bibfnamefont {B.~P.}\ \bibnamefont
  {Abbott}} \emph {et~al.} (\bibinfo {collaboration} {LIGO Scientific,
  Virgo}),\ }\href {\doibase 10.1103/PhysRevX.9.031040} {\bibfield  {journal}
  {\bibinfo  {journal} {Phys. Rev. X}\ }\textbf {\bibinfo {volume} {9}},\
  \bibinfo {pages} {031040} (\bibinfo {year} {2019}{\natexlab{a}})},\ \Eprint
  {http://arxiv.org/abs/1811.12907} {arXiv:1811.12907 [astro-ph.HE]}
  \BibitemShut {NoStop}%
\bibitem [{\citenamefont {Abbott}\ \emph
  {et~al.}(2021{\natexlab{a}})\citenamefont {Abbott} \emph
  {et~al.}}]{LIGOScientific:2020ibl}%
  \BibitemOpen
  \bibfield  {author} {\bibinfo {author} {\bibfnamefont {R.}~\bibnamefont
  {Abbott}} \emph {et~al.} (\bibinfo {collaboration} {LIGO Scientific,
  Virgo}),\ }\href {\doibase 10.1103/PhysRevX.11.021053} {\bibfield  {journal}
  {\bibinfo  {journal} {Phys. Rev. X}\ }\textbf {\bibinfo {volume} {11}},\
  \bibinfo {pages} {021053} (\bibinfo {year} {2021}{\natexlab{a}})},\ \Eprint
  {http://arxiv.org/abs/2010.14527} {arXiv:2010.14527 [gr-qc]} \BibitemShut
  {NoStop}%
\bibitem [{\citenamefont {Abbott}\ \emph
  {et~al.}(2021{\natexlab{b}})\citenamefont {Abbott} \emph
  {et~al.}}]{LIGOScientific:2021djp}%
  \BibitemOpen
  \bibfield  {author} {\bibinfo {author} {\bibfnamefont {R.}~\bibnamefont
  {Abbott}} \emph {et~al.} (\bibinfo {collaboration} {LIGO Scientific, VIRGO,
  KAGRA}),\ }\href@noop {} {\  (\bibinfo {year} {2021}{\natexlab{b}})},\
  \Eprint {http://arxiv.org/abs/2111.03606} {arXiv:2111.03606 [gr-qc]}
  \BibitemShut {NoStop}%
\bibitem [{\citenamefont {Abbott}\ \emph
  {et~al.}(2019{\natexlab{b}})\citenamefont {Abbott} \emph
  {et~al.}}]{LIGOScientific:2018jsj}%
  \BibitemOpen
  \bibfield  {author} {\bibinfo {author} {\bibfnamefont {B.~P.}\ \bibnamefont
  {Abbott}} \emph {et~al.} (\bibinfo {collaboration} {LIGO Scientific,
  Virgo}),\ }\href {\doibase 10.3847/2041-8213/ab3800} {\bibfield  {journal}
  {\bibinfo  {journal} {Astrophys. J. Lett.}\ }\textbf {\bibinfo {volume}
  {882}},\ \bibinfo {pages} {L24} (\bibinfo {year} {2019}{\natexlab{b}})},\
  \Eprint {http://arxiv.org/abs/1811.12940} {arXiv:1811.12940 [astro-ph.HE]}
  \BibitemShut {NoStop}%
\bibitem [{\citenamefont {Abbott}\ \emph
  {et~al.}(2021{\natexlab{c}})\citenamefont {Abbott} \emph
  {et~al.}}]{LIGOScientific:2020kqk}%
  \BibitemOpen
  \bibfield  {author} {\bibinfo {author} {\bibfnamefont {R.}~\bibnamefont
  {Abbott}} \emph {et~al.} (\bibinfo {collaboration} {LIGO Scientific,
  Virgo}),\ }\href {\doibase 10.3847/2041-8213/abe949} {\bibfield  {journal}
  {\bibinfo  {journal} {Astrophys. J. Lett.}\ }\textbf {\bibinfo {volume}
  {913}},\ \bibinfo {pages} {L7} (\bibinfo {year} {2021}{\natexlab{c}})},\
  \Eprint {http://arxiv.org/abs/2010.14533} {arXiv:2010.14533 [astro-ph.HE]}
  \BibitemShut {NoStop}%
\bibitem [{\citenamefont {Abbott}\ \emph
  {et~al.}(2021{\natexlab{d}})\citenamefont {Abbott} \emph
  {et~al.}}]{LIGOScientific:2021psn}%
  \BibitemOpen
  \bibfield  {author} {\bibinfo {author} {\bibfnamefont {R.}~\bibnamefont
  {Abbott}} \emph {et~al.} (\bibinfo {collaboration} {LIGO Scientific, VIRGO,
  KAGRA}),\ }\href@noop {} {\  (\bibinfo {year} {2021}{\natexlab{d}})},\
  \Eprint {http://arxiv.org/abs/2111.03634} {arXiv:2111.03634 [astro-ph.HE]}
  \BibitemShut {NoStop}%
\bibitem [{\citenamefont {Abbott}\ \emph
  {et~al.}(2019{\natexlab{c}})\citenamefont {Abbott} \emph
  {et~al.}}]{LIGOScientific:2019fpa}%
  \BibitemOpen
  \bibfield  {author} {\bibinfo {author} {\bibfnamefont {B.~P.}\ \bibnamefont
  {Abbott}} \emph {et~al.} (\bibinfo {collaboration} {LIGO Scientific,
  Virgo}),\ }\href {\doibase 10.1103/PhysRevD.100.104036} {\bibfield  {journal}
  {\bibinfo  {journal} {Phys. Rev. D}\ }\textbf {\bibinfo {volume} {100}},\
  \bibinfo {pages} {104036} (\bibinfo {year} {2019}{\natexlab{c}})},\ \Eprint
  {http://arxiv.org/abs/1903.04467} {arXiv:1903.04467 [gr-qc]} \BibitemShut
  {NoStop}%
\bibitem [{\citenamefont {Abbott}\ \emph
  {et~al.}(2021{\natexlab{e}})\citenamefont {Abbott} \emph
  {et~al.}}]{LIGOScientific:2020tif}%
  \BibitemOpen
  \bibfield  {author} {\bibinfo {author} {\bibfnamefont {R.}~\bibnamefont
  {Abbott}} \emph {et~al.} (\bibinfo {collaboration} {LIGO Scientific,
  Virgo}),\ }\href {\doibase 10.1103/PhysRevD.103.122002} {\bibfield  {journal}
  {\bibinfo  {journal} {Phys. Rev. D}\ }\textbf {\bibinfo {volume} {103}},\
  \bibinfo {pages} {122002} (\bibinfo {year} {2021}{\natexlab{e}})},\ \Eprint
  {http://arxiv.org/abs/2010.14529} {arXiv:2010.14529 [gr-qc]} \BibitemShut
  {NoStop}%
\bibitem [{\citenamefont {Abbott}\ \emph
  {et~al.}(2021{\natexlab{f}})\citenamefont {Abbott} \emph
  {et~al.}}]{LIGOScientific:2021sio}%
  \BibitemOpen
  \bibfield  {author} {\bibinfo {author} {\bibfnamefont {R.}~\bibnamefont
  {Abbott}} \emph {et~al.} (\bibinfo {collaboration} {LIGO Scientific, VIRGO,
  KAGRA}),\ }\href@noop {} {\  (\bibinfo {year} {2021}{\natexlab{f}})},\
  \Eprint {http://arxiv.org/abs/2112.06861} {arXiv:2112.06861 [gr-qc]}
  \BibitemShut {NoStop}%
\bibitem [{\citenamefont {Blanchet}\ and\ \citenamefont
  {Damour}(1992)}]{Blanchet:1992br}%
  \BibitemOpen
  \bibfield  {author} {\bibinfo {author} {\bibfnamefont {L.}~\bibnamefont
  {Blanchet}}\ and\ \bibinfo {author} {\bibfnamefont {T.}~\bibnamefont
  {Damour}},\ }\href {\doibase 10.1103/PhysRevD.46.4304} {\bibfield  {journal}
  {\bibinfo  {journal} {Phys. Rev. D}\ }\textbf {\bibinfo {volume} {46}},\
  \bibinfo {pages} {4304} (\bibinfo {year} {1992})}\BibitemShut {NoStop}%
\bibitem [{\citenamefont {Blanchet}(2014)}]{Blanchet:2013haa}%
  \BibitemOpen
  \bibfield  {author} {\bibinfo {author} {\bibfnamefont {L.}~\bibnamefont
  {Blanchet}},\ }\href {\doibase 10.12942/lrr-2014-2} {\bibfield  {journal}
  {\bibinfo  {journal} {Living Rev. Rel.}\ }\textbf {\bibinfo {volume} {17}},\
  \bibinfo {pages} {2} (\bibinfo {year} {2014})},\ \Eprint
  {http://arxiv.org/abs/1310.1528} {arXiv:1310.1528 [gr-qc]} \BibitemShut
  {NoStop}%
\bibitem [{\citenamefont {Zel'dovich}\ and\ \citenamefont
  {Polnarev}(1974)}]{zel1974}%
  \BibitemOpen
  \bibfield  {author} {\bibinfo {author} {\bibfnamefont {Y.~B.}\ \bibnamefont
  {Zel'dovich}}\ and\ \bibinfo {author} {\bibfnamefont {A.~G.}\ \bibnamefont
  {Polnarev}},\ }\href@noop {} {\bibfield  {journal} {\bibinfo  {journal} {Sov.
  Astron. 18}\ }\textbf {\bibinfo {volume} {17}} (\bibinfo {year}
  {1974})}\BibitemShut {NoStop}%
\bibitem [{\citenamefont {Braginsky}\ and\ \citenamefont
  {Throne}(1987)}]{throne1987}%
  \BibitemOpen
  \bibfield  {author} {\bibinfo {author} {\bibfnamefont {V.~B.}\ \bibnamefont
  {Braginsky}}\ and\ \bibinfo {author} {\bibfnamefont {K.~S.}\ \bibnamefont
  {Throne}},\ }\href@noop {} {\bibfield  {journal} {\bibinfo  {journal} {Nature
  (London) 327}\ }\textbf {\bibinfo {volume} {123}} (\bibinfo {year}
  {1987})}\BibitemShut {NoStop}%
\bibitem [{\citenamefont {Christodoulou}(1991)}]{PhysRevLett.67.1486}%
  \BibitemOpen
  \bibfield  {author} {\bibinfo {author} {\bibfnamefont {D.}~\bibnamefont
  {Christodoulou}},\ }\href {\doibase 10.1103/PhysRevLett.67.1486} {\bibfield
  {journal} {\bibinfo  {journal} {Phys. Rev. Lett.}\ }\textbf {\bibinfo
  {volume} {67}},\ \bibinfo {pages} {1486} (\bibinfo {year}
  {1991})}\BibitemShut {NoStop}%
\bibitem [{\citenamefont {Bieri}\ and\ \citenamefont
  {Garfinkle}(2014)}]{Bieri:2013ada}%
  \BibitemOpen
  \bibfield  {author} {\bibinfo {author} {\bibfnamefont {L.}~\bibnamefont
  {Bieri}}\ and\ \bibinfo {author} {\bibfnamefont {D.}~\bibnamefont
  {Garfinkle}},\ }\href {\doibase 10.1103/PhysRevD.89.084039} {\bibfield
  {journal} {\bibinfo  {journal} {Phys. Rev. D}\ }\textbf {\bibinfo {volume}
  {89}},\ \bibinfo {pages} {084039} (\bibinfo {year} {2014})},\ \Eprint
  {http://arxiv.org/abs/1312.6871} {arXiv:1312.6871 [gr-qc]} \BibitemShut
  {NoStop}%
\bibitem [{\citenamefont {Nichols}(2018)}]{Nichols:2018qac}%
  \BibitemOpen
  \bibfield  {author} {\bibinfo {author} {\bibfnamefont {D.~A.}\ \bibnamefont
  {Nichols}},\ }\href {\doibase 10.1103/PhysRevD.98.064032} {\bibfield
  {journal} {\bibinfo  {journal} {Phys. Rev. D}\ }\textbf {\bibinfo {volume}
  {98}},\ \bibinfo {pages} {064032} (\bibinfo {year} {2018})},\ \Eprint
  {http://arxiv.org/abs/1807.08767} {arXiv:1807.08767 [gr-qc]} \BibitemShut
  {NoStop}%
\bibitem [{\citenamefont {Bondi}\ \emph {et~al.}(1962)\citenamefont {Bondi},
  \citenamefont {van~der Burg},\ and\ \citenamefont {Metzner}}]{Bondi:1962px}%
  \BibitemOpen
  \bibfield  {author} {\bibinfo {author} {\bibfnamefont {H.}~\bibnamefont
  {Bondi}}, \bibinfo {author} {\bibfnamefont {M.~G.~J.}\ \bibnamefont {van~der
  Burg}}, \ and\ \bibinfo {author} {\bibfnamefont {A.~W.~K.}\ \bibnamefont
  {Metzner}},\ }\href {\doibase 10.1098/rspa.1962.0161} {\bibfield  {journal}
  {\bibinfo  {journal} {Proc. Roy. Soc. Lond. A}\ }\textbf {\bibinfo {volume}
  {269}},\ \bibinfo {pages} {21} (\bibinfo {year} {1962})}\BibitemShut
  {NoStop}%
\bibitem [{\citenamefont {Sachs}(1962)}]{Sachs:1962wk}%
  \BibitemOpen
  \bibfield  {author} {\bibinfo {author} {\bibfnamefont {R.~K.}\ \bibnamefont
  {Sachs}},\ }\href {\doibase 10.1098/rspa.1962.0206} {\bibfield  {journal}
  {\bibinfo  {journal} {Proc. Roy. Soc. Lond. A}\ }\textbf {\bibinfo {volume}
  {270}},\ \bibinfo {pages} {103} (\bibinfo {year} {1962})}\BibitemShut
  {NoStop}%
\bibitem [{\citenamefont {de~Boer}\ and\ \citenamefont
  {Solodukhin}(2003)}]{deBoer:2003vf}%
  \BibitemOpen
  \bibfield  {author} {\bibinfo {author} {\bibfnamefont {J.}~\bibnamefont
  {de~Boer}}\ and\ \bibinfo {author} {\bibfnamefont {S.~N.}\ \bibnamefont
  {Solodukhin}},\ }\href {\doibase 10.1016/S0550-3213(03)00494-2} {\bibfield
  {journal} {\bibinfo  {journal} {Nucl. Phys. B}\ }\textbf {\bibinfo {volume}
  {665}},\ \bibinfo {pages} {545} (\bibinfo {year} {2003})},\ \Eprint
  {http://arxiv.org/abs/hep-th/0303006} {arXiv:hep-th/0303006} \BibitemShut
  {NoStop}%
\bibitem [{\citenamefont {Barnich}\ and\ \citenamefont
  {Troessaert}(2010{\natexlab{a}})}]{Barnich:2009se}%
  \BibitemOpen
  \bibfield  {author} {\bibinfo {author} {\bibfnamefont {G.}~\bibnamefont
  {Barnich}}\ and\ \bibinfo {author} {\bibfnamefont {C.}~\bibnamefont
  {Troessaert}},\ }\href {\doibase 10.1103/PhysRevLett.105.111103} {\bibfield
  {journal} {\bibinfo  {journal} {Phys. Rev. Lett.}\ }\textbf {\bibinfo
  {volume} {105}},\ \bibinfo {pages} {111103} (\bibinfo {year}
  {2010}{\natexlab{a}})},\ \Eprint {http://arxiv.org/abs/0909.2617}
  {arXiv:0909.2617 [gr-qc]} \BibitemShut {NoStop}%
\bibitem [{\citenamefont {Barnich}\ and\ \citenamefont
  {Troessaert}(2010{\natexlab{b}})}]{Barnich:2010ojg}%
  \BibitemOpen
  \bibfield  {author} {\bibinfo {author} {\bibfnamefont {G.}~\bibnamefont
  {Barnich}}\ and\ \bibinfo {author} {\bibfnamefont {C.}~\bibnamefont
  {Troessaert}},\ }\href {\doibase 10.22323/1.127.0010} {\bibfield  {journal}
  {\bibinfo  {journal} {PoS}\ }\textbf {\bibinfo {volume} {CNCFG2010}},\
  \bibinfo {pages} {010} (\bibinfo {year} {2010}{\natexlab{b}})},\ \Eprint
  {http://arxiv.org/abs/1102.4632} {arXiv:1102.4632 [gr-qc]} \BibitemShut
  {NoStop}%
\bibitem [{\citenamefont {Kapec}\ \emph {et~al.}(2014)\citenamefont {Kapec},
  \citenamefont {Lysov}, \citenamefont {Pasterski},\ and\ \citenamefont
  {Strominger}}]{Kapec:2014opa}%
  \BibitemOpen
  \bibfield  {author} {\bibinfo {author} {\bibfnamefont {D.}~\bibnamefont
  {Kapec}}, \bibinfo {author} {\bibfnamefont {V.}~\bibnamefont {Lysov}},
  \bibinfo {author} {\bibfnamefont {S.}~\bibnamefont {Pasterski}}, \ and\
  \bibinfo {author} {\bibfnamefont {A.}~\bibnamefont {Strominger}},\ }\href
  {\doibase 10.1007/JHEP08(2014)058} {\bibfield  {journal} {\bibinfo  {journal}
  {JHEP}\ }\textbf {\bibinfo {volume} {08}},\ \bibinfo {pages} {058} (\bibinfo
  {year} {2014})},\ \Eprint {http://arxiv.org/abs/1406.3312} {arXiv:1406.3312
  [hep-th]} \BibitemShut {NoStop}%
\bibitem [{\citenamefont {Kapec}\ \emph {et~al.}(2017)\citenamefont {Kapec},
  \citenamefont {Mitra}, \citenamefont {Raclariu},\ and\ \citenamefont
  {Strominger}}]{Kapec:2016jld}%
  \BibitemOpen
  \bibfield  {author} {\bibinfo {author} {\bibfnamefont {D.}~\bibnamefont
  {Kapec}}, \bibinfo {author} {\bibfnamefont {P.}~\bibnamefont {Mitra}},
  \bibinfo {author} {\bibfnamefont {A.-M.}\ \bibnamefont {Raclariu}}, \ and\
  \bibinfo {author} {\bibfnamefont {A.}~\bibnamefont {Strominger}},\ }\href
  {\doibase 10.1103/PhysRevLett.119.121601} {\bibfield  {journal} {\bibinfo
  {journal} {Phys. Rev. Lett.}\ }\textbf {\bibinfo {volume} {119}},\ \bibinfo
  {pages} {121601} (\bibinfo {year} {2017})},\ \Eprint
  {http://arxiv.org/abs/1609.00282} {arXiv:1609.00282 [hep-th]} \BibitemShut
  {NoStop}%
\bibitem [{\citenamefont {He}\ \emph {et~al.}(2017)\citenamefont {He},
  \citenamefont {Kapec}, \citenamefont {Raclariu},\ and\ \citenamefont
  {Strominger}}]{He:2017fsb}%
  \BibitemOpen
  \bibfield  {author} {\bibinfo {author} {\bibfnamefont {T.}~\bibnamefont
  {He}}, \bibinfo {author} {\bibfnamefont {D.}~\bibnamefont {Kapec}}, \bibinfo
  {author} {\bibfnamefont {A.-M.}\ \bibnamefont {Raclariu}}, \ and\ \bibinfo
  {author} {\bibfnamefont {A.}~\bibnamefont {Strominger}},\ }\href {\doibase
  10.1007/JHEP08(2017)050} {\bibfield  {journal} {\bibinfo  {journal} {JHEP}\
  }\textbf {\bibinfo {volume} {08}},\ \bibinfo {pages} {050} (\bibinfo {year}
  {2017})},\ \Eprint {http://arxiv.org/abs/1701.00496} {arXiv:1701.00496
  [hep-th]} \BibitemShut {NoStop}%
\bibitem [{\citenamefont {Strominger}\ and\ \citenamefont
  {Zhiboedov}(2016)}]{Strominger:2014pwa}%
  \BibitemOpen
  \bibfield  {author} {\bibinfo {author} {\bibfnamefont {A.}~\bibnamefont
  {Strominger}}\ and\ \bibinfo {author} {\bibfnamefont {A.}~\bibnamefont
  {Zhiboedov}},\ }\href {\doibase 10.1007/JHEP01(2016)086} {\bibfield
  {journal} {\bibinfo  {journal} {JHEP}\ }\textbf {\bibinfo {volume} {01}},\
  \bibinfo {pages} {086} (\bibinfo {year} {2016})},\ \Eprint
  {http://arxiv.org/abs/1411.5745} {arXiv:1411.5745 [hep-th]} \BibitemShut
  {NoStop}%
\bibitem [{\citenamefont {Pasterski}\ \emph {et~al.}(2016)\citenamefont
  {Pasterski}, \citenamefont {Strominger},\ and\ \citenamefont
  {Zhiboedov}}]{Pasterski:2015tva}%
  \BibitemOpen
  \bibfield  {author} {\bibinfo {author} {\bibfnamefont {S.}~\bibnamefont
  {Pasterski}}, \bibinfo {author} {\bibfnamefont {A.}~\bibnamefont
  {Strominger}}, \ and\ \bibinfo {author} {\bibfnamefont {A.}~\bibnamefont
  {Zhiboedov}},\ }\href {\doibase 10.1007/JHEP12(2016)053} {\bibfield
  {journal} {\bibinfo  {journal} {JHEP}\ }\textbf {\bibinfo {volume} {12}},\
  \bibinfo {pages} {053} (\bibinfo {year} {2016})},\ \Eprint
  {http://arxiv.org/abs/1502.06120} {arXiv:1502.06120 [hep-th]} \BibitemShut
  {NoStop}%
\bibitem [{\citenamefont {Comp\`ere}\ \emph {et~al.}(2020)\citenamefont
  {Comp\`ere}, \citenamefont {Oliveri},\ and\ \citenamefont
  {Seraj}}]{Compere:2019gft}%
  \BibitemOpen
  \bibfield  {author} {\bibinfo {author} {\bibfnamefont {G.}~\bibnamefont
  {Comp\`ere}}, \bibinfo {author} {\bibfnamefont {R.}~\bibnamefont {Oliveri}},
  \ and\ \bibinfo {author} {\bibfnamefont {A.}~\bibnamefont {Seraj}},\ }\href
  {\doibase 10.1007/JHEP10(2020)116} {\bibfield  {journal} {\bibinfo  {journal}
  {JHEP}\ }\textbf {\bibinfo {volume} {10}},\ \bibinfo {pages} {116} (\bibinfo
  {year} {2020})},\ \Eprint {http://arxiv.org/abs/1912.03164} {arXiv:1912.03164
  [gr-qc]} \BibitemShut {NoStop}%
\bibitem [{\citenamefont {Barnich}\ and\ \citenamefont
  {Troessaert}(2011)}]{Barnich:2011mi}%
  \BibitemOpen
  \bibfield  {author} {\bibinfo {author} {\bibfnamefont {G.}~\bibnamefont
  {Barnich}}\ and\ \bibinfo {author} {\bibfnamefont {C.}~\bibnamefont
  {Troessaert}},\ }\href {\doibase 10.1007/JHEP12(2011)105} {\bibfield
  {journal} {\bibinfo  {journal} {JHEP}\ }\textbf {\bibinfo {volume} {12}},\
  \bibinfo {pages} {105} (\bibinfo {year} {2011})},\ \Eprint
  {http://arxiv.org/abs/1106.0213} {arXiv:1106.0213 [hep-th]} \BibitemShut
  {NoStop}%
\bibitem [{\citenamefont {Flanagan}\ and\ \citenamefont
  {Nichols}(2017)}]{Flanagan:2015pxa}%
  \BibitemOpen
  \bibfield  {author} {\bibinfo {author} {\bibfnamefont {E.~E.}\ \bibnamefont
  {Flanagan}}\ and\ \bibinfo {author} {\bibfnamefont {D.~A.}\ \bibnamefont
  {Nichols}},\ }\href {\doibase 10.1103/PhysRevD.95.044002} {\bibfield
  {journal} {\bibinfo  {journal} {Phys. Rev. D}\ }\textbf {\bibinfo {volume}
  {95}},\ \bibinfo {pages} {044002} (\bibinfo {year} {2017})},\ \Eprint
  {http://arxiv.org/abs/1510.03386} {arXiv:1510.03386 [hep-th]} \BibitemShut
  {NoStop}%
\bibitem [{\citenamefont {Nichols}(2017)}]{Nichols:2017rqr}%
  \BibitemOpen
  \bibfield  {author} {\bibinfo {author} {\bibfnamefont {D.~A.}\ \bibnamefont
  {Nichols}},\ }\href {\doibase 10.1103/PhysRevD.95.084048} {\bibfield
  {journal} {\bibinfo  {journal} {Phys. Rev. D}\ }\textbf {\bibinfo {volume}
  {95}},\ \bibinfo {pages} {084048} (\bibinfo {year} {2017})},\ \Eprint
  {http://arxiv.org/abs/1702.03300} {arXiv:1702.03300 [gr-qc]} \BibitemShut
  {NoStop}%
\bibitem [{\citenamefont {Mitman}\ \emph {et~al.}(2020)\citenamefont {Mitman},
  \citenamefont {Moxon}, \citenamefont {Scheel}, \citenamefont {Teukolsky},
  \citenamefont {Boyle}, \citenamefont {Deppe}, \citenamefont {Kidder},\ and\
  \citenamefont {Throwe}}]{Mitman:2020pbt}%
  \BibitemOpen
  \bibfield  {author} {\bibinfo {author} {\bibfnamefont {K.}~\bibnamefont
  {Mitman}}, \bibinfo {author} {\bibfnamefont {J.}~\bibnamefont {Moxon}},
  \bibinfo {author} {\bibfnamefont {M.~A.}\ \bibnamefont {Scheel}}, \bibinfo
  {author} {\bibfnamefont {S.~A.}\ \bibnamefont {Teukolsky}}, \bibinfo {author}
  {\bibfnamefont {M.}~\bibnamefont {Boyle}}, \bibinfo {author} {\bibfnamefont
  {N.}~\bibnamefont {Deppe}}, \bibinfo {author} {\bibfnamefont {L.~E.}\
  \bibnamefont {Kidder}}, \ and\ \bibinfo {author} {\bibfnamefont
  {W.}~\bibnamefont {Throwe}},\ }\href {\doibase 10.1103/PhysRevD.102.104007}
  {\bibfield  {journal} {\bibinfo  {journal} {Phys. Rev. D}\ }\textbf {\bibinfo
  {volume} {102}},\ \bibinfo {pages} {104007} (\bibinfo {year} {2020})},\
  \Eprint {http://arxiv.org/abs/2007.11562} {arXiv:2007.11562 [gr-qc]}
  \BibitemShut {NoStop}%
\bibitem [{\citenamefont {Mitman}\ \emph {et~al.}(2021)\citenamefont {Mitman}
  \emph {et~al.}}]{Mitman:2020bjf}%
  \BibitemOpen
  \bibfield  {author} {\bibinfo {author} {\bibfnamefont {K.}~\bibnamefont
  {Mitman}} \emph {et~al.},\ }\href {\doibase 10.1103/PhysRevD.103.024031}
  {\bibfield  {journal} {\bibinfo  {journal} {Phys. Rev. D}\ }\textbf {\bibinfo
  {volume} {103}},\ \bibinfo {pages} {024031} (\bibinfo {year} {2021})},\
  \Eprint {http://arxiv.org/abs/2011.01309} {arXiv:2011.01309 [gr-qc]}
  \BibitemShut {NoStop}%
\bibitem [{\citenamefont {Jenkins}\ and\ \citenamefont
  {Sakellariadou}(2021)}]{Jenkins:2021kcj}%
  \BibitemOpen
  \bibfield  {author} {\bibinfo {author} {\bibfnamefont {A.~C.}\ \bibnamefont
  {Jenkins}}\ and\ \bibinfo {author} {\bibfnamefont {M.}~\bibnamefont
  {Sakellariadou}},\ }\href {\doibase 10.1088/1361-6382/ac1084} {\bibfield
  {journal} {\bibinfo  {journal} {Class. Quant. Grav.}\ }\textbf {\bibinfo
  {volume} {38}},\ \bibinfo {pages} {165004} (\bibinfo {year} {2021})},\
  \Eprint {http://arxiv.org/abs/2102.12487} {arXiv:2102.12487 [gr-qc]}
  \BibitemShut {NoStop}%
\bibitem [{\citenamefont {Mukhopadhyay}\ \emph {et~al.}(2021)\citenamefont
  {Mukhopadhyay}, \citenamefont {Cardona},\ and\ \citenamefont
  {Lunardini}}]{Mukhopadhyay:2021zbt}%
  \BibitemOpen
  \bibfield  {author} {\bibinfo {author} {\bibfnamefont {M.}~\bibnamefont
  {Mukhopadhyay}}, \bibinfo {author} {\bibfnamefont {C.}~\bibnamefont
  {Cardona}}, \ and\ \bibinfo {author} {\bibfnamefont {C.}~\bibnamefont
  {Lunardini}},\ }\href {\doibase 10.1088/1475-7516/2021/07/055} {\bibfield
  {journal} {\bibinfo  {journal} {JCAP}\ }\textbf {\bibinfo {volume} {07}},\
  \bibinfo {pages} {055} (\bibinfo {year} {2021})},\ \Eprint
  {http://arxiv.org/abs/2105.05862} {arXiv:2105.05862 [astro-ph.HE]}
  \BibitemShut {NoStop}%
\bibitem [{\citenamefont {Du}\ and\ \citenamefont
  {Nishizawa}(2016)}]{Du:2016hww}%
  \BibitemOpen
  \bibfield  {author} {\bibinfo {author} {\bibfnamefont {S.~M.}\ \bibnamefont
  {Du}}\ and\ \bibinfo {author} {\bibfnamefont {A.}~\bibnamefont {Nishizawa}},\
  }\href {\doibase 10.1103/PhysRevD.94.104063} {\bibfield  {journal} {\bibinfo
  {journal} {Phys. Rev. D}\ }\textbf {\bibinfo {volume} {94}},\ \bibinfo
  {pages} {104063} (\bibinfo {year} {2016})},\ \Eprint
  {http://arxiv.org/abs/1609.09825} {arXiv:1609.09825 [gr-qc]} \BibitemShut
  {NoStop}%
\bibitem [{\citenamefont {Seraj}(2021)}]{Seraj:2021qja}%
  \BibitemOpen
  \bibfield  {author} {\bibinfo {author} {\bibfnamefont {A.}~\bibnamefont
  {Seraj}},\ }\href {\doibase 10.1007/JHEP05(2021)283} {\bibfield  {journal}
  {\bibinfo  {journal} {JHEP}\ }\textbf {\bibinfo {volume} {05}},\ \bibinfo
  {pages} {283} (\bibinfo {year} {2021})},\ \Eprint
  {http://arxiv.org/abs/2103.12185} {arXiv:2103.12185 [hep-th]} \BibitemShut
  {NoStop}%
\bibitem [{\citenamefont {Tahura}\ \emph {et~al.}(2021)\citenamefont {Tahura},
  \citenamefont {Nichols},\ and\ \citenamefont {Yagi}}]{Tahura:2021hbk}%
  \BibitemOpen
  \bibfield  {author} {\bibinfo {author} {\bibfnamefont {S.}~\bibnamefont
  {Tahura}}, \bibinfo {author} {\bibfnamefont {D.~A.}\ \bibnamefont {Nichols}},
  \ and\ \bibinfo {author} {\bibfnamefont {K.}~\bibnamefont {Yagi}},\ }\href
  {\doibase 10.1103/PhysRevD.104.104010} {\bibfield  {journal} {\bibinfo
  {journal} {Phys. Rev. D}\ }\textbf {\bibinfo {volume} {104}},\ \bibinfo
  {pages} {104010} (\bibinfo {year} {2021})},\ \Eprint
  {http://arxiv.org/abs/2107.02208} {arXiv:2107.02208 [gr-qc]} \BibitemShut
  {NoStop}%
\bibitem [{\citenamefont {Hou}\ \emph {et~al.}(2022)\citenamefont {Hou},
  \citenamefont {Zhu},\ and\ \citenamefont {Zhu}}]{Hou:2021bxz}%
  \BibitemOpen
  \bibfield  {author} {\bibinfo {author} {\bibfnamefont {S.}~\bibnamefont
  {Hou}}, \bibinfo {author} {\bibfnamefont {T.}~\bibnamefont {Zhu}}, \ and\
  \bibinfo {author} {\bibfnamefont {Z.-H.}\ \bibnamefont {Zhu}},\ }\href
  {\doibase 10.1088/1475-7516/2022/04/032} {\bibfield  {journal} {\bibinfo
  {journal} {JCAP}\ }\textbf {\bibinfo {volume} {04}},\ \bibinfo {pages} {032}
  (\bibinfo {year} {2022})},\ \Eprint {http://arxiv.org/abs/2112.13049}
  {arXiv:2112.13049 [gr-qc]} \BibitemShut {NoStop}%
\bibitem [{\citenamefont {Koyama}(2020)}]{Koyama:2020vfc}%
  \BibitemOpen
  \bibfield  {author} {\bibinfo {author} {\bibfnamefont {K.}~\bibnamefont
  {Koyama}},\ }\href {\doibase 10.1103/PhysRevD.102.021502} {\bibfield
  {journal} {\bibinfo  {journal} {Phys. Rev. D}\ }\textbf {\bibinfo {volume}
  {102}},\ \bibinfo {pages} {021502} (\bibinfo {year} {2020})},\ \Eprint
  {http://arxiv.org/abs/2006.15914} {arXiv:2006.15914 [gr-qc]} \BibitemShut
  {NoStop}%
\bibitem [{\citenamefont {Braginsky}\ and\ \citenamefont
  {Grishchuk}(1985)}]{zel1985}%
  \BibitemOpen
  \bibfield  {author} {\bibinfo {author} {\bibfnamefont {V.~B.}\ \bibnamefont
  {Braginsky}}\ and\ \bibinfo {author} {\bibfnamefont {L.~P.}\ \bibnamefont
  {Grishchuk}},\ }\href@noop {} {\bibfield  {journal} {\bibinfo  {journal}
  {Sov. Phys. JETP 62}\ }\textbf {\bibinfo {volume} {427}} (\bibinfo {year}
  {1985})}\BibitemShut {NoStop}%
\bibitem [{\citenamefont {Abbott}\ \emph {et~al.}(2016)\citenamefont {Abbott}
  \emph {et~al.}}]{LIGOScientific:2016aoc}%
  \BibitemOpen
  \bibfield  {author} {\bibinfo {author} {\bibfnamefont {B.~P.}\ \bibnamefont
  {Abbott}} \emph {et~al.} (\bibinfo {collaboration} {LIGO Scientific,
  Virgo}),\ }\href {\doibase 10.1103/PhysRevLett.116.061102} {\bibfield
  {journal} {\bibinfo  {journal} {Phys. Rev. Lett.}\ }\textbf {\bibinfo
  {volume} {116}},\ \bibinfo {pages} {061102} (\bibinfo {year} {2016})},\
  \Eprint {http://arxiv.org/abs/1602.03837} {arXiv:1602.03837 [gr-qc]}
  \BibitemShut {NoStop}%
\bibitem [{\citenamefont {Seto}(2009)}]{10.1111/j.1745-3933.2009.00758.x}%
  \BibitemOpen
  \bibfield  {author} {\bibinfo {author} {\bibfnamefont {N.}~\bibnamefont
  {Seto}},\ }\href {\doibase 10.1111/j.1745-3933.2009.00758.x} {\bibfield
  {journal} {\bibinfo  {journal} {Mon. Not. Roy. Astron. Soc.}\ }\textbf
  {\bibinfo {volume} {400}},\ \bibinfo {pages} {L38} (\bibinfo {year}
  {2009})},\ \Eprint {http://arxiv.org/abs/0909.1379} {arXiv:0909.1379
  [astro-ph.CO]} \BibitemShut {NoStop}%
\bibitem [{\citenamefont {van Haasteren}\ and\ \citenamefont
  {Levin}(2010)}]{vanHaasteren:2009fy}%
  \BibitemOpen
  \bibfield  {author} {\bibinfo {author} {\bibfnamefont {R.}~\bibnamefont {van
  Haasteren}}\ and\ \bibinfo {author} {\bibfnamefont {Y.}~\bibnamefont
  {Levin}},\ }\href {\doibase 10.1111/j.1365-2966.2009.15885.x} {\bibfield
  {journal} {\bibinfo  {journal} {Mon. Not. Roy. Astron. Soc.}\ }\textbf
  {\bibinfo {volume} {401}},\ \bibinfo {pages} {2372} (\bibinfo {year}
  {2010})},\ \Eprint {http://arxiv.org/abs/0909.0954} {arXiv:0909.0954
  [astro-ph.IM]} \BibitemShut {NoStop}%
\bibitem [{\citenamefont {Pshirkov}\ \emph {et~al.}(2010)\citenamefont
  {Pshirkov}, \citenamefont {Baskaran},\ and\ \citenamefont
  {Postnov}}]{Pshirkov:2009ak}%
  \BibitemOpen
  \bibfield  {author} {\bibinfo {author} {\bibfnamefont {M.~S.}\ \bibnamefont
  {Pshirkov}}, \bibinfo {author} {\bibfnamefont {D.}~\bibnamefont {Baskaran}},
  \ and\ \bibinfo {author} {\bibfnamefont {K.~A.}\ \bibnamefont {Postnov}},\
  }\href {\doibase 10.1111/j.1365-2966.2009.15887.x} {\bibfield  {journal}
  {\bibinfo  {journal} {Mon. Not. Roy. Astron. Soc.}\ }\textbf {\bibinfo
  {volume} {402}},\ \bibinfo {pages} {417} (\bibinfo {year} {2010})},\ \Eprint
  {http://arxiv.org/abs/0909.0742} {arXiv:0909.0742 [astro-ph.CO]} \BibitemShut
  {NoStop}%
\bibitem [{\citenamefont {Cordes}\ and\ \citenamefont
  {Jenet}(2012)}]{Cordes:2012zz}%
  \BibitemOpen
  \bibfield  {author} {\bibinfo {author} {\bibfnamefont {J.~M.}\ \bibnamefont
  {Cordes}}\ and\ \bibinfo {author} {\bibfnamefont {F.~A.}\ \bibnamefont
  {Jenet}},\ }\href {\doibase 10.1088/0004-637X/752/1/54} {\bibfield  {journal}
  {\bibinfo  {journal} {Astrophys. J.}\ }\textbf {\bibinfo {volume} {752}},\
  \bibinfo {pages} {54} (\bibinfo {year} {2012})}\BibitemShut {NoStop}%
\bibitem [{\citenamefont {Madison}\ \emph {et~al.}(2014)\citenamefont
  {Madison}, \citenamefont {Cordes},\ and\ \citenamefont
  {Chatterjee}}]{madison}%
  \BibitemOpen
  \bibfield  {author} {\bibinfo {author} {\bibfnamefont {D.~R.}\ \bibnamefont
  {Madison}}, \bibinfo {author} {\bibfnamefont {J.~M.}\ \bibnamefont {Cordes}},
  \ and\ \bibinfo {author} {\bibfnamefont {S.}~\bibnamefont {Chatterjee}},\
  }\href {\doibase 10.1088/0004-637X/752/1/54} {\bibfield  {journal} {\bibinfo
  {journal} {Astrophys. J.}\ }\textbf {\bibinfo {volume} {788}},\ \bibinfo
  {pages} {741} (\bibinfo {year} {2014})}\BibitemShut {NoStop}%
\bibitem [{\citenamefont {Arzoumanian}\ \emph {et~al.}(2015)\citenamefont
  {Arzoumanian} \emph {et~al.}}]{NANOGrav:2015xuc}%
  \BibitemOpen
  \bibfield  {author} {\bibinfo {author} {\bibfnamefont {Z.}~\bibnamefont
  {Arzoumanian}} \emph {et~al.} (\bibinfo {collaboration} {NANOGrav}),\ }\href
  {\doibase 10.1088/0004-637X/810/2/150} {\bibfield  {journal} {\bibinfo
  {journal} {Astrophys. J.}\ }\textbf {\bibinfo {volume} {810}},\ \bibinfo
  {pages} {150} (\bibinfo {year} {2015})},\ \Eprint
  {http://arxiv.org/abs/1501.05343} {arXiv:1501.05343 [astro-ph.GA]}
  \BibitemShut {NoStop}%
\bibitem [{\citenamefont {Lasky}\ \emph {et~al.}(2016)\citenamefont {Lasky},
  \citenamefont {Thrane}, \citenamefont {Levin}, \citenamefont {Blackman},\
  and\ \citenamefont {Chen}}]{Lasky:2016knh}%
  \BibitemOpen
  \bibfield  {author} {\bibinfo {author} {\bibfnamefont {P.~D.}\ \bibnamefont
  {Lasky}}, \bibinfo {author} {\bibfnamefont {E.}~\bibnamefont {Thrane}},
  \bibinfo {author} {\bibfnamefont {Y.}~\bibnamefont {Levin}}, \bibinfo
  {author} {\bibfnamefont {J.}~\bibnamefont {Blackman}}, \ and\ \bibinfo
  {author} {\bibfnamefont {Y.}~\bibnamefont {Chen}},\ }\href {\doibase
  10.1103/PhysRevLett.117.061102} {\bibfield  {journal} {\bibinfo  {journal}
  {Phys. Rev. Lett.}\ }\textbf {\bibinfo {volume} {117}},\ \bibinfo {pages}
  {061102} (\bibinfo {year} {2016})},\ \Eprint
  {http://arxiv.org/abs/1605.01415} {arXiv:1605.01415 [astro-ph.HE]}
  \BibitemShut {NoStop}%
\bibitem [{\citenamefont {McNeill}\ \emph {et~al.}(2017)\citenamefont
  {McNeill}, \citenamefont {Thrane},\ and\ \citenamefont
  {Lasky}}]{McNeill:2017uvq}%
  \BibitemOpen
  \bibfield  {author} {\bibinfo {author} {\bibfnamefont {L.~O.}\ \bibnamefont
  {McNeill}}, \bibinfo {author} {\bibfnamefont {E.}~\bibnamefont {Thrane}}, \
  and\ \bibinfo {author} {\bibfnamefont {P.~D.}\ \bibnamefont {Lasky}},\ }\href
  {\doibase 10.1103/PhysRevLett.118.181103} {\bibfield  {journal} {\bibinfo
  {journal} {Phys. Rev. Lett.}\ }\textbf {\bibinfo {volume} {118}},\ \bibinfo
  {pages} {181103} (\bibinfo {year} {2017})},\ \Eprint
  {http://arxiv.org/abs/1702.01759} {arXiv:1702.01759 [astro-ph.IM]}
  \BibitemShut {NoStop}%
\bibitem [{\citenamefont {Divakarla}\ \emph {et~al.}(2020)\citenamefont
  {Divakarla}, \citenamefont {Thrane}, \citenamefont {Lasky},\ and\
  \citenamefont {Whiting}}]{Divakarla:2019zjj}%
  \BibitemOpen
  \bibfield  {author} {\bibinfo {author} {\bibfnamefont {A.~K.}\ \bibnamefont
  {Divakarla}}, \bibinfo {author} {\bibfnamefont {E.}~\bibnamefont {Thrane}},
  \bibinfo {author} {\bibfnamefont {P.~D.}\ \bibnamefont {Lasky}}, \ and\
  \bibinfo {author} {\bibfnamefont {B.~F.}\ \bibnamefont {Whiting}},\ }\href
  {\doibase 10.1103/PhysRevD.102.023010} {\bibfield  {journal} {\bibinfo
  {journal} {Phys. Rev. D}\ }\textbf {\bibinfo {volume} {102}},\ \bibinfo
  {pages} {023010} (\bibinfo {year} {2020})},\ \Eprint
  {http://arxiv.org/abs/1911.07998} {arXiv:1911.07998 [gr-qc]} \BibitemShut
  {NoStop}%
\bibitem [{\citenamefont {Boersma}\ \emph {et~al.}(2020)\citenamefont
  {Boersma}, \citenamefont {Nichols},\ and\ \citenamefont
  {Schmidt}}]{Boersma:2020gxx}%
  \BibitemOpen
  \bibfield  {author} {\bibinfo {author} {\bibfnamefont {O.~M.}\ \bibnamefont
  {Boersma}}, \bibinfo {author} {\bibfnamefont {D.~A.}\ \bibnamefont
  {Nichols}}, \ and\ \bibinfo {author} {\bibfnamefont {P.}~\bibnamefont
  {Schmidt}},\ }\href {\doibase 10.1103/PhysRevD.101.083026} {\bibfield
  {journal} {\bibinfo  {journal} {Phys. Rev. D}\ }\textbf {\bibinfo {volume}
  {101}},\ \bibinfo {pages} {083026} (\bibinfo {year} {2020})},\ \Eprint
  {http://arxiv.org/abs/2002.01821} {arXiv:2002.01821 [astro-ph.HE]}
  \BibitemShut {NoStop}%
\bibitem [{\citenamefont {Khera}\ \emph {et~al.}(2021)\citenamefont {Khera},
  \citenamefont {Krishnan}, \citenamefont {Ashtekar},\ and\ \citenamefont
  {De~Lorenzo}}]{Khera:2020mcz}%
  \BibitemOpen
  \bibfield  {author} {\bibinfo {author} {\bibfnamefont {N.}~\bibnamefont
  {Khera}}, \bibinfo {author} {\bibfnamefont {B.}~\bibnamefont {Krishnan}},
  \bibinfo {author} {\bibfnamefont {A.}~\bibnamefont {Ashtekar}}, \ and\
  \bibinfo {author} {\bibfnamefont {T.}~\bibnamefont {De~Lorenzo}},\ }\href
  {\doibase 10.1103/PhysRevD.103.044012} {\bibfield  {journal} {\bibinfo
  {journal} {Phys. Rev. D}\ }\textbf {\bibinfo {volume} {103}},\ \bibinfo
  {pages} {044012} (\bibinfo {year} {2021})},\ \Eprint
  {http://arxiv.org/abs/2009.06351} {arXiv:2009.06351 [gr-qc]} \BibitemShut
  {NoStop}%
\bibitem [{\citenamefont {H\"ubner}\ \emph {et~al.}(2020)\citenamefont
  {H\"ubner}, \citenamefont {Talbot}, \citenamefont {Lasky},\ and\
  \citenamefont {Thrane}}]{Hubner:2019sly}%
  \BibitemOpen
  \bibfield  {author} {\bibinfo {author} {\bibfnamefont {M.}~\bibnamefont
  {H\"ubner}}, \bibinfo {author} {\bibfnamefont {C.}~\bibnamefont {Talbot}},
  \bibinfo {author} {\bibfnamefont {P.~D.}\ \bibnamefont {Lasky}}, \ and\
  \bibinfo {author} {\bibfnamefont {E.}~\bibnamefont {Thrane}},\ }\href
  {\doibase 10.1103/PhysRevD.101.023011} {\bibfield  {journal} {\bibinfo
  {journal} {Phys. Rev. D}\ }\textbf {\bibinfo {volume} {101}},\ \bibinfo
  {pages} {023011} (\bibinfo {year} {2020})},\ \Eprint
  {http://arxiv.org/abs/1911.12496} {arXiv:1911.12496 [astro-ph.HE]}
  \BibitemShut {NoStop}%
\bibitem [{\citenamefont {H\"ubner}\ \emph {et~al.}(2021)\citenamefont
  {H\"ubner}, \citenamefont {Lasky},\ and\ \citenamefont
  {Thrane}}]{Hubner:2021amk}%
  \BibitemOpen
  \bibfield  {author} {\bibinfo {author} {\bibfnamefont {M.}~\bibnamefont
  {H\"ubner}}, \bibinfo {author} {\bibfnamefont {P.}~\bibnamefont {Lasky}}, \
  and\ \bibinfo {author} {\bibfnamefont {E.}~\bibnamefont {Thrane}},\ }\href
  {\doibase 10.1103/PhysRevD.104.023004} {\bibfield  {journal} {\bibinfo
  {journal} {Phys. Rev. D}\ }\textbf {\bibinfo {volume} {104}},\ \bibinfo
  {pages} {023004} (\bibinfo {year} {2021})},\ \Eprint
  {http://arxiv.org/abs/2105.02879} {arXiv:2105.02879 [gr-qc]} \BibitemShut
  {NoStop}%
\bibitem [{\citenamefont {Islam}\ \emph {et~al.}(2021)\citenamefont {Islam},
  \citenamefont {Field}, \citenamefont {Khanna},\ and\ \citenamefont
  {Warburton}}]{Islam:2021old}%
  \BibitemOpen
  \bibfield  {author} {\bibinfo {author} {\bibfnamefont {T.}~\bibnamefont
  {Islam}}, \bibinfo {author} {\bibfnamefont {S.~E.}\ \bibnamefont {Field}},
  \bibinfo {author} {\bibfnamefont {G.}~\bibnamefont {Khanna}}, \ and\ \bibinfo
  {author} {\bibfnamefont {N.}~\bibnamefont {Warburton}},\ }\href@noop {} {\
  (\bibinfo {year} {2021})},\ \Eprint {http://arxiv.org/abs/2109.00754}
  {arXiv:2109.00754 [gr-qc]} \BibitemShut {NoStop}%
\bibitem [{\citenamefont {Zhao}\ \emph {et~al.}(2021)\citenamefont {Zhao},
  \citenamefont {Liu}, \citenamefont {Cao},\ and\ \citenamefont
  {He}}]{Zhao:2021hmx}%
  \BibitemOpen
  \bibfield  {author} {\bibinfo {author} {\bibfnamefont {Z.-C.}\ \bibnamefont
  {Zhao}}, \bibinfo {author} {\bibfnamefont {X.}~\bibnamefont {Liu}}, \bibinfo
  {author} {\bibfnamefont {Z.}~\bibnamefont {Cao}}, \ and\ \bibinfo {author}
  {\bibfnamefont {X.}~\bibnamefont {He}},\ }\href {\doibase
  10.1103/PhysRevD.104.064056} {\bibfield  {journal} {\bibinfo  {journal}
  {Phys. Rev. D}\ }\textbf {\bibinfo {volume} {104}},\ \bibinfo {pages}
  {064056} (\bibinfo {year} {2021})}\BibitemShut {NoStop}%
\bibitem [{\citenamefont {Abbott}\ \emph {et~al.}(2017)\citenamefont {Abbott}
  \emph {et~al.}}]{LIGOScientific:2016wof}%
  \BibitemOpen
  \bibfield  {author} {\bibinfo {author} {\bibfnamefont {B.~P.}\ \bibnamefont
  {Abbott}} \emph {et~al.} (\bibinfo {collaboration} {LIGO Scientific}),\
  }\href {\doibase 10.1088/1361-6382/aa51f4} {\bibfield  {journal} {\bibinfo
  {journal} {Class. Quant. Grav.}\ }\textbf {\bibinfo {volume} {34}},\ \bibinfo
  {pages} {044001} (\bibinfo {year} {2017})},\ \Eprint
  {http://arxiv.org/abs/1607.08697} {arXiv:1607.08697 [astro-ph.IM]}
  \BibitemShut {NoStop}%
\bibitem [{\citenamefont {Punturo}\ \emph {et~al.}(2010)\citenamefont {Punturo}
  \emph {et~al.}}]{Punturo:2010zz}%
  \BibitemOpen
  \bibfield  {author} {\bibinfo {author} {\bibfnamefont {M.}~\bibnamefont
  {Punturo}} \emph {et~al.},\ }\href {\doibase 10.1088/0264-9381/27/19/194002}
  {\bibfield  {journal} {\bibinfo  {journal} {Class. Quant. Grav.}\ }\textbf
  {\bibinfo {volume} {27}},\ \bibinfo {pages} {194002} (\bibinfo {year}
  {2010})}\BibitemShut {NoStop}%
\bibitem [{\citenamefont {Yang}\ and\ \citenamefont
  {Martynov}(2018)}]{Yang:2018ceq}%
  \BibitemOpen
  \bibfield  {author} {\bibinfo {author} {\bibfnamefont {H.}~\bibnamefont
  {Yang}}\ and\ \bibinfo {author} {\bibfnamefont {D.}~\bibnamefont
  {Martynov}},\ }\href {\doibase 10.1103/PhysRevLett.121.071102} {\bibfield
  {journal} {\bibinfo  {journal} {Phys. Rev. Lett.}\ }\textbf {\bibinfo
  {volume} {121}},\ \bibinfo {pages} {071102} (\bibinfo {year} {2018})},\
  \Eprint {http://arxiv.org/abs/1803.02429} {arXiv:1803.02429 [gr-qc]}
  \BibitemShut {NoStop}%
\bibitem [{\citenamefont {Islo}\ \emph {et~al.}(2019)\citenamefont {Islo},
  \citenamefont {Simon}, \citenamefont {Burke-Spolaor},\ and\ \citenamefont
  {Siemens}}]{Islo:2019qht}%
  \BibitemOpen
  \bibfield  {author} {\bibinfo {author} {\bibfnamefont {K.}~\bibnamefont
  {Islo}}, \bibinfo {author} {\bibfnamefont {J.}~\bibnamefont {Simon}},
  \bibinfo {author} {\bibfnamefont {S.}~\bibnamefont {Burke-Spolaor}}, \ and\
  \bibinfo {author} {\bibfnamefont {X.}~\bibnamefont {Siemens}},\ }\href@noop
  {} {\  (\bibinfo {year} {2019})},\ \Eprint {http://arxiv.org/abs/1906.11936}
  {arXiv:1906.11936 [astro-ph.HE]} \BibitemShut {NoStop}%
\bibitem [{\citenamefont {Luo}\ \emph {et~al.}(2016)\citenamefont {Luo} \emph
  {et~al.}}]{TianQin:2015yph}%
  \BibitemOpen
  \bibfield  {author} {\bibinfo {author} {\bibfnamefont {J.}~\bibnamefont
  {Luo}} \emph {et~al.} (\bibinfo {collaboration} {TianQin}),\ }\href {\doibase
  10.1088/0264-9381/33/3/035010} {\bibfield  {journal} {\bibinfo  {journal}
  {Class. Quant. Grav.}\ }\textbf {\bibinfo {volume} {33}},\ \bibinfo {pages}
  {035010} (\bibinfo {year} {2016})},\ \Eprint
  {http://arxiv.org/abs/1512.02076} {arXiv:1512.02076 [astro-ph.IM]}
  \BibitemShut {NoStop}%
\bibitem [{\citenamefont {Mei}\ \emph {et~al.}(2021)\citenamefont {Mei} \emph
  {et~al.}}]{TianQin:2020hid}%
  \BibitemOpen
  \bibfield  {author} {\bibinfo {author} {\bibfnamefont {J.}~\bibnamefont
  {Mei}} \emph {et~al.} (\bibinfo {collaboration} {TianQin}),\ }\href {\doibase
  10.1093/ptep/ptaa114} {\bibfield  {journal} {\bibinfo  {journal} {PTEP}\
  }\textbf {\bibinfo {volume} {2021}},\ \bibinfo {pages} {05A107} (\bibinfo
  {year} {2021})},\ \Eprint {http://arxiv.org/abs/2008.10332} {arXiv:2008.10332
  [gr-qc]} \BibitemShut {NoStop}%
\bibitem [{\citenamefont {Hu}\ \emph {et~al.}(2017)\citenamefont {Hu},
  \citenamefont {Mei},\ and\ \citenamefont {Luo}}]{Hu:2017yoc}%
  \BibitemOpen
  \bibfield  {author} {\bibinfo {author} {\bibfnamefont {Y.-M.}\ \bibnamefont
  {Hu}}, \bibinfo {author} {\bibfnamefont {J.}~\bibnamefont {Mei}}, \ and\
  \bibinfo {author} {\bibfnamefont {J.}~\bibnamefont {Luo}},\ }\href {\doibase
  10.1093/nsr/nwx115} {\bibfield  {journal} {\bibinfo  {journal} {Natl. Sci.
  Rev.}\ }\textbf {\bibinfo {volume} {4}},\ \bibinfo {pages} {683} (\bibinfo
  {year} {2017})}\BibitemShut {NoStop}%
\bibitem [{\citenamefont {Fan}\ \emph {et~al.}(2020)\citenamefont {Fan},
  \citenamefont {Hu}, \citenamefont {Barausse}, \citenamefont {Sesana},
  \citenamefont {Zhang}, \citenamefont {Zhang}, \citenamefont {Zi},\ and\
  \citenamefont {Mei}}]{Fan:2020zhy}%
  \BibitemOpen
  \bibfield  {author} {\bibinfo {author} {\bibfnamefont {H.-M.}\ \bibnamefont
  {Fan}}, \bibinfo {author} {\bibfnamefont {Y.-M.}\ \bibnamefont {Hu}},
  \bibinfo {author} {\bibfnamefont {E.}~\bibnamefont {Barausse}}, \bibinfo
  {author} {\bibfnamefont {A.}~\bibnamefont {Sesana}}, \bibinfo {author}
  {\bibfnamefont {J.-d.}\ \bibnamefont {Zhang}}, \bibinfo {author}
  {\bibfnamefont {X.}~\bibnamefont {Zhang}}, \bibinfo {author} {\bibfnamefont
  {T.-G.}\ \bibnamefont {Zi}}, \ and\ \bibinfo {author} {\bibfnamefont
  {J.}~\bibnamefont {Mei}},\ }\href {\doibase 10.1103/PhysRevD.102.063016}
  {\bibfield  {journal} {\bibinfo  {journal} {Phys. Rev. D}\ }\textbf {\bibinfo
  {volume} {102}},\ \bibinfo {pages} {063016} (\bibinfo {year} {2020})},\
  \Eprint {http://arxiv.org/abs/2005.08212} {arXiv:2005.08212 [astro-ph.HE]}
  \BibitemShut {NoStop}%
\bibitem [{\citenamefont {Liu}\ \emph {et~al.}(2020)\citenamefont {Liu},
  \citenamefont {Hu}, \citenamefont {Zhang},\ and\ \citenamefont
  {Mei}}]{Liu:2020eko}%
  \BibitemOpen
  \bibfield  {author} {\bibinfo {author} {\bibfnamefont {S.}~\bibnamefont
  {Liu}}, \bibinfo {author} {\bibfnamefont {Y.-M.}\ \bibnamefont {Hu}},
  \bibinfo {author} {\bibfnamefont {J.-d.}\ \bibnamefont {Zhang}}, \ and\
  \bibinfo {author} {\bibfnamefont {J.}~\bibnamefont {Mei}},\ }\href {\doibase
  10.1103/PhysRevD.101.103027} {\bibfield  {journal} {\bibinfo  {journal}
  {Phys. Rev. D}\ }\textbf {\bibinfo {volume} {101}},\ \bibinfo {pages}
  {103027} (\bibinfo {year} {2020})},\ \Eprint
  {http://arxiv.org/abs/2004.14242} {arXiv:2004.14242 [astro-ph.HE]}
  \BibitemShut {NoStop}%
\bibitem [{\citenamefont {Liu}\ \emph {et~al.}(2022)\citenamefont {Liu},
  \citenamefont {Zhu}, \citenamefont {Hu}, \citenamefont {Zhang},\ and\
  \citenamefont {Ji}}]{Liu:2021yoy}%
  \BibitemOpen
  \bibfield  {author} {\bibinfo {author} {\bibfnamefont {S.}~\bibnamefont
  {Liu}}, \bibinfo {author} {\bibfnamefont {L.-G.}\ \bibnamefont {Zhu}},
  \bibinfo {author} {\bibfnamefont {Y.-M.}\ \bibnamefont {Hu}}, \bibinfo
  {author} {\bibfnamefont {J.-d.}\ \bibnamefont {Zhang}}, \ and\ \bibinfo
  {author} {\bibfnamefont {M.-J.}\ \bibnamefont {Ji}},\ }\href {\doibase
  10.1103/PhysRevD.105.023019} {\bibfield  {journal} {\bibinfo  {journal}
  {Phys. Rev. D}\ }\textbf {\bibinfo {volume} {105}},\ \bibinfo {pages}
  {023019} (\bibinfo {year} {2022})},\ \Eprint
  {http://arxiv.org/abs/2110.05248} {arXiv:2110.05248 [astro-ph.HE]}
  \BibitemShut {NoStop}%
\bibitem [{\citenamefont {Huang}\ \emph {et~al.}(2020)\citenamefont {Huang},
  \citenamefont {Hu}, \citenamefont {Korol}, \citenamefont {Li}, \citenamefont
  {Liang}, \citenamefont {Lu}, \citenamefont {Wang}, \citenamefont {Yu},\ and\
  \citenamefont {Mei}}]{Huang:2020rjf}%
  \BibitemOpen
  \bibfield  {author} {\bibinfo {author} {\bibfnamefont {S.-J.}\ \bibnamefont
  {Huang}}, \bibinfo {author} {\bibfnamefont {Y.-M.}\ \bibnamefont {Hu}},
  \bibinfo {author} {\bibfnamefont {V.}~\bibnamefont {Korol}}, \bibinfo
  {author} {\bibfnamefont {P.-C.}\ \bibnamefont {Li}}, \bibinfo {author}
  {\bibfnamefont {Z.-C.}\ \bibnamefont {Liang}}, \bibinfo {author}
  {\bibfnamefont {Y.}~\bibnamefont {Lu}}, \bibinfo {author} {\bibfnamefont
  {H.-T.}\ \bibnamefont {Wang}}, \bibinfo {author} {\bibfnamefont
  {S.}~\bibnamefont {Yu}}, \ and\ \bibinfo {author} {\bibfnamefont
  {J.}~\bibnamefont {Mei}},\ }\href {\doibase 10.1103/PhysRevD.102.063021}
  {\bibfield  {journal} {\bibinfo  {journal} {Phys. Rev. D}\ }\textbf {\bibinfo
  {volume} {102}},\ \bibinfo {pages} {063021} (\bibinfo {year} {2020})},\
  \Eprint {http://arxiv.org/abs/2005.07889} {arXiv:2005.07889 [astro-ph.HE]}
  \BibitemShut {NoStop}%
\bibitem [{\citenamefont {Liang}\ \emph {et~al.}()\citenamefont {Liang},
  \citenamefont {Hu}, \citenamefont {Jiang}, \citenamefont {Cheng},
  \citenamefont {Zhang},\ and\ \citenamefont {Mei}}]{Liang:2021bde}%
  \BibitemOpen
  \bibfield  {author} {\bibinfo {author} {\bibfnamefont {Z.-C.}\ \bibnamefont
  {Liang}}, \bibinfo {author} {\bibfnamefont {Y.-M.}\ \bibnamefont {Hu}},
  \bibinfo {author} {\bibfnamefont {Y.}~\bibnamefont {Jiang}}, \bibinfo
  {author} {\bibfnamefont {J.}~\bibnamefont {Cheng}}, \bibinfo {author}
  {\bibfnamefont {J.-d.}\ \bibnamefont {Zhang}}, \ and\ \bibinfo {author}
  {\bibfnamefont {J.}~\bibnamefont {Mei}},\ }\href@noop {} {\bibfield
  {journal} {\bibinfo  {journal} {arXiv: 2107.08643}\ }}\Eprint
  {http://arxiv.org/abs/2107.08643} {arXiv:2107.08643 [astro-ph.CO]}
  \BibitemShut {NoStop}%
\bibitem [{\citenamefont {Shi}\ \emph {et~al.}(2019)\citenamefont {Shi},
  \citenamefont {Bao}, \citenamefont {Wang}, \citenamefont {Zhang},
  \citenamefont {Hu}, \citenamefont {Sesana}, \citenamefont {Barausse},
  \citenamefont {Mei},\ and\ \citenamefont {Luo}}]{Shi:2019hqa}%
  \BibitemOpen
  \bibfield  {author} {\bibinfo {author} {\bibfnamefont {C.}~\bibnamefont
  {Shi}}, \bibinfo {author} {\bibfnamefont {J.}~\bibnamefont {Bao}}, \bibinfo
  {author} {\bibfnamefont {H.}~\bibnamefont {Wang}}, \bibinfo {author}
  {\bibfnamefont {J.-d.}\ \bibnamefont {Zhang}}, \bibinfo {author}
  {\bibfnamefont {Y.}~\bibnamefont {Hu}}, \bibinfo {author} {\bibfnamefont
  {A.}~\bibnamefont {Sesana}}, \bibinfo {author} {\bibfnamefont
  {E.}~\bibnamefont {Barausse}}, \bibinfo {author} {\bibfnamefont
  {J.}~\bibnamefont {Mei}}, \ and\ \bibinfo {author} {\bibfnamefont
  {J.}~\bibnamefont {Luo}},\ }\href {\doibase 10.1103/PhysRevD.100.044036}
  {\bibfield  {journal} {\bibinfo  {journal} {Phys. Rev. D}\ }\textbf {\bibinfo
  {volume} {100}},\ \bibinfo {pages} {044036} (\bibinfo {year} {2019})},\
  \Eprint {http://arxiv.org/abs/1902.08922} {arXiv:1902.08922 [gr-qc]}
  \BibitemShut {NoStop}%
\bibitem [{\citenamefont {Bao}\ \emph {et~al.}(2019)\citenamefont {Bao},
  \citenamefont {Shi}, \citenamefont {Wang}, \citenamefont {Zhang},
  \citenamefont {Hu}, \citenamefont {Mei},\ and\ \citenamefont
  {Luo}}]{Bao:2019kgt}%
  \BibitemOpen
  \bibfield  {author} {\bibinfo {author} {\bibfnamefont {J.}~\bibnamefont
  {Bao}}, \bibinfo {author} {\bibfnamefont {C.}~\bibnamefont {Shi}}, \bibinfo
  {author} {\bibfnamefont {H.}~\bibnamefont {Wang}}, \bibinfo {author}
  {\bibfnamefont {J.-d.}\ \bibnamefont {Zhang}}, \bibinfo {author}
  {\bibfnamefont {Y.}~\bibnamefont {Hu}}, \bibinfo {author} {\bibfnamefont
  {J.}~\bibnamefont {Mei}}, \ and\ \bibinfo {author} {\bibfnamefont
  {J.}~\bibnamefont {Luo}},\ }\href {\doibase 10.1103/PhysRevD.100.084024}
  {\bibfield  {journal} {\bibinfo  {journal} {Phys. Rev. D}\ }\textbf {\bibinfo
  {volume} {100}},\ \bibinfo {pages} {084024} (\bibinfo {year} {2019})},\
  \Eprint {http://arxiv.org/abs/1905.11674} {arXiv:1905.11674 [gr-qc]}
  \BibitemShut {NoStop}%
\bibitem [{\citenamefont {Zi}\ \emph {et~al.}(2021)\citenamefont {Zi},
  \citenamefont {Zhang}, \citenamefont {Fan}, \citenamefont {Zhang},
  \citenamefont {Hu}, \citenamefont {Shi},\ and\ \citenamefont
  {Mei}}]{Zi:2021pdp}%
  \BibitemOpen
  \bibfield  {author} {\bibinfo {author} {\bibfnamefont {T.-G.}\ \bibnamefont
  {Zi}}, \bibinfo {author} {\bibfnamefont {J.-D.}\ \bibnamefont {Zhang}},
  \bibinfo {author} {\bibfnamefont {H.-M.}\ \bibnamefont {Fan}}, \bibinfo
  {author} {\bibfnamefont {X.-T.}\ \bibnamefont {Zhang}}, \bibinfo {author}
  {\bibfnamefont {Y.-M.}\ \bibnamefont {Hu}}, \bibinfo {author} {\bibfnamefont
  {C.}~\bibnamefont {Shi}}, \ and\ \bibinfo {author} {\bibfnamefont
  {J.}~\bibnamefont {Mei}},\ }\href {\doibase 10.1103/PhysRevD.104.064008}
  {\bibfield  {journal} {\bibinfo  {journal} {Phys. Rev. D}\ }\textbf {\bibinfo
  {volume} {104}},\ \bibinfo {pages} {064008} (\bibinfo {year} {2021})},\
  \Eprint {http://arxiv.org/abs/2104.06047} {arXiv:2104.06047 [gr-qc]}
  \BibitemShut {NoStop}%
\bibitem [{\citenamefont {Zhu}\ \emph {et~al.}(2022{\natexlab{a}})\citenamefont
  {Zhu}, \citenamefont {Xie}, \citenamefont {Hu}, \citenamefont {Liu},
  \citenamefont {Li}, \citenamefont {Napolitano}, \citenamefont {Tang},
  \citenamefont {Zhang},\ and\ \citenamefont {Mei}}]{Zhu:2021bpp}%
  \BibitemOpen
  \bibfield  {author} {\bibinfo {author} {\bibfnamefont {L.-G.}\ \bibnamefont
  {Zhu}}, \bibinfo {author} {\bibfnamefont {L.-H.}\ \bibnamefont {Xie}},
  \bibinfo {author} {\bibfnamefont {Y.-M.}\ \bibnamefont {Hu}}, \bibinfo
  {author} {\bibfnamefont {S.}~\bibnamefont {Liu}}, \bibinfo {author}
  {\bibfnamefont {E.-K.}\ \bibnamefont {Li}}, \bibinfo {author} {\bibfnamefont
  {N.~R.}\ \bibnamefont {Napolitano}}, \bibinfo {author} {\bibfnamefont
  {B.-T.}\ \bibnamefont {Tang}}, \bibinfo {author} {\bibfnamefont {J.-d.}\
  \bibnamefont {Zhang}}, \ and\ \bibinfo {author} {\bibfnamefont
  {J.}~\bibnamefont {Mei}},\ }\href {\doibase 10.1007/s11433-021-1859-9}
  {\bibfield  {journal} {\bibinfo  {journal} {Sci. China Phys. Mech. Astron.}\
  }\textbf {\bibinfo {volume} {65}},\ \bibinfo {pages} {259811} (\bibinfo
  {year} {2022}{\natexlab{a}})},\ \Eprint {http://arxiv.org/abs/2110.05224}
  {arXiv:2110.05224 [astro-ph.CO]} \BibitemShut {NoStop}%
\bibitem [{\citenamefont {Shi}\ \emph {et~al.}(2022)\citenamefont {Shi},
  \citenamefont {Ji}, \citenamefont {Zhang},\ and\ \citenamefont
  {Mei}}]{Shi:2022qno}%
  \BibitemOpen
  \bibfield  {author} {\bibinfo {author} {\bibfnamefont {C.}~\bibnamefont
  {Shi}}, \bibinfo {author} {\bibfnamefont {M.}~\bibnamefont {Ji}}, \bibinfo
  {author} {\bibfnamefont {J.-d.}\ \bibnamefont {Zhang}}, \ and\ \bibinfo
  {author} {\bibfnamefont {J.}~\bibnamefont {Mei}},\ }\href@noop {} {\
  (\bibinfo {year} {2022})},\ \Eprint {http://arxiv.org/abs/2210.13006}
  {arXiv:2210.13006 [gr-qc]} \BibitemShut {NoStop}%
\bibitem [{\citenamefont {Xie}\ \emph {et~al.}(2022)\citenamefont {Xie},
  \citenamefont {Zhang}, \citenamefont {Huang}, \citenamefont {Hu},\ and\
  \citenamefont {Mei}}]{Xie:2022wkx}%
  \BibitemOpen
  \bibfield  {author} {\bibinfo {author} {\bibfnamefont {N.}~\bibnamefont
  {Xie}}, \bibinfo {author} {\bibfnamefont {J.-d.}\ \bibnamefont {Zhang}},
  \bibinfo {author} {\bibfnamefont {S.-J.}\ \bibnamefont {Huang}}, \bibinfo
  {author} {\bibfnamefont {Y.-M.}\ \bibnamefont {Hu}}, \ and\ \bibinfo {author}
  {\bibfnamefont {J.}~\bibnamefont {Mei}},\ }\href@noop {} {\  (\bibinfo {year}
  {2022})},\ \Eprint {http://arxiv.org/abs/2208.10831} {arXiv:2208.10831
  [gr-qc]} \BibitemShut {NoStop}%
\bibitem [{\citenamefont {Wang}\ \emph {et~al.}(2019)\citenamefont {Wang} \emph
  {et~al.}}]{haitian}%
  \BibitemOpen
  \bibfield  {author} {\bibinfo {author} {\bibfnamefont {H.-T.}\ \bibnamefont
  {Wang}} \emph {et~al.},\ }\href {\doibase 10.1103/PhysRevD.100.043003}
  {\bibfield  {journal} {\bibinfo  {journal} {Phys. Rev. D}\ }\textbf {\bibinfo
  {volume} {100}},\ \bibinfo {pages} {043003} (\bibinfo {year} {2019})},\
  \Eprint {http://arxiv.org/abs/1902.04423} {arXiv:1902.04423 [astro-ph.HE]}
  \BibitemShut {NoStop}%
\bibitem [{\citenamefont {Feng}\ \emph {et~al.}(2019)\citenamefont {Feng},
  \citenamefont {Wang}, \citenamefont {Hu}, \citenamefont {Hu},\ and\
  \citenamefont {Wang}}]{Feng:2019wgq}%
  \BibitemOpen
  \bibfield  {author} {\bibinfo {author} {\bibfnamefont {W.-F.}\ \bibnamefont
  {Feng}}, \bibinfo {author} {\bibfnamefont {H.-T.}\ \bibnamefont {Wang}},
  \bibinfo {author} {\bibfnamefont {X.-C.}\ \bibnamefont {Hu}}, \bibinfo
  {author} {\bibfnamefont {Y.-M.}\ \bibnamefont {Hu}}, \ and\ \bibinfo {author}
  {\bibfnamefont {Y.}~\bibnamefont {Wang}},\ }\href {\doibase
  10.1103/PhysRevD.99.123002} {\bibfield  {journal} {\bibinfo  {journal} {Phys.
  Rev. D}\ }\textbf {\bibinfo {volume} {99}},\ \bibinfo {pages} {123002}
  (\bibinfo {year} {2019})},\ \Eprint {http://arxiv.org/abs/1901.02159}
  {arXiv:1901.02159 [astro-ph.IM]} \BibitemShut {NoStop}%
\bibitem [{\citenamefont {Favata}(2009{\natexlab{a}})}]{Favata:2008yd}%
  \BibitemOpen
  \bibfield  {author} {\bibinfo {author} {\bibfnamefont {M.}~\bibnamefont
  {Favata}},\ }\href {\doibase 10.1103/PhysRevD.80.024002} {\bibfield
  {journal} {\bibinfo  {journal} {Phys. Rev. D}\ }\textbf {\bibinfo {volume}
  {80}},\ \bibinfo {pages} {024002} (\bibinfo {year} {2009}{\natexlab{a}})},\
  \Eprint {http://arxiv.org/abs/0812.0069} {arXiv:0812.0069 [gr-qc]}
  \BibitemShut {NoStop}%
\bibitem [{\citenamefont {Favata}(2009{\natexlab{b}})}]{Favata:2009ii}%
  \BibitemOpen
  \bibfield  {author} {\bibinfo {author} {\bibfnamefont {M.}~\bibnamefont
  {Favata}},\ }\href {\doibase 10.1088/0004-637X/696/2/L159} {\bibfield
  {journal} {\bibinfo  {journal} {Astrophys. J. Lett.}\ }\textbf {\bibinfo
  {volume} {696}},\ \bibinfo {pages} {L159} (\bibinfo {year}
  {2009}{\natexlab{b}})},\ \Eprint {http://arxiv.org/abs/0902.3660}
  {arXiv:0902.3660 [astro-ph.SR]} \BibitemShut {NoStop}%
\bibitem [{\citenamefont {Favata}(2010)}]{Favata:2010zu}%
  \BibitemOpen
  \bibfield  {author} {\bibinfo {author} {\bibfnamefont {M.}~\bibnamefont
  {Favata}},\ }\href {\doibase 10.1088/0264-9381/27/8/084036} {\bibfield
  {journal} {\bibinfo  {journal} {Class. Quant. Grav.}\ }\textbf {\bibinfo
  {volume} {27}},\ \bibinfo {pages} {084036} (\bibinfo {year} {2010})},\
  \Eprint {http://arxiv.org/abs/1003.3486} {arXiv:1003.3486 [gr-qc]}
  \BibitemShut {NoStop}%
\bibitem [{\citenamefont {Talbot}\ \emph {et~al.}(2018)\citenamefont {Talbot},
  \citenamefont {Thrane}, \citenamefont {Lasky},\ and\ \citenamefont
  {Lin}}]{Talbot:2018sgr}%
  \BibitemOpen
  \bibfield  {author} {\bibinfo {author} {\bibfnamefont {C.}~\bibnamefont
  {Talbot}}, \bibinfo {author} {\bibfnamefont {E.}~\bibnamefont {Thrane}},
  \bibinfo {author} {\bibfnamefont {P.~D.}\ \bibnamefont {Lasky}}, \ and\
  \bibinfo {author} {\bibfnamefont {F.}~\bibnamefont {Lin}},\ }\href {\doibase
  10.1103/PhysRevD.98.064031} {\bibfield  {journal} {\bibinfo  {journal} {Phys.
  Rev. D}\ }\textbf {\bibinfo {volume} {98}},\ \bibinfo {pages} {064031}
  (\bibinfo {year} {2018})},\ \Eprint {http://arxiv.org/abs/1807.00990}
  {arXiv:1807.00990 [astro-ph.HE]} \BibitemShut {NoStop}%
\bibitem [{\citenamefont {Comp\`ere}\ and\ \citenamefont
  {Fiorucci}(2018)}]{Compere:2018aar}%
  \BibitemOpen
  \bibfield  {author} {\bibinfo {author} {\bibfnamefont {G.}~\bibnamefont
  {Comp\`ere}}\ and\ \bibinfo {author} {\bibfnamefont {A.}~\bibnamefont
  {Fiorucci}},\ }\href@noop {} {\  (\bibinfo {year} {2018})},\ \Eprint
  {http://arxiv.org/abs/1801.07064} {arXiv:1801.07064 [hep-th]} \BibitemShut
  {NoStop}%
\bibitem [{\citenamefont {Barnich}\ and\ \citenamefont
  {Troessaert}(2010{\natexlab{c}})}]{Barnich:2010eb}%
  \BibitemOpen
  \bibfield  {author} {\bibinfo {author} {\bibfnamefont {G.}~\bibnamefont
  {Barnich}}\ and\ \bibinfo {author} {\bibfnamefont {C.}~\bibnamefont
  {Troessaert}},\ }\href {\doibase 10.1007/JHEP05(2010)062} {\bibfield
  {journal} {\bibinfo  {journal} {JHEP}\ }\textbf {\bibinfo {volume} {05}},\
  \bibinfo {pages} {062} (\bibinfo {year} {2010}{\natexlab{c}})},\ \Eprint
  {http://arxiv.org/abs/1001.1541} {arXiv:1001.1541 [hep-th]} \BibitemShut
  {NoStop}%
\bibitem [{sxs()}]{sxs}%
  \BibitemOpen
  \href@noop {} {\enquote {\bibinfo {title} {{SXS Gravitational Waveform
  Database}},}\ }\bibinfo {howpublished}
  {\url{http://www.black-holes.org/waveforms}}\BibitemShut {NoStop}%
\bibitem [{\citenamefont {Boyle}\ \emph {et~al.}(2019)\citenamefont {Boyle}
  \emph {et~al.}}]{Boyle:2019kee}%
  \BibitemOpen
  \bibfield  {author} {\bibinfo {author} {\bibfnamefont {M.}~\bibnamefont
  {Boyle}} \emph {et~al.},\ }\href {\doibase 10.1088/1361-6382/ab34e2}
  {\bibfield  {journal} {\bibinfo  {journal} {Class. Quant. Grav.}\ }\textbf
  {\bibinfo {volume} {36}},\ \bibinfo {pages} {195006} (\bibinfo {year}
  {2019})},\ \Eprint {http://arxiv.org/abs/1904.04831} {arXiv:1904.04831
  [gr-qc]} \BibitemShut {NoStop}%
\bibitem [{\citenamefont {Moxon}\ \emph {et~al.}(2020)\citenamefont {Moxon},
  \citenamefont {Scheel},\ and\ \citenamefont {Teukolsky}}]{Moxon:2020gha}%
  \BibitemOpen
  \bibfield  {author} {\bibinfo {author} {\bibfnamefont {J.}~\bibnamefont
  {Moxon}}, \bibinfo {author} {\bibfnamefont {M.~A.}\ \bibnamefont {Scheel}}, \
  and\ \bibinfo {author} {\bibfnamefont {S.~A.}\ \bibnamefont {Teukolsky}},\
  }\href {\doibase 10.1103/PhysRevD.102.044052} {\bibfield  {journal} {\bibinfo
   {journal} {Phys. Rev. D}\ }\textbf {\bibinfo {volume} {102}},\ \bibinfo
  {pages} {044052} (\bibinfo {year} {2020})},\ \Eprint
  {http://arxiv.org/abs/2007.01339} {arXiv:2007.01339 [gr-qc]} \BibitemShut
  {NoStop}%
\bibitem [{\citenamefont {Ezra}\ and\ \citenamefont {Roger}(1962)}]{roger1}%
  \BibitemOpen
  \bibfield  {author} {\bibinfo {author} {\bibfnamefont {N.}~\bibnamefont
  {Ezra}}\ and\ \bibinfo {author} {\bibfnamefont {P.}~\bibnamefont {Roger}},\
  }\href {\doibase https://doi.org/10.1063/1.1724257} {\bibfield  {journal}
  {\bibinfo  {journal} {J. Math. Phys.}\ }\textbf {\bibinfo {volume} {3}},\
  \bibinfo {pages} {566} (\bibinfo {year} {1962})},\ \Eprint
  {http://arxiv.org/abs/1803.01944} {arXiv:1803.01944 [astro-ph.HE]}
  \BibitemShut {NoStop}%
\bibitem [{\citenamefont {Garc\'\i{}a-Quir\'os}\ \emph
  {et~al.}(2020)\citenamefont {Garc\'\i{}a-Quir\'os}, \citenamefont {Colleoni},
  \citenamefont {Husa}, \citenamefont {Estell\'es}, \citenamefont {Pratten},
  \citenamefont {Ramos-Buades}, \citenamefont {Mateu-Lucena},\ and\
  \citenamefont {Jaume}}]{Garcia-Quiros:2020qpx}%
  \BibitemOpen
  \bibfield  {author} {\bibinfo {author} {\bibfnamefont {C.}~\bibnamefont
  {Garc\'\i{}a-Quir\'os}}, \bibinfo {author} {\bibfnamefont {M.}~\bibnamefont
  {Colleoni}}, \bibinfo {author} {\bibfnamefont {S.}~\bibnamefont {Husa}},
  \bibinfo {author} {\bibfnamefont {H.}~\bibnamefont {Estell\'es}}, \bibinfo
  {author} {\bibfnamefont {G.}~\bibnamefont {Pratten}}, \bibinfo {author}
  {\bibfnamefont {A.}~\bibnamefont {Ramos-Buades}}, \bibinfo {author}
  {\bibfnamefont {M.}~\bibnamefont {Mateu-Lucena}}, \ and\ \bibinfo {author}
  {\bibfnamefont {R.}~\bibnamefont {Jaume}},\ }\href {\doibase
  10.1103/PhysRevD.102.064002} {\bibfield  {journal} {\bibinfo  {journal}
  {Phys. Rev. D}\ }\textbf {\bibinfo {volume} {102}},\ \bibinfo {pages}
  {064002} (\bibinfo {year} {2020})},\ \Eprint
  {http://arxiv.org/abs/2001.10914} {arXiv:2001.10914 [gr-qc]} \BibitemShut
  {NoStop}%
\bibitem [{\citenamefont {Colleoni}\ \emph {et~al.}(2021)\citenamefont
  {Colleoni}, \citenamefont {Mateu-Lucena}, \citenamefont {Estell\'es},
  \citenamefont {Garc\'\i{}a-Quir\'os}, \citenamefont {Keitel}, \citenamefont
  {Pratten}, \citenamefont {Ramos-Buades},\ and\ \citenamefont
  {Husa}}]{Colleoni:2020tgc}%
  \BibitemOpen
  \bibfield  {author} {\bibinfo {author} {\bibfnamefont {M.}~\bibnamefont
  {Colleoni}}, \bibinfo {author} {\bibfnamefont {M.}~\bibnamefont
  {Mateu-Lucena}}, \bibinfo {author} {\bibfnamefont {H.}~\bibnamefont
  {Estell\'es}}, \bibinfo {author} {\bibfnamefont {C.}~\bibnamefont
  {Garc\'\i{}a-Quir\'os}}, \bibinfo {author} {\bibfnamefont {D.}~\bibnamefont
  {Keitel}}, \bibinfo {author} {\bibfnamefont {G.}~\bibnamefont {Pratten}},
  \bibinfo {author} {\bibfnamefont {A.}~\bibnamefont {Ramos-Buades}}, \ and\
  \bibinfo {author} {\bibfnamefont {S.}~\bibnamefont {Husa}},\ }\href {\doibase
  10.1103/PhysRevD.103.024029} {\bibfield  {journal} {\bibinfo  {journal}
  {Phys. Rev. D}\ }\textbf {\bibinfo {volume} {103}},\ \bibinfo {pages}
  {024029} (\bibinfo {year} {2021})},\ \Eprint
  {http://arxiv.org/abs/2010.05830} {arXiv:2010.05830 [gr-qc]} \BibitemShut
  {NoStop}%
\bibitem [{\citenamefont {Pratten}\ \emph {et~al.}(2020)\citenamefont
  {Pratten}, \citenamefont {Husa}, \citenamefont {Garcia-Quiros}, \citenamefont
  {Colleoni}, \citenamefont {Ramos-Buades}, \citenamefont {Estelles},\ and\
  \citenamefont {Jaume}}]{Pratten:2020fqn}%
  \BibitemOpen
  \bibfield  {author} {\bibinfo {author} {\bibfnamefont {G.}~\bibnamefont
  {Pratten}}, \bibinfo {author} {\bibfnamefont {S.}~\bibnamefont {Husa}},
  \bibinfo {author} {\bibfnamefont {C.}~\bibnamefont {Garcia-Quiros}}, \bibinfo
  {author} {\bibfnamefont {M.}~\bibnamefont {Colleoni}}, \bibinfo {author}
  {\bibfnamefont {A.}~\bibnamefont {Ramos-Buades}}, \bibinfo {author}
  {\bibfnamefont {H.}~\bibnamefont {Estelles}}, \ and\ \bibinfo {author}
  {\bibfnamefont {R.}~\bibnamefont {Jaume}},\ }\href {\doibase
  10.1103/PhysRevD.102.064001} {\bibfield  {journal} {\bibinfo  {journal}
  {Phys. Rev. D}\ }\textbf {\bibinfo {volume} {102}},\ \bibinfo {pages}
  {064001} (\bibinfo {year} {2020})},\ \Eprint
  {http://arxiv.org/abs/2001.11412} {arXiv:2001.11412 [gr-qc]} \BibitemShut
  {NoStop}%
\bibitem [{\citenamefont {{LIGO Scientific Collaboration}}(2018)}]{LALSuite}%
  \BibitemOpen
  \bibfield  {author} {\bibinfo {author} {\bibnamefont {{LIGO Scientific
  Collaboration}}},\ }\href {\doibase 10.7935/GT1W-FZ16} {\enquote {\bibinfo
  {title} {{LIGO} {A}lgorithm {L}ibrary - {LALS}uite},}\ }\bibinfo
  {howpublished} {free software (GPL)} (\bibinfo {year} {2018})\BibitemShut
  {NoStop}%
\bibitem [{\citenamefont {McKechan}\ \emph {et~al.}(2010)\citenamefont
  {McKechan}, \citenamefont {Robinson},\ and\ \citenamefont
  {Sathyaprakash}}]{McKechan:2010kp}%
  \BibitemOpen
  \bibfield  {author} {\bibinfo {author} {\bibfnamefont {D.~J.~A.}\
  \bibnamefont {McKechan}}, \bibinfo {author} {\bibfnamefont {C.}~\bibnamefont
  {Robinson}}, \ and\ \bibinfo {author} {\bibfnamefont {B.~S.}\ \bibnamefont
  {Sathyaprakash}},\ }\href {\doibase 10.1088/0264-9381/27/8/084020} {\bibfield
   {journal} {\bibinfo  {journal} {Class. Quant. Grav.}\ }\textbf {\bibinfo
  {volume} {27}},\ \bibinfo {pages} {084020} (\bibinfo {year} {2010})},\
  \Eprint {http://arxiv.org/abs/1003.2939} {arXiv:1003.2939 [gr-qc]}
  \BibitemShut {NoStop}%
\bibitem [{\citenamefont {Maggiore}(2007)}]{Maggiore:2007ulw}%
  \BibitemOpen
  \bibfield  {author} {\bibinfo {author} {\bibfnamefont {M.}~\bibnamefont
  {Maggiore}},\ }\href@noop {} {\emph {\bibinfo {title} {{Gravitational Waves.
  Vol. 1: Theory and Experiments}}}},\ Oxford Master Series in Physics\
  (\bibinfo  {publisher} {Oxford University Press},\ \bibinfo {year}
  {2007})\BibitemShut {NoStop}%
\bibitem [{\citenamefont {Cornish}\ and\ \citenamefont
  {Robson}(2017)}]{Cornish:2017vip}%
  \BibitemOpen
  \bibfield  {author} {\bibinfo {author} {\bibfnamefont {N.}~\bibnamefont
  {Cornish}}\ and\ \bibinfo {author} {\bibfnamefont {T.}~\bibnamefont
  {Robson}},\ }\href {\doibase 10.1088/1742-6596/840/1/012024} {\bibfield
  {journal} {\bibinfo  {journal} {J. Phys. Conf. Ser.}\ }\textbf {\bibinfo
  {volume} {840}},\ \bibinfo {pages} {012024} (\bibinfo {year} {2017})},\
  \Eprint {http://arxiv.org/abs/1703.09858} {arXiv:1703.09858 [astro-ph.IM]}
  \BibitemShut {NoStop}%
\bibitem [{\citenamefont {Robson}\ \emph {et~al.}(2019)\citenamefont {Robson},
  \citenamefont {Cornish},\ and\ \citenamefont {Liu}}]{lisasen}%
  \BibitemOpen
  \bibfield  {author} {\bibinfo {author} {\bibfnamefont {T.}~\bibnamefont
  {Robson}}, \bibinfo {author} {\bibfnamefont {N.~J.}\ \bibnamefont {Cornish}},
  \ and\ \bibinfo {author} {\bibfnamefont {C.}~\bibnamefont {Liu}},\ }\href
  {\doibase 10.1088/1361-6382/ab1101} {\bibfield  {journal} {\bibinfo
  {journal} {Class. Quant. Grav.}\ }\textbf {\bibinfo {volume} {36}},\ \bibinfo
  {pages} {105011} (\bibinfo {year} {2019})},\ \Eprint
  {http://arxiv.org/abs/1803.01944} {arXiv:1803.01944 [astro-ph.HE]}
  \BibitemShut {NoStop}%
\bibitem [{\citenamefont {Kennefick}(1994)}]{Kennefick:1994nw}%
  \BibitemOpen
  \bibfield  {author} {\bibinfo {author} {\bibfnamefont {D.}~\bibnamefont
  {Kennefick}},\ }\href {\doibase 10.1103/PhysRevD.50.3587} {\bibfield
  {journal} {\bibinfo  {journal} {Phys. Rev. D}\ }\textbf {\bibinfo {volume}
  {50}},\ \bibinfo {pages} {3587} (\bibinfo {year} {1994})}\BibitemShut
  {NoStop}%
\bibitem [{\citenamefont {Johnson}\ \emph {et~al.}(2019)\citenamefont
  {Johnson}, \citenamefont {Kapadia}, \citenamefont {Osborne}, \citenamefont
  {Hixon},\ and\ \citenamefont {Kennefick}}]{Johnson:2018xly}%
  \BibitemOpen
  \bibfield  {author} {\bibinfo {author} {\bibfnamefont {A.~D.}\ \bibnamefont
  {Johnson}}, \bibinfo {author} {\bibfnamefont {S.~J.}\ \bibnamefont
  {Kapadia}}, \bibinfo {author} {\bibfnamefont {A.}~\bibnamefont {Osborne}},
  \bibinfo {author} {\bibfnamefont {A.}~\bibnamefont {Hixon}}, \ and\ \bibinfo
  {author} {\bibfnamefont {D.}~\bibnamefont {Kennefick}},\ }\href {\doibase
  10.1103/PhysRevD.99.044045} {\bibfield  {journal} {\bibinfo  {journal} {Phys.
  Rev. D}\ }\textbf {\bibinfo {volume} {99}},\ \bibinfo {pages} {044045}
  (\bibinfo {year} {2019})},\ \Eprint {http://arxiv.org/abs/1810.09563}
  {arXiv:1810.09563 [gr-qc]} \BibitemShut {NoStop}%
\bibitem [{\citenamefont {Reisswig}\ \emph {et~al.}(2009)\citenamefont
  {Reisswig}, \citenamefont {Husa}, \citenamefont {Rezzolla}, \citenamefont
  {Dorband}, \citenamefont {Pollney},\ and\ \citenamefont
  {Seiler}}]{Reisswig:2009vc}%
  \BibitemOpen
  \bibfield  {author} {\bibinfo {author} {\bibfnamefont {C.}~\bibnamefont
  {Reisswig}}, \bibinfo {author} {\bibfnamefont {S.}~\bibnamefont {Husa}},
  \bibinfo {author} {\bibfnamefont {L.}~\bibnamefont {Rezzolla}}, \bibinfo
  {author} {\bibfnamefont {E.~N.}\ \bibnamefont {Dorband}}, \bibinfo {author}
  {\bibfnamefont {D.}~\bibnamefont {Pollney}}, \ and\ \bibinfo {author}
  {\bibfnamefont {J.}~\bibnamefont {Seiler}},\ }\href {\doibase
  10.1103/PhysRevD.80.124026} {\bibfield  {journal} {\bibinfo  {journal} {Phys.
  Rev. D}\ }\textbf {\bibinfo {volume} {80}},\ \bibinfo {pages} {124026}
  (\bibinfo {year} {2009})},\ \Eprint {http://arxiv.org/abs/0907.0462}
  {arXiv:0907.0462 [gr-qc]} \BibitemShut {NoStop}%
\bibitem [{\citenamefont {Lousto}\ \emph {et~al.}(2010)\citenamefont {Lousto},
  \citenamefont {Campanelli}, \citenamefont {Zlochower},\ and\ \citenamefont
  {Nakano}}]{Lousto:2009mf}%
  \BibitemOpen
  \bibfield  {author} {\bibinfo {author} {\bibfnamefont {C.~O.}\ \bibnamefont
  {Lousto}}, \bibinfo {author} {\bibfnamefont {M.}~\bibnamefont {Campanelli}},
  \bibinfo {author} {\bibfnamefont {Y.}~\bibnamefont {Zlochower}}, \ and\
  \bibinfo {author} {\bibfnamefont {H.}~\bibnamefont {Nakano}},\ }\href
  {\doibase 10.1088/0264-9381/27/11/114006} {\bibfield  {journal} {\bibinfo
  {journal} {Class. Quant. Grav.}\ }\textbf {\bibinfo {volume} {27}},\ \bibinfo
  {pages} {114006} (\bibinfo {year} {2010})},\ \Eprint
  {http://arxiv.org/abs/0904.3541} {arXiv:0904.3541 [gr-qc]} \BibitemShut
  {NoStop}%
\bibitem [{\citenamefont {Ajith}\ \emph {et~al.}(2011)\citenamefont {Ajith}
  \emph {et~al.}}]{Ajith:2009bn}%
  \BibitemOpen
  \bibfield  {author} {\bibinfo {author} {\bibfnamefont {P.}~\bibnamefont
  {Ajith}} \emph {et~al.},\ }\href {\doibase 10.1103/PhysRevLett.106.241101}
  {\bibfield  {journal} {\bibinfo  {journal} {Phys. Rev. Lett.}\ }\textbf
  {\bibinfo {volume} {106}},\ \bibinfo {pages} {241101} (\bibinfo {year}
  {2011})},\ \Eprint {http://arxiv.org/abs/0909.2867} {arXiv:0909.2867 [gr-qc]}
  \BibitemShut {NoStop}%
\bibitem [{\citenamefont {Santamaria}\ \emph {et~al.}(2010)\citenamefont
  {Santamaria} \emph {et~al.}}]{Santamaria:2010yb}%
  \BibitemOpen
  \bibfield  {author} {\bibinfo {author} {\bibfnamefont {L.}~\bibnamefont
  {Santamaria}} \emph {et~al.},\ }\href {\doibase 10.1103/PhysRevD.82.064016}
  {\bibfield  {journal} {\bibinfo  {journal} {Phys. Rev. D}\ }\textbf {\bibinfo
  {volume} {82}},\ \bibinfo {pages} {064016} (\bibinfo {year} {2010})},\
  \Eprint {http://arxiv.org/abs/1005.3306} {arXiv:1005.3306 [gr-qc]}
  \BibitemShut {NoStop}%
\bibitem [{\citenamefont {Barausse}(2012)}]{Barausse:2012fy}%
  \BibitemOpen
  \bibfield  {author} {\bibinfo {author} {\bibfnamefont {E.}~\bibnamefont
  {Barausse}},\ }\href {\doibase 10.1111/j.1365-2966.2012.21057.x} {\bibfield
  {journal} {\bibinfo  {journal} {Mon. Not. Roy. Astron. Soc.}\ }\textbf
  {\bibinfo {volume} {423}},\ \bibinfo {pages} {2533} (\bibinfo {year}
  {2012})},\ \Eprint {http://arxiv.org/abs/1201.5888} {arXiv:1201.5888
  [astro-ph.CO]} \BibitemShut {NoStop}%
\bibitem [{\citenamefont {Sesana}\ \emph {et~al.}(2014)\citenamefont {Sesana},
  \citenamefont {Barausse}, \citenamefont {Dotti},\ and\ \citenamefont
  {Rossi}}]{Sesana:2014bea}%
  \BibitemOpen
  \bibfield  {author} {\bibinfo {author} {\bibfnamefont {A.}~\bibnamefont
  {Sesana}}, \bibinfo {author} {\bibfnamefont {E.}~\bibnamefont {Barausse}},
  \bibinfo {author} {\bibfnamefont {M.}~\bibnamefont {Dotti}}, \ and\ \bibinfo
  {author} {\bibfnamefont {E.~M.}\ \bibnamefont {Rossi}},\ }\href {\doibase
  10.1088/0004-637X/794/2/104} {\bibfield  {journal} {\bibinfo  {journal}
  {Astrophys. J.}\ }\textbf {\bibinfo {volume} {794}},\ \bibinfo {pages} {104}
  (\bibinfo {year} {2014})},\ \Eprint {http://arxiv.org/abs/1402.7088}
  {arXiv:1402.7088 [astro-ph.CO]} \BibitemShut {NoStop}%
\bibitem [{\citenamefont {Antonini}\ \emph {et~al.}(2015)\citenamefont
  {Antonini}, \citenamefont {Barausse},\ and\ \citenamefont
  {Silk}}]{Antonini:2015sza}%
  \BibitemOpen
  \bibfield  {author} {\bibinfo {author} {\bibfnamefont {F.}~\bibnamefont
  {Antonini}}, \bibinfo {author} {\bibfnamefont {E.}~\bibnamefont {Barausse}},
  \ and\ \bibinfo {author} {\bibfnamefont {J.}~\bibnamefont {Silk}},\ }\href
  {\doibase 10.1088/0004-637X/812/1/72} {\bibfield  {journal} {\bibinfo
  {journal} {Astrophys. J.}\ }\textbf {\bibinfo {volume} {812}},\ \bibinfo
  {pages} {72} (\bibinfo {year} {2015})},\ \Eprint
  {http://arxiv.org/abs/1506.02050} {arXiv:1506.02050 [astro-ph.GA]}
  \BibitemShut {NoStop}%
\bibitem [{\citenamefont {Madau}\ and\ \citenamefont
  {Rees}(2001)}]{Madau:2001sc}%
  \BibitemOpen
  \bibfield  {author} {\bibinfo {author} {\bibfnamefont {P.}~\bibnamefont
  {Madau}}\ and\ \bibinfo {author} {\bibfnamefont {M.~J.}\ \bibnamefont
  {Rees}},\ }\href {\doibase 10.1086/319848} {\bibfield  {journal} {\bibinfo
  {journal} {Astrophys. J. Lett.}\ }\textbf {\bibinfo {volume} {551}},\
  \bibinfo {pages} {L27} (\bibinfo {year} {2001})},\ \Eprint
  {http://arxiv.org/abs/astro-ph/0101223} {arXiv:astro-ph/0101223} \BibitemShut
  {NoStop}%
\bibitem [{\citenamefont {Bromm}\ and\ \citenamefont
  {Loeb}(2003)}]{Bromm:2002hb}%
  \BibitemOpen
  \bibfield  {author} {\bibinfo {author} {\bibfnamefont {V.}~\bibnamefont
  {Bromm}}\ and\ \bibinfo {author} {\bibfnamefont {A.}~\bibnamefont {Loeb}},\
  }\href {\doibase 10.1086/377529} {\bibfield  {journal} {\bibinfo  {journal}
  {Astrophys. J.}\ }\textbf {\bibinfo {volume} {596}},\ \bibinfo {pages} {34}
  (\bibinfo {year} {2003})},\ \Eprint {http://arxiv.org/abs/astro-ph/0212400}
  {arXiv:astro-ph/0212400} \BibitemShut {NoStop}%
\bibitem [{\citenamefont {Begelman}\ \emph {et~al.}(2006)\citenamefont
  {Begelman}, \citenamefont {Volonteri},\ and\ \citenamefont
  {Rees}}]{Begelman:2006db}%
  \BibitemOpen
  \bibfield  {author} {\bibinfo {author} {\bibfnamefont {M.~C.}\ \bibnamefont
  {Begelman}}, \bibinfo {author} {\bibfnamefont {M.}~\bibnamefont {Volonteri}},
  \ and\ \bibinfo {author} {\bibfnamefont {M.~J.}\ \bibnamefont {Rees}},\
  }\href {\doibase 10.1111/j.1365-2966.2006.10467.x} {\bibfield  {journal}
  {\bibinfo  {journal} {Mon. Not. Roy. Astron. Soc.}\ }\textbf {\bibinfo
  {volume} {370}},\ \bibinfo {pages} {289} (\bibinfo {year} {2006})},\ \Eprint
  {http://arxiv.org/abs/astro-ph/0602363} {arXiv:astro-ph/0602363} \BibitemShut
  {NoStop}%
\bibitem [{\citenamefont {Lodato}\ and\ \citenamefont
  {Natarajan}(2006)}]{Lodato:2006hw}%
  \BibitemOpen
  \bibfield  {author} {\bibinfo {author} {\bibfnamefont {G.}~\bibnamefont
  {Lodato}}\ and\ \bibinfo {author} {\bibfnamefont {P.}~\bibnamefont
  {Natarajan}},\ }\href {\doibase 10.1111/j.1365-2966.2006.10801.x} {\bibfield
  {journal} {\bibinfo  {journal} {Mon. Not. Roy. Astron. Soc.}\ }\textbf
  {\bibinfo {volume} {371}},\ \bibinfo {pages} {1813} (\bibinfo {year}
  {2006})},\ \Eprint {http://arxiv.org/abs/astro-ph/0606159}
  {arXiv:astro-ph/0606159} \BibitemShut {NoStop}%
\bibitem [{\citenamefont {Zhu}\ \emph {et~al.}(2022{\natexlab{b}})\citenamefont
  {Zhu}, \citenamefont {Hu}, \citenamefont {Wang}, \citenamefont {Zhang},
  \citenamefont {Li}, \citenamefont {Hendry},\ and\ \citenamefont
  {Mei}}]{Zhu:2021aat}%
  \BibitemOpen
  \bibfield  {author} {\bibinfo {author} {\bibfnamefont {L.-G.}\ \bibnamefont
  {Zhu}}, \bibinfo {author} {\bibfnamefont {Y.-M.}\ \bibnamefont {Hu}},
  \bibinfo {author} {\bibfnamefont {H.-T.}\ \bibnamefont {Wang}}, \bibinfo
  {author} {\bibfnamefont {J.-d.}\ \bibnamefont {Zhang}}, \bibinfo {author}
  {\bibfnamefont {X.-D.}\ \bibnamefont {Li}}, \bibinfo {author} {\bibfnamefont
  {M.}~\bibnamefont {Hendry}}, \ and\ \bibinfo {author} {\bibfnamefont
  {J.}~\bibnamefont {Mei}},\ }\href {\doibase 10.1103/PhysRevResearch.4.013247}
  {\bibfield  {journal} {\bibinfo  {journal} {Phys. Rev. Res.}\ }\textbf
  {\bibinfo {volume} {4}},\ \bibinfo {pages} {013247} (\bibinfo {year}
  {2022}{\natexlab{b}})},\ \Eprint {http://arxiv.org/abs/2104.11956}
  {arXiv:2104.11956 [astro-ph.CO]} \BibitemShut {NoStop}%
\bibitem [{\citenamefont {Vallisneri}(2008)}]{Vallisneri:2007ev}%
  \BibitemOpen
  \bibfield  {author} {\bibinfo {author} {\bibfnamefont {M.}~\bibnamefont
  {Vallisneri}},\ }\href {\doibase 10.1103/PhysRevD.77.042001} {\bibfield
  {journal} {\bibinfo  {journal} {Phys. Rev. D}\ }\textbf {\bibinfo {volume}
  {77}},\ \bibinfo {pages} {042001} (\bibinfo {year} {2008})},\ \Eprint
  {http://arxiv.org/abs/gr-qc/0703086} {arXiv:gr-qc/0703086} \BibitemShut
  {NoStop}%
\bibitem [{\citenamefont {Chatziioannou}\ \emph {et~al.}(2017)\citenamefont
  {Chatziioannou}, \citenamefont {Klein}, \citenamefont {Yunes},\ and\
  \citenamefont {Cornish}}]{Chatziioannou:2017tdw}%
  \BibitemOpen
  \bibfield  {author} {\bibinfo {author} {\bibfnamefont {K.}~\bibnamefont
  {Chatziioannou}}, \bibinfo {author} {\bibfnamefont {A.}~\bibnamefont
  {Klein}}, \bibinfo {author} {\bibfnamefont {N.}~\bibnamefont {Yunes}}, \ and\
  \bibinfo {author} {\bibfnamefont {N.}~\bibnamefont {Cornish}},\ }\href
  {\doibase 10.1103/PhysRevD.95.104004} {\bibfield  {journal} {\bibinfo
  {journal} {Phys. Rev. D}\ }\textbf {\bibinfo {volume} {95}},\ \bibinfo
  {pages} {104004} (\bibinfo {year} {2017})},\ \Eprint
  {http://arxiv.org/abs/1703.03967} {arXiv:1703.03967 [gr-qc]} \BibitemShut
  {NoStop}%
\bibitem [{\citenamefont {Mangiagli}\ \emph {et~al.}(2019)\citenamefont
  {Mangiagli}, \citenamefont {Klein}, \citenamefont {Sesana}, \citenamefont
  {Barausse},\ and\ \citenamefont {Colpi}}]{Mangiagli:2018kpu}%
  \BibitemOpen
  \bibfield  {author} {\bibinfo {author} {\bibfnamefont {A.}~\bibnamefont
  {Mangiagli}}, \bibinfo {author} {\bibfnamefont {A.}~\bibnamefont {Klein}},
  \bibinfo {author} {\bibfnamefont {A.}~\bibnamefont {Sesana}}, \bibinfo
  {author} {\bibfnamefont {E.}~\bibnamefont {Barausse}}, \ and\ \bibinfo
  {author} {\bibfnamefont {M.}~\bibnamefont {Colpi}},\ }\href {\doibase
  10.1103/PhysRevD.99.064056} {\bibfield  {journal} {\bibinfo  {journal} {Phys.
  Rev. D}\ }\textbf {\bibinfo {volume} {99}},\ \bibinfo {pages} {064056}
  (\bibinfo {year} {2019})},\ \Eprint {http://arxiv.org/abs/1811.01805}
  {arXiv:1811.01805 [gr-qc]} \BibitemShut {NoStop}%
\bibitem [{\citenamefont {Baird}\ \emph {et~al.}(2013)\citenamefont {Baird},
  \citenamefont {Fairhurst}, \citenamefont {Hannam},\ and\ \citenamefont
  {Murphy}}]{Baird:2012cu}%
  \BibitemOpen
  \bibfield  {author} {\bibinfo {author} {\bibfnamefont {E.}~\bibnamefont
  {Baird}}, \bibinfo {author} {\bibfnamefont {S.}~\bibnamefont {Fairhurst}},
  \bibinfo {author} {\bibfnamefont {M.}~\bibnamefont {Hannam}}, \ and\ \bibinfo
  {author} {\bibfnamefont {P.}~\bibnamefont {Murphy}},\ }\href {\doibase
  10.1103/PhysRevD.87.024035} {\bibfield  {journal} {\bibinfo  {journal} {Phys.
  Rev. D}\ }\textbf {\bibinfo {volume} {87}},\ \bibinfo {pages} {024035}
  (\bibinfo {year} {2013})},\ \Eprint {http://arxiv.org/abs/1211.0546}
  {arXiv:1211.0546 [gr-qc]} \BibitemShut {NoStop}%
\bibitem [{\citenamefont {Boyle}(2020)}]{sxsgit}%
  \BibitemOpen
  \bibfield  {author} {\bibinfo {author} {\bibfnamefont {M.}~\bibnamefont
  {Boyle}},\ }\href {\doibase 10.5281/zenodo.4361521} {\enquote {\bibinfo
  {title} {sxs-collaboration/sxs: Release v2020.12.0},}\ } (\bibinfo {year}
  {2020})\BibitemShut {NoStop}%
\bibitem [{\citenamefont {van~der Walt}\ \emph {et~al.}(2011)\citenamefont
  {van~der Walt}, \citenamefont {Colbert},\ and\ \citenamefont
  {Varoquaux}}]{vanderWalt:2011bqk}%
  \BibitemOpen
  \bibfield  {author} {\bibinfo {author} {\bibfnamefont {S.}~\bibnamefont
  {van~der Walt}}, \bibinfo {author} {\bibfnamefont {S.~C.}\ \bibnamefont
  {Colbert}}, \ and\ \bibinfo {author} {\bibfnamefont {G.}~\bibnamefont
  {Varoquaux}},\ }\href {\doibase 10.1109/MCSE.2011.37} {\bibfield  {journal}
  {\bibinfo  {journal} {Comput. Sci. Eng.}\ }\textbf {\bibinfo {volume} {13}},\
  \bibinfo {pages} {22} (\bibinfo {year} {2011})},\ \Eprint
  {http://arxiv.org/abs/1102.1523} {arXiv:1102.1523 [cs.MS]} \BibitemShut
  {NoStop}%
\bibitem [{\citenamefont {Virtanen}\ \emph {et~al.}(2020)\citenamefont
  {Virtanen} \emph {et~al.}}]{Virtanen:2019joe}%
  \BibitemOpen
  \bibfield  {author} {\bibinfo {author} {\bibfnamefont {P.}~\bibnamefont
  {Virtanen}} \emph {et~al.},\ }\href {\doibase 10.1038/s41592-019-0686-2}
  {\bibfield  {journal} {\bibinfo  {journal} {Nature Meth.}\ }\textbf {\bibinfo
  {volume} {17}},\ \bibinfo {pages} {261} (\bibinfo {year} {2020})},\ \Eprint
  {http://arxiv.org/abs/1907.10121} {arXiv:1907.10121 [cs.MS]} \BibitemShut
  {NoStop}%
\bibitem [{\citenamefont {Hunter}(2007)}]{Hunter:2007ouj}%
  \BibitemOpen
  \bibfield  {author} {\bibinfo {author} {\bibfnamefont {J.~D.}\ \bibnamefont
  {Hunter}},\ }\href {\doibase 10.1109/MCSE.2007.55} {\bibfield  {journal}
  {\bibinfo  {journal} {Comput. Sci. Eng.}\ }\textbf {\bibinfo {volume} {9}},\
  \bibinfo {pages} {90} (\bibinfo {year} {2007})}\BibitemShut {NoStop}%
\end{thebibliography}%

\end{document}